\documentclass[aps,showpacs,amsfonts,amsmath, amssymb,twocolumn,floatfix,superscriptaddress]{revtex4}

\usepackage[final]{graphicx}
\usepackage{color} 
\usepackage{blindtext}
\usepackage{epstopdf}
\usepackage{units}
\usepackage{braket}

\newcommand{\vct}[1]{\mathbf{#1}}

\newcommand{\G}{\mathbb{G}} 
\newcommand{\T}{\mathbb{T}} 

\renewcommand\Re{\operatorname{Re}}
\renewcommand\Im{\operatorname{Im}}
\newcommand\Tr{{\rm Tr}}

\newcommand{\be}{\begin{equation}}
\newcommand{\ee}{\end{equation}}

\allowdisplaybreaks

\DeclareSymbolFont{bbgreek}{U}{bbold}{m}{n}
\DeclareMathSymbol{\bbmu}{\mathbb}{bbgreek}{'26}
\DeclareMathSymbol{\bbeps}{\mathbb}{bbgreek}{'17}

\graphicspath{{figures/}}

\begin{document}
\title{Many body heat radiation and heat transfer in the presence of a non-absorbing background medium}

\date{\today}

\author{Boris M\"uller}
\affiliation{4th Institute for Theoretical Physics, Universit\"at Stuttgart, Germany}
\affiliation{Max Planck Institute for Intelligent Systems, 70569 Stuttgart, Germany}

\author{Roberta Incardone}
\affiliation{4th Institute for Theoretical Physics, Universit\"at Stuttgart, Germany}
\affiliation{Max Planck Institute for Intelligent Systems, 70569 Stuttgart, Germany}

\author{Mauro Antezza}
\affiliation{Laboratoire Charles Coulomb (L2C), UMR 5221 CNRS-Universit\'e de Montpellier, F-34095 Montpellier, France}
\affiliation{Institut Universitaire de France, 1 rue Descartes, F-75231 Paris Cedex 05, France}

\author{Thorsten Emig}
\affiliation{MultiScale Materials Science
for Energy and Environment, Joint MIT-CNRS Laboratory (UMI 3466),
Massachusetts Institute of Technology,  Cambridge, Massachusetts 02139, USA}
\affiliation{Laboratoire de Physique
Th\'eorique et Mod\`eles Statistiques, CNRS UMR 8626,
Universit\'e Paris-Saclay, 91405 Orsay cedex, France}

\author{Matthias Kr\"uger}
\affiliation{4th Institute for Theoretical Physics, Universit\"at Stuttgart, Germany}
\affiliation{Max Planck Institute for Intelligent Systems, 70569 Stuttgart, Germany}

\begin{abstract}
Heat radiation and near-field radiative heat transfer can be strongly manipulated by adjusting geometrical shapes, optical properties, or the relative positions of the objects involved. Typically these objects are considered as embedded in vacuum. By applying the methods of fluctuational electrodynamics, we derive general closed-form expressions for heat radiation and heat transfer in a system of $N$ arbitrary objects embedded in a passive non-absorbing background medium. Taking into account the principle of reciprocity, we explicitly prove the symmetry and positivity of transfer in any such system. Regarding applications, we find that the heat radiation of a sphere as well as the heat transfer between two parallel plates is strongly enhanced by the presence of a background medium. Regarding near- and far-field transfer through a gas like air, we show that a microscopic model (based on gas particles) and a  macroscopic model (using a dielectric contrast) yield identical results. We also compare the radiative transfer through a medium like air and the energy transfer found from kinetic gas theory.
\end{abstract}

\maketitle

\nopagebreak

\section{Introduction}\label{sec:Introduction}
The derivation of Planck's law of thermal radiation more than a century ago set off a thorough and still ongoing development in the field of heat radiation and radiative heat transfer between objects in thermal non-equilibrium \cite{planck1901law}. 
Over the years, and following the seminal works by Rytov \cite{rytov1958correlation,rytov1989elements}, considerable physical insight has been gained after conceptually relating these phenomena to the presence of charge and current fluctuations inside the objects, or to the fluctuating electromagnetic field in such systems. The Rytov theory, based on the fluctuation-dissipation theorem, relates the electromagnetic field radiated by an object at a given temperature to its sources, i.e.\ to the fluctuating electric currents inside it. It allows us to have a clear physical intuition of several phenomena, like the Casimir-Lifshitz dispersion forces \cite{casimir1948attraction,casimir1948influence,dzyaloshinskii1961the} occurring between any polarizable objects. These forces, in the Rytov's spirit, have been recently extended to non-equilibrium systems \cite{henkel2002radiation,antezza2005new,antezza2008casimir,bimonte2009scattering,messina2011casimir,krueger2011non,messina2011scattering,kruger2012trace,golyk2012casimir,narayanaswamy2014a,messina2014three,mueller2016anisotropic,bimonte2016non}, for which lately experimental tests have been done using trapped atomic Bose-Einstein condensates \cite{antezza2004effect,obrecht2007measurement}, and have been proposed in other systems \cite{bimonte2015observing}.
Another important outcome of the Rytov theory is the development of a more general framework for heat radiation and for near-field heat transfer \cite{messina2011casimir,kruger2011nonequilibrium}.

While Planck's theory of black body radiation offers a precise theoretical description for the heat transfer on macroscopic length scales, it was found by Polder and Van Hove in the early 1970s that the situation on the submicron or even nanoscale can be fundamentally different \cite{polder1971theory}. The two theoreticians were prompted to reinvestigate the theory of radiative heat transfer between closely spaced bodies after being confronted with new experimental measurement results for the heat transfer between two chromium layers at the time \cite{hargreaves1969anomalous}. In their famous work, they presented a general formalism based on the fluctuation-dissipation theorem for the heat transfer between macroscopic planar bodies of arbitrary dielectric properties. Their ground-breaking results for the heat transfer across a vacuum gap revealed a strong increase of many orders of magnitude for diminishing the gap width due to evanescent wave contribution in good agreement to experimental data. 

In recent years, the field of heat radiation and heat transfer has regained considerable interest due to significant progress on the theoretical side, including improved general formalisms \cite{volokitin2001radiative,volokitin2007nearfield,bimonte2009scattering,messina2011casimir,kruger2011nonequilibrium,messina2011scattering,kruger2012trace,narayanaswamy2014a,messina2014three,bimonte2016non} and new powerful numerical methods \cite{rodriguez2011frequency,mccauley2012modeling,rodriguez2013fluctuating} also for systems with temperature gradients \cite{polimeridis2015fluctuating}. The common concept underlying all these works is the fluctuation-dissipation theorem \cite{eckhardt1984macroscopic}, which is used to describe the correlations of the fluctuating electromagnetic field of the radiating bodies in thermal non-equilibrium. Despite the generality of the available theories, a closed-form expression for the heat transfer between three objects in vacuum has only recently been given \cite{messina2014three}, and has already provided several interesting applications to heat transfer amplification \cite{messina2012three} and guiding \cite{messina2016hyperbolic}. Besides the advances in general formalism, various applications have been investigated in recent non-equilibrium studies \cite{chapuis2008effects,biehs2011modulation,lussange2012radiative,guerout2012enhanced,incardone2014heat,zhu2014near,guo2014fluctuational,chen2015heat,chen2016near}. On the experimental side, the development of high precision measurement devices has verified many theoretical predictions concerning radiative heat transfer \cite{kittel2005near,shen2009surface,rousseau2009radiative,ottens2011nearfield,yusuke2011infrared,kim2015radiative}. The mentioned theories share the similarity that they are restricted to the case of objects being embedded in a vacuum environment.

In this work, we revisit and extend the scattering formalism for radiative heat transfer between objects in vacuum that was presented in Ref.~\cite{kruger2012trace} to be applicable to an arrangement of $N$ objects embedded in a passive non-absorbing background medium (e.g.\ a fluid or gas). Specifically, we find compact trace formulas for the radiation in $N$ body systems, which are valid in either vacuum, or in a passive background medium. We show that the symmetry and positivity of heat transfer, known for two bodies \cite{rodriguez2013fluctuating,kruger2012trace}, is also valid for three or more bodies, and in the presence of the passive medium. We give several examples, including the radiation of a sphere in a background medium, as well as the near-field heat transfer between two parallel plates separated by a gap filled with a passive medium. In both cases, the background medium can drastically change the energy transfer. We also analyze the transfer between parallel plates in the presence of a dilute gas, both directly (by considering scattering from individual gas particles) as well as effectively (by assigning a dielectric function to the gas). The two approaches share a common limit of dilution.

\section{Discussion of the setup and experimental relevance}\label{sec:Discussion}
\begin{figure}[b]
\includegraphics[width=.6\linewidth]{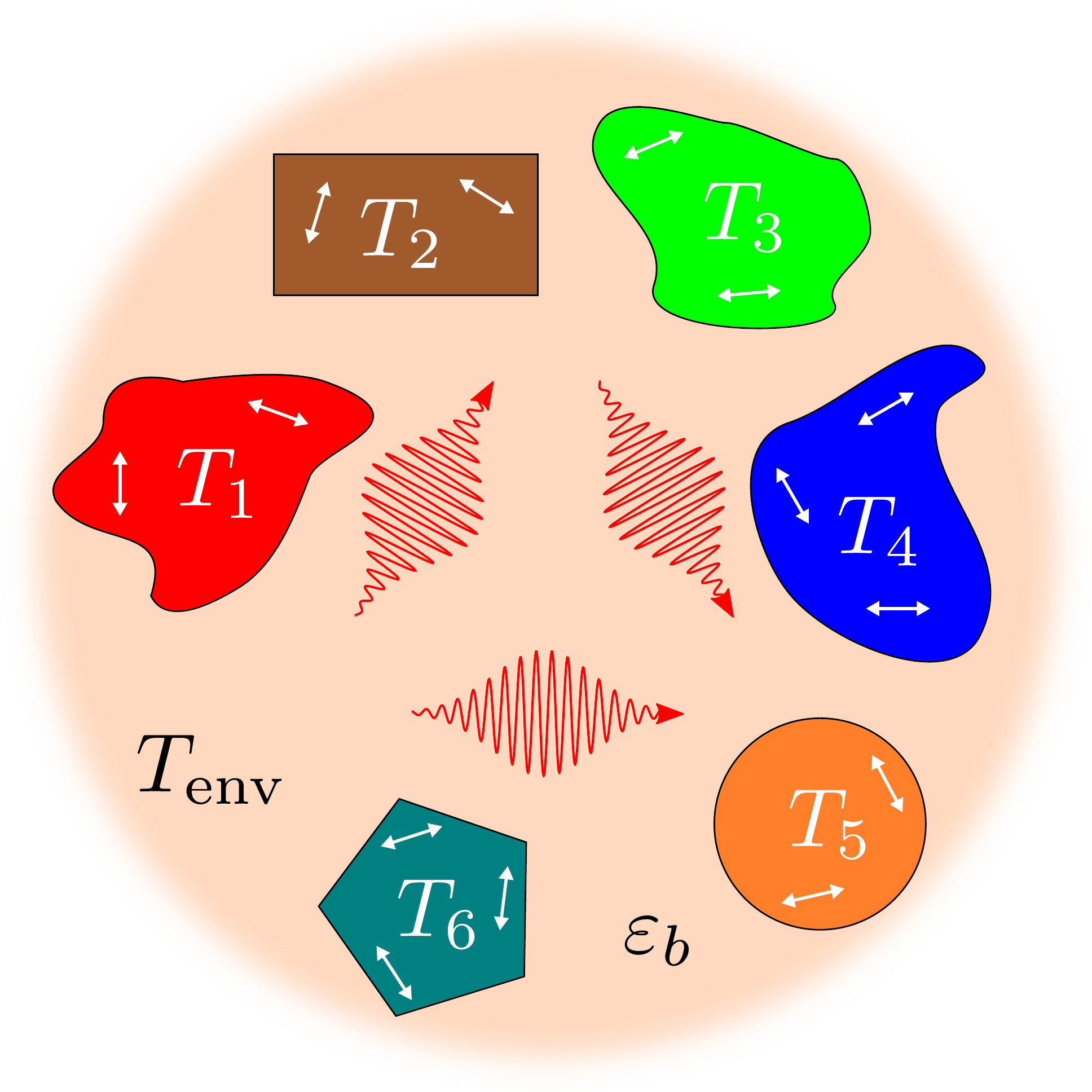}
\caption{\label{fig:setup} The system of $N$ arbitrary objects embedded in a passive non-absorbing background medium in thermal non-equilibrium as considered in this manuscript. The objects are subject to thermal charge and current fluctuations (indicated by the white arrows). The resulting electromagnetic field fluctuations (mimicked by the red arrows) give rise to typical non-equilibrium phenomena, such as heat radiation and heat transfer between objects.}
\end{figure}

The setup under study consists of $N$ objects labeled by $\alpha=1\dots N$ at time-independent, homogeneous temperatures $\{T_\alpha\}$ in a passive non-absorbing background medium as schematically illustrated in Fig.~\ref{fig:setup}. The requirement of vanishing absorptance of the background medium results in appreciable technical simplifications regarding the following theoretical derivations. In practice, we expect the derived predictions to be valid as long as absorption in the background medium is negligible. We thus neglect absorption and emission of the medium \cite{kirchhoff1860ueber}. Technically, we treat the background medium as an enclosing passive body occupying the infinite space complementary to the arrangement of objects. In this respect, the medium can, in principle, be any kind of weakly absorbing liquid or gas. Our general considerations are also valid if the background medium is inhomogeneous, as e.g.\ in density gradients or through adsorption near object surfaces. For specific examples, we take it to be homogeneous, isotropic and local. Despite being non-absorbing, the background medium contains the famous environment dust \cite{eckhardt1984macroscopic}, so that eventually, all radiation is absorbed by it at far distances. Due to this, as far as radiation is concerned, only the temperature of environment far away from all objects is relevant (denoted $T_\mathrm{env}$ and assumed homogeneous). This is important, as the liquid or gas may acquire spatially dependent temperatures near the objects, which, again, are irrelevant for the electromagnetic field radiation. 

In an exemplary experimental realization, the medium as well as the  ensemble of $N$ objects, may be considered as inside surrounding material walls, which are kept at fixed temperature $T_\mathrm{env}$ \cite{antezza2005new,antezza2006surface,messina2011casimir}. The surrounding walls should be far away from the $N$-body system so that only far-field waves emitted by the walls impinge on the $N$-body system. The walls should also be non-regular (black) to produce an isotropic radiation at the $N$-body location, and to allow us to neglect the backscattered radiation generated by the $N$-body system impinging on the walls. To avoid such multiple reflections, a tiny absorption in the background medium is advantageous. Such requirements allow us to treat the radiation produced by the surrounding walls as an isotropic blackbody radiation, independent of the dielectric permittivity of the  walls, with temperature $T_\mathrm{env}$. The mathematical description  of such a setup is identical to the one using the environment dust \cite{eckhardt1984macroscopic}. 
As the temperature of the background medium may spatially vary in the considered setup, it is worth stressing that $T_\mathrm{env}$ is not equal to its  temperature in the vicinity of the objects. Since it is non-absorbing, its temperature in the vicinity of the objects is irrelevant. 

It is worth commenting here that the energy transport via electromagnetic fields is in competition with the energy flux evoked by thermal conduction of the medium. We suspect metamaterials to be suitable candidates to reduce the latter contribution to the total net energy flux, as these kinds of structures can be successfully thermally insulated \cite{narayana2013transient,davisn2014nanophonic}. In particular, we think of structural defects in the composition of the medium, e.g.\ tiny vacuum gaps milled into a slab made of metamaterial, to inhibit the propagation of phonons through the medium and thus promoting the idea of thermal isolation. 

\section{Fluctuational electrodynamics in a passive background medium}\label{sec:General_concepts}

In this section, we generalize the formalism introduced in
Refs.~\cite{kruger2011nonequilibrium,kruger2012trace} to be applicable to
an arrangement of objects that is surrounded by a passive non-absorbing
background medium.
In the given non-equilibrium situation, each object is assumed to be at
local equilibrium, such that the spontaneous charge and current
fluctuations within the individual object satisfy the fluctuation
dissipation theorem at the appropriate temperature (see
Appendix~\ref{app:Electric_field_correlator} for details). The system
under study is characterized in terms of its macroscopic material
properties, i.e.\ through its electric and magnetic response
$\bbeps(\omega;\vct{r},\vct{r}')$ and $\bbmu(\omega;\vct{r},\vct{r}')$,
which can in general be nonlocal complex tensors,
$\bbeps(\omega;\vct{r},\vct{r}')=\varepsilon_{ij}(\omega;\vct{r},\vct{r}')$. Operator products generally include a matrix multiplication in $3\times 3$ space, as well as an integral in $\mathbb{R}^3$ of the common spatial argument. 
The response functions constitute the potential operator of the system
$\mathbb{V}=\frac{\omega^2}{c^2}(\bbeps-\mathbb{I})+\nabla\times\left(\mathbb{I}-\frac{1}{\bbmu}\right)\nabla\times$
as defined in the classical Helmholtz equation \cite{jackson1999classical,tsang2004scattering,rahi2009scattering}
\be\label{eq:GreenFull}
\left[
\mathbb{H}_0-\mathbb{V}-\frac{\omega^2}{c^2}\mathbb{I}\right]\G(\vct{r},\vct{r}')=\mathbb{I}\delta^{(3)}(\vct{r}-\vct{r}')\,,
\ee
where $\mathbb{H}_0=\nabla\times\nabla\times$ describes free space and
$\G$ is the Green's function of the system.
In contrast to the vacuum case
\cite{kruger2012trace}, the potential operator is nonzero everywhere in
space and we need to make a distinction of cases

\be\label{eq:Full_Potential}
\mathbb{V}(\vct{r},\vct{r}')=\begin{cases}
          \mathbb{V}_\alpha(\vct{r},\vct{r}'), & \vct{r},\vct{r}' \in
V_\alpha\\
          \mathbb{V}_b(\vct{r},\vct{r}'), & \vct{r},\vct{r}' \notin
\{V_\alpha\}\\
           \end{cases}\,.
\ee
Here, we introduced the  potential
$\mathbb{V}_\alpha(\vct{r},\vct{r}')$ of the individual object $\alpha$,
which is confined to its volume $V_\alpha$ and associated with the
response functions $\bbeps_\alpha(\omega;\vct{r},\vct{r}')$ and
$\bbmu_\alpha(\omega;\vct{r},\vct{r}')$. For simplicity, we restrict to the case where the potential $\mathbb V$ does not connect points in the background medium with those in the objects, to allow for sharp boundaries between medium
and objects. The background medium, described by the potential $\mathbb{V}_b$ and electric and magnetic responses  $\bbeps_b(\omega;\vct{r},\vct{r}')$ and
$\bbmu_b(\omega;\vct{r},\vct{r}')$, respectively, is assumed non-absorbing. This assumption, together with
symmetries of microreversibility \cite {eckhardt1984macroscopic} make its
electric and magnetic response Hermitian \cite{stancil2012theory},
\begin{align}\label{eq:ResponsesBulk}
\begin{split}
\bbeps_b&=\bbeps_b^\dagger\,,\\
\bbmu_b&=\bbmu_b^\dagger\,.
\end{split}
\end{align}
Furthermore, we introduce $\G_b$, which is the Green's function in Eq.\ \eqref{eq:GreenFull} for $\mathbb{V}=\mathbb{V}_b$. It takes the role which the free Green's function has for objects placed in vacuum. For mathematical reasons, we define the potential $\mathbb{V}_b$ also inside $\{V_\alpha\}$, where it can take any (non-absorbing) form in order for the following theoretical steps to be valid. In practice, one can choose a convenient form of $\mathbb{V}_b$ inside $\{V_\alpha\}$. As done in the examples provided below, for the case where the background is homogeneous and local, $\mathbb{V}_b(\omega;\vct{r})=\mathbb{V}_b \delta^{(3)}(\vct{r}-\vct{r}')$, we will naturally choose that the potential takes the same value inside the objects as outside. 

Making use of the identities $\Im[\G]=-\G\Im[\G^{-1}]\G^*$ and $\Im[\mathbb{V}-\mathbb{V}_b]=-\Im[\G^{-1}-\G^{-1}_b]$, which can be directly found from Eq.\ \eqref{eq:GreenFull}, we can rewrite the equilibrium correlator given in Appendix~\ref{app:Electric_field_correlator}, Eq.\ \eqref{eq:FDT}, according to its originating thermal sources \cite{kruger2011nonequilibrium,kruger2012trace}
\be\label{eq:Corr_Eq}
\mathbb{C}^\mathrm{eq}=\langle\vct{E}\otimes\vct{E}^*\rangle^0_\omega+\sum_\alpha\mathbb{C}^\mathrm{sc}_\alpha(T)+\mathbb{C}^\mathrm{env}(T)\,.
\ee
In this equation, we separated the zero-point term $\langle\vct{E}\otimes\vct{E}^*\rangle^0_\omega\equiv a_0\Im[\G]$ and the contributions of object $\alpha$ and the environment are identified respectively with
\begin{align}\label{eq:Corr_Sc}
 \mathbb{C}^\mathrm{sc}_\alpha(T)&=a(T)\G\Im[\Delta \mathbb{V}_\alpha]\G^*\,,\\
 \label{eq:Corr_Env}
\mathbb{C}^\mathrm{env}(T)&=-a(T)\G\Im[\G_b^{-1}]\G^*\,.
\end{align}
The amplitude factors $a(T)$ and $a_0$ are given in Appendix~\ref{app:Electric_field_correlator}, Eqs.~\eqref{eq:Amplitude_T} and \eqref{eq:Amplitude_Zero}. 
We introduced the potential difference $\Delta\mathbb{V}\equiv \mathbb{V}-\mathbb{V}_b=\sum_\alpha\Delta\mathbb{V}_\alpha$, which is only nonzero inside the objects. For object $\alpha$, it reads specifically,
\be\label{eq:Potential_Difference_PositionSpace}
\Delta\mathbb{V}_\alpha(\vct{r},\vct{r}')=\begin{cases}
          \mathbb{V}_\alpha(\vct{r},\vct{r}')- \mathbb{V}_b(\vct{r},\vct{r}'), & \vct{r},\vct{r}' \in V_\alpha\\
         0, & \mathrm{else}\\
           \end{cases}\,.
\ee
Having identified the different sources of radiation in Eq.~\eqref{eq:Corr_Eq}, we use the key assumption of local equilibrium in fluctuational electrodynamics, such that we can change the temperatures of the different sources independently to arrive at the field correlator in the non-equilibrium situation \cite{kruger2011nonequilibrium,kruger2012trace}
\be\label{eq:Corr_Neq2}
\mathbb{C}^\mathrm{neq}(\{T_\alpha\},T_\mathrm{env})=\mathbb{C}^\mathrm{eq}(T_\mathrm{env})+\sum_\alpha\left[\mathbb{C}^\mathrm{sc}_\alpha(T_\alpha)-\mathbb{C}^\mathrm{sc}_\alpha(T_\mathrm{env})\right]\,.
\ee
In this step, we eliminated the environment contribution in Eq.\ \eqref{eq:Corr_Eq} by introducing the equilibrium correlator $\mathbb{C}^\mathrm{eq}$. It is again important to note that the environment temperature $T_\mathrm{env}$ is measured far away from all objects. In other words, the value of the medium temperature near the objects is irrelevant. This is seen most explicitly in Eq.\ \eqref{eq:Corr_Env}, which denotes the radiation from sources in the environment: Due to the infinitesimal character of $\Im[\G_b^{-1}]$ (the ``dust''), only (the large) regions far away contribute.   

While the fluctuating electromagnetic field is subject to the laws of quantum mechanics [compare the appearence of the reduced Planck constant $\hbar$ in Eq.\ \eqref{eq:FDT}], its scattering at the objects obeys classical scattering theory. The scattering by arbitrarily-shaped particles is described by the $T$-matrix approach introduced in detail in Ref.~\cite{tsang2004scattering}. 
For the objects in the background medium, the $\mathbb{T}$-operator is defined slightly differently compared to the vacuum case. We derive it by starting from the Lippmann-Schwinger equation \cite{lippmann1950variational}
\be\label{eq:Lippmann_Schwinger}
\vct{E}_b^\mathrm{sc}=\vct{E}_b+\G_b\Delta\mathbb{V}\vct{E}_b^\mathrm{sc}\,.
\ee
Eq.\ \eqref{eq:Lippmann_Schwinger} expresses the general solution $\vct{E}^\mathrm{sc}_b$ of the Helmholtz equation with the objects present in the background medium \cite{tsang2004scattering},
\be\label{eq:Helmholtz_Eq}
\left[ \mathbb{H}_0-\mathbb{V}-\frac{\omega^2}{c^2}\mathbb{I}\right]\vct{E}^\mathrm{sc}_b=0\,.
\ee
Note that $\vct{E}_b$ is the solution to the Helmholtz equation in the absence of any object, i.e., where $\mathbb{V}=\mathbb{V}_b$.
Starting from Eq.\ \eqref{eq:Lippmann_Schwinger}, we can iteratively substitute for $\vct{E}^\mathrm{sc}_b$ to obtain the formal expression
\begin{align}
\begin{split}\label{eq:TOperator_Lippmann}
\vct{E}^\mathrm{sc}_b&=\vct{E}_b+\G_b\Delta\mathrm{V}\vct{E}_b+\G_b\Delta\mathrm{V}\G_b\Delta\mathbb{V}\vct{E}^\mathrm{sc}+\dots\\
&=\vct{E}_b+\G_b\T\vct{E}_b\,.
\end{split}
\end{align}
Solving Eq.\ \eqref{eq:Lippmann_Schwinger} and \eqref{eq:TOperator_Lippmann} for $\mathbb{T}$, we find for the $\T$ operator of objects in the background medium,
\be\label{eq:TOperator_medium}
\T=\Delta\mathbb{V}\frac{1}{1-\G_b\Delta\mathbb{V}}\,.
\ee
It is related to the Green's function $\G$ via
\be\label{eq:GreenBackground_T_Medium}
\G=\G_b+\G_b\T\G_b\,.
\ee
Since any Green's function is symmetric \cite{eckhardt1984macroscopic}, the $\T$ operator retains its symmetry. Furthermore, we point out that $\T$ is the scattering operator of the entire collection of objects in the background medium, whereas we shall use $\T_\alpha$ for object $\alpha$ in isolation. In Table~\ref{table:1}, we provide the equivalent quantities of a system in vacuum to those in a non-absorbing background medium to easily convert any formula between these two. 

In particular, Eqs.~\eqref{eq:TOperator_medium} and \eqref{eq:GreenBackground_T_Medium} approach the known vacuum expressions for $\mathbb{V}_b=0$.

\begin{table}[t]
\begin{ruledtabular}
\begin{tabular}{l|lcl}
\multicolumn{4}{c}{Quantities and equivalences}\\\hline
Vacuum & $\G_0$ & $\mathbb{V}$ & $\T=\mathbb{V}\frac{1}{1-\G_0\mathbb{V}}$ \\\hline
Medium & $\G_b$ & $\Delta\mathbb{V}$ & $\T=\Delta\mathbb{V}\frac{1}{1-\G_b\Delta\mathbb{V}}$ \\
\end{tabular}
\end{ruledtabular}
\caption{\label{table:1} Overview of equivalent quantities for a system in vacuum as presented in Ref.~\cite{kruger2012trace} and in a non-absorbing background medium as discussed here.}
\end{table}

\section{Three or more objects: Total heat absorption in a medium}\label{sec:Three_or_more_objects_heat_absorption}
Application of the formalism presented in the previous section allows us to specify the energy fluxes in a many body system of $N$ arbitrary objects in a passive non-absorbing background medium. The many body heat radiation and heat transfer in a medium will constitute the main results of this paper. One measurable quantity in the system is the total heat $H^{(\beta)}$ absorbed by object $\beta$ in the presence of all other objects. 
Using the representation of the non-equilibrium field correlator in Eq.\ \eqref{eq:Corr_Neq2}, we can write \cite{kruger2012trace}
\be\label{eq:TotalHeat_absorbed}
H^{(\beta)}(\{T_\alpha\},T_\mathrm{env})=\sum_\alpha\left(H_\alpha^{(\beta)}(T_\alpha)-H_\alpha^{(\beta)}(T_\mathrm{env})\right)\,.
\ee
Note that $H_\alpha^{(\beta)}$ (the heat transfer rate) is the component of radiation emitted by object $\alpha$ and absorbed by another object $\beta$ in a many body system. On the other hand, $H_\beta^{(\beta)}$ (the ``self''-emission) describes the heat emitted by object $\beta$ in the presence of all other objects. The subtractive term in Eq.\ \eqref{eq:TotalHeat_absorbed} implicitly reflects the principle of detailed balance: At global thermal equilibrium, i.e.\ $\{T_\alpha\}=T_\mathrm{env}$, all radiative fluxes cancel each other.
In order to evaluate the total heat absorbed by object $\beta$, we require general expressions for these quantities which will be given in the following subsections.

\subsection{Heat transfer rate $H_\alpha^{(\beta)}$}
\begin{figure}[b]
\includegraphics[width=.6\linewidth]{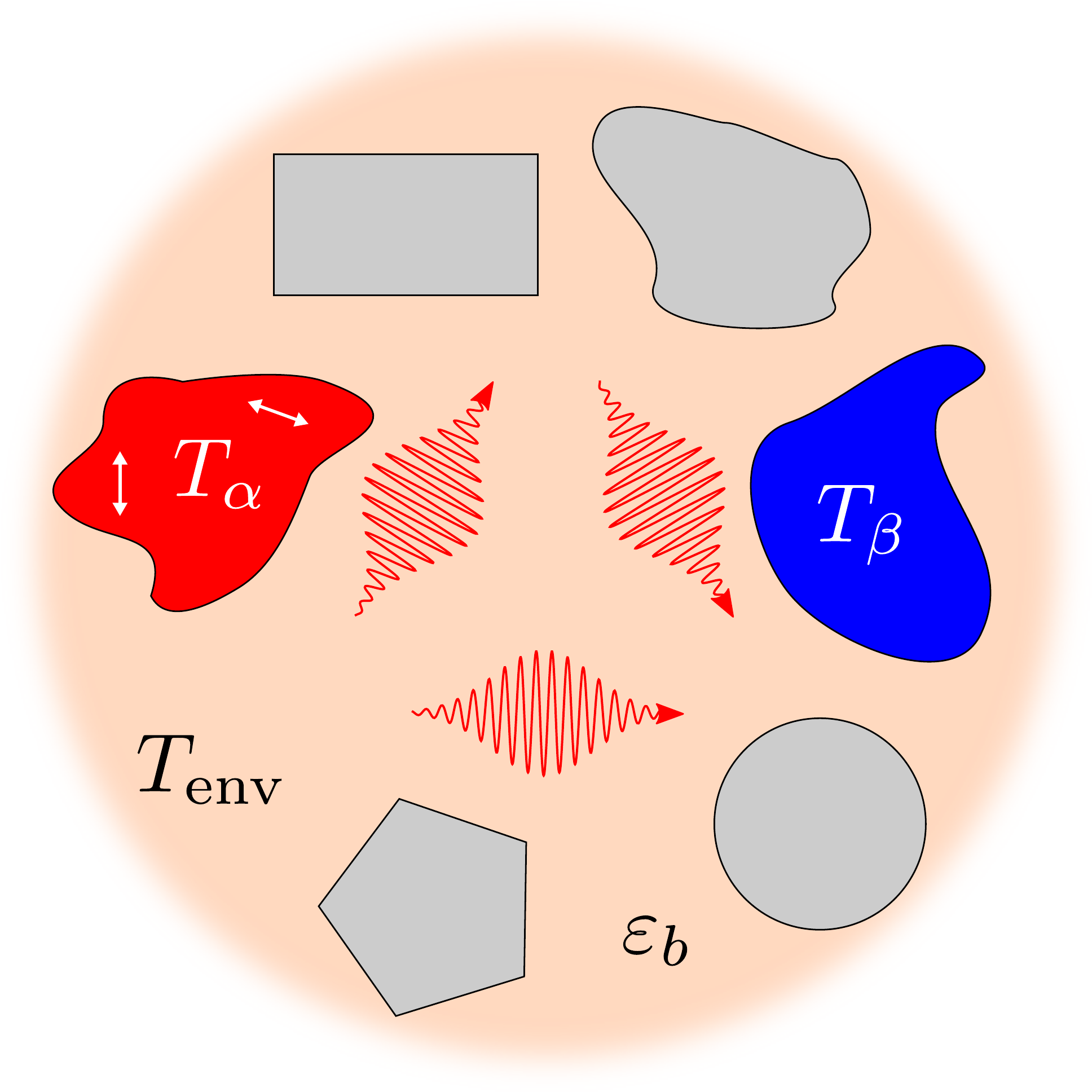}
\caption{\label{fig:Heat_transfer} Illustration of the quantity $H_\alpha^{(\beta)}$ in a many body system embedded in a passive non-absorbing background medium. The heat radiation emitted by object $\alpha$ (red) at temperature $T_\alpha$ is partially absorbed by object $\beta$ (blue) at temperature $T_\beta$. The residual objects are marked by a gray color. }
\end{figure}
We first derive the generalized heat transfer rate $H_\alpha^{(\beta)}$ for $\alpha\neq\beta$ in a many body system of $N$ objects. The situation is schematically illustrated in Fig.~\ref{fig:Heat_transfer}. 
In the following, we attach the single scattering operator $\T_\alpha$ to object $\alpha$ and $\T_\beta$ to object $\beta$.
Moreover, $\T_{\overline{\alpha\beta}}$ will denote the composite $T$ operator of the residual $N-2$ objects. (In a three-body system, it is the operator of the third object.) Using multiple scatterings (see Appendix~\ref{app:Redefinied_operators}), we place the $N-2$ objects described by $\T_{\overline{\alpha\beta}}$ into the field $\vct{E}_{b,\alpha}^\mathrm{iso}$ radiated by the isolated object $\alpha$:
\be\label{eq:Sc_Field_N-1}
\vct{\tilde{E}}^\mathrm{sc}_{b,\alpha\beta}=(1+\G_b\T_{\overline{\alpha\beta}})\frac{1}{1-\G_b\T_\alpha\G_b\T_{\overline{\alpha\beta}}}\vct{E}_{b,\alpha}^\mathrm{iso}\,.
\ee
Finally, we add object $\beta$ into this field and obtain the full scattering solution
\be\label{eq:Sc_Field_total}
\vct{E}^\mathrm{sc}_{b,\alpha}=(1+\G_b\T_\beta)\frac{1}{1-\G_b\T_{\overline{\beta}}\G_b\T_\beta}\vct{\tilde{E}}^\mathrm{sc}_{b,\alpha\beta}\,.
\ee
In this equation, $\T_{\overline{\beta}}$ is the composite $T$ operator of all objects except object $\beta$. Insertion of Eq.\ \eqref{eq:Sc_Field_N-1} into Eq.\ \eqref{eq:Sc_Field_total} yields an explicit form of the multiple scattering operator $\mathbb{O}_\alpha$,
\begin{align}
\vct{E}^\mathrm{sc}_{b,\alpha}&=\mathbb{O}_{\alpha}\vct{E}^\mathrm{iso}_{b,\alpha}\,,\\\label{eq:Multiple_scattering2}
\begin{split}
\mathbb{O}_{\alpha} &=(1+\G_b\T_\beta)\frac{1}{1-\G_b\T_{\overline{\beta}}\G_b\T_\beta}\\
&\times(1+\G_b\T_{\overline{\alpha\beta}})\frac{1}{1-\G_b\T_\alpha\G_b\T_{\overline{\alpha\beta}}}\,.
\end{split}
\end{align}
This is identical (but more explicit) to the representation of $\mathbb{O}_\alpha$ in Appendix~\ref{app:Redefinied_operators}, Eq.\ \eqref{eq:Multiple_scattering}.
Note that the operator does not depend on the specific choice of object $\beta$ as can be directly seen from Eq.\ \eqref{eq:Multiple_scattering}.
The representation of the multiple scattering operator in Eq.\ \eqref{eq:Multiple_scattering2} can be used to derive a general trace formula for the heat transfer rate $H_\alpha^{(\beta)}$ in a many body system. Following the derivation for two objects in vacuum in Ref.~\cite{kruger2012trace}, we can express $H_\alpha^{(\beta)}$ in terms of three well-known quantities: the radiation operator $\mathbb{R}_\alpha$, the multiple scattering operator $\mathbb{O}_\alpha$, and the free Green's function $\G_0$. The radiation operator describes the emitted field by object $\alpha$ embedded in the background medium and is defined according to
\be\label{eq:Radiation_operator}
\mathbb{R}_\alpha\equiv\G_b\left[\Im[\T_\alpha]-\T_\alpha\Im[\G_b]\T_\alpha^*\right]\G_b^*\,.
\ee
Explicitly, the heat transfer rate is given by an integral over the volume of object $\beta$ as 
\begin{align}\label{eq:Heat_Transfer_NObjects}
\begin{split}
H_\alpha^{(\beta)}&=-\frac{2\hbar}{\pi}\int_0^\infty\mathrm{d}\omega\frac{\omega}{e^{\frac{\hbar\omega}{k_BT_\alpha}}-1}\\
&\times\Im\sum_i\int_{V_\beta}\mathrm{d}^3r(\mathbb{O}_\alpha\mathbb{R}_\alpha\mathbb{O}_\alpha^\dagger\G_0^{-1*})_{ii}(\vct{r},\vct{r})\,.
\end{split}
\end{align}
This expression directly finds the absorbed energy by integrating over the object's volume, as the integrand gives the dissipated energy ($\mathbb{G}_0^{-1}$ enters from the conversion of electric field to current density \cite{kruger2012trace}). It can also be reverted to a surface integral of the surface normal of the Poynting vector, evaluated in the background medium. Because of the non-absorptive properties of the background medium, this Poynting vector is well defined.    

With regard to Eq.\ \eqref{eq:Multiple_scattering2}, $\mathbb{O}_\alpha\mathbb{R}_\alpha\mathbb{O}_\alpha^\dagger\G_0^{-1*}(\vct{r},\vct{r})$ will carry the operator product $\G_b^*\G_0^{-1*}$ on its rightmost position. Using the identity $\G_0^{-1}=\G_b^{-1}+\mathbb{V}_b$, which can be directly found from Eq.\ \eqref{eq:GreenFull}, we can rewrite this product according to
\be
\G_b^*\G_0^{-1*}=1+\G_b^*\mathbb{V}_b^*\,.
\ee
As a result, we can split up the operator to be integrated over the volume of object $\beta$ into a sum of three terms
\be\label{eq:SplitUp}
\mathbb{O}_\alpha\mathbb{R}_\alpha\mathbb{O}_\alpha^\dagger\G_0^{-1*}=\mathbb{O}_\alpha\mathbb{R}_\alpha\mathbb{O}_{\alpha,\beta}^\dagger+\mathbb{O}_\alpha\mathbb{R}_\alpha\mathbb{O}_{\alpha,\overline{\beta}}^\dagger+\mathbb{O}_\alpha\mathbb{R}_\alpha\mathbb{O}_\alpha^\dagger\mathbb{V}_b^*\,,
\ee
where we defined 
\begin{align}\label{eq:SelfPart}
\begin{split}
\mathbb{O}_\alpha\mathbb{R}_\alpha\mathbb{O}_{\alpha,\beta}^\dagger&=\mathbb{O}_\alpha\mathbb{R}_\alpha\frac{1}{1-\T^*_{\overline{\alpha\beta}}\G_b^*\T^*_\alpha\G^*_b}(1+\T^*_{\overline{\alpha\beta}}\G^*_b)\\
&\times\frac{1}{1-\T^*_\beta\G^*_b\T^*_{\overline{\beta}}\G^*_b}\T^*_\beta\,,
\end{split}\\
\begin{split}
\mathbb{O}_\alpha\mathbb{R}_\alpha\mathbb{O}_{\alpha,\overline{\beta}}^\dagger&=\mathbb{O}_\alpha\mathbb{R}_\alpha\frac{1}{1-\T^*_{\overline{\alpha\beta}}\G_b^*\T^*_\alpha\G^*_b}(1+\T^*_{\overline{\alpha\beta}}\G^*_b)\\
&\times\frac{1}{1-\T^*_\beta\G^*_b\T^*_{\overline{\beta}}\G^*_b}\G_b^{-1*}\,.
\end{split}
\end{align}
The two parts differ precisely by the operator in the rightmost position,
\begin{align}
\mathbb{O}_\alpha\mathbb{R}_\alpha\mathbb{O}_{\alpha,\beta}^\dagger&=\dots\T_\beta^*\,,\\
\mathbb{O}_\alpha\mathbb{R}_\alpha\mathbb{O}_{\alpha,\overline{\beta}}^\dagger&=\dots\T_{\gamma}^* \qquad (\gamma\neq\beta)\,.
\end{align}
Note that $\T_\gamma$ can be the scattering operator of any object except for object $\beta$. As a consequence, $\mathbb{O}_\alpha\mathbb{R}_\alpha\mathbb{O}_{\alpha,\overline{\beta}}^\dagger(\vct{r},\vct{r})$ is identically zero if $\vct{r}$ is located inside object $\beta$ and does not contribute to the volume integral in Eq.\ \eqref{eq:Heat_Transfer_NObjects}. This is because $\T_\gamma(\vct{r},\vct{r}')$ can only be nonzero if both arguments are located within the volume $V_{\overline{\beta}}$. Furthermore, in Appendix~\ref{app:Nonabsorbing_background_medium}, we show that the third term in Eq.\ \eqref{eq:SplitUp} does not contribute either for the case of a non-absorbing background medium. 
Thus, the heat transfer rate $H_\alpha^{(\beta)}$ in Eq.\ \eqref{eq:Heat_Transfer_NObjects} is given in terms of $\mathbb{O}_\alpha\mathbb{R}_\alpha\mathbb{O}_{\alpha,\beta}^\dagger$ only. Because this term is only nonzero if $\vct{r}$  is located inside $V_\beta$, we can extend the integration range over all space to obtain a trace
\be\label{eq:HeatTransferRate}
H_\alpha^{(\beta)}(T_\alpha)=-\frac{2\hbar}{\pi}\int_0^\infty\mathrm{d}\omega\frac{\omega}{e^{\frac{\hbar\omega}{k_BT_\alpha}}-1}\Im\Tr[\mathbb{O}_\alpha\mathbb{R}_\alpha\mathbb{O}_{\alpha,\beta}^\dagger]\,.
\ee
This trace is now understood over spatial coordinate $\vct{r}$ as well as over matrix indices $i$.  Using the cyclic properties of the trace, we give the final representation of the heat transfer rate $H_\alpha^{(\beta)}$ in explicit form
\begin{widetext}
\begin{align}\label{eq:Heat_Transfer_NObjects2}
\begin{split}
H_\alpha^{(\beta)}&=\frac{2\hbar}{\pi}\int_0^\infty\mathrm{d}\omega\frac{\omega}{e^{\frac{\hbar\omega}{k_BT_\alpha}}-1}\Tr\Bigg\{[\Im[\T_\beta]-\T_\beta^*\Im[\G_b]\T_\beta]\frac{1}{1-\G_b\T_{\overline{\beta}}\G_b\T_\beta}(1+\G_b\T_{\overline{\alpha\beta}})\frac{1}{1-\G_b\T_\alpha\G_b\T_{\overline{\alpha\beta}}}\\
&\times\G_b[\Im[\T_\alpha]-\T_\alpha\Im[\G_b]\T_\alpha^*]\frac{1}{1-\G_b^*\T^*_{\overline{\alpha\beta}}\G_b^*\T_\alpha^*}(1+\G^*_b\T_{\overline{\alpha\beta}}^*)\G_b^*\frac{1}{1-\T_\beta^*\G_b^*\T_{\overline{\beta}}\G_b^*}\Bigg\}\,.
\end{split}
\end{align}
\end{widetext}
The basis independent trace formula is completely determined by the scattering properties of the $N$ objects and the Green's function of the background medium. As a consistency check for our formula, we compare it to the case of two objects in vacuum as derived in Ref.~\cite{kruger2012trace}. By setting $\T_{\overline{\alpha\beta}}=0$ and using the conversion in Table~\ref{table:1}, the formula reduces exactly to the known result.

\subsection{Positivity and symmetry of transfer in a many body system}
The positivity and symmetry of transfer already known for two objects \cite{rodriguez2013fluctuating,kruger2012trace} can be generalized to a many body system by means of Eq.\ \eqref{eq:Heat_Transfer_NObjects2}. In this section, we provide these properties, while the technical details are given in Appendix \ref{app:Positivity_and_symmetry_of_transfer}.
As naturally expected, the heat transfer rate $H_\alpha^{\beta}$ is a non-negative number
\be
H_\alpha^{(\beta)}\ge 0\,.
\ee
Based on the principle of reciprocity, the function also obeys the symmetry relation
\be
H_\alpha^{(\beta)}(T)=H_\beta^{(\alpha)}(T)\,,
\ee
stating that the roles of emitter and the absorber can be interchanged in the presence of an arbitrary number of passive scatterers. The symmetry and positivity of the heat transfer can be used to show that the net heat flux between a warm object $\alpha$ and a cold object $\beta$ is always positive, i.e.\ that heat is always transferred from the warmer object to the colder one:
\be
H_\alpha^{(\beta)}(T_\alpha)-H_\alpha^{(\beta)}(T_\beta)\ge 0 \quad \mathrm{if\;} T_\alpha\ge T_\beta\,.
\ee
The statement is a direct confirmation of the second law of thermodynamics. Our derivation relies on the assumption that the objects' potentials $\{\mathbb{V}_\alpha\}$ do not depend on temperature. 

\subsection{``Self''-emission $H_\beta^{(\beta)}$}
\begin{figure}[b]
\includegraphics[width=.6\linewidth]{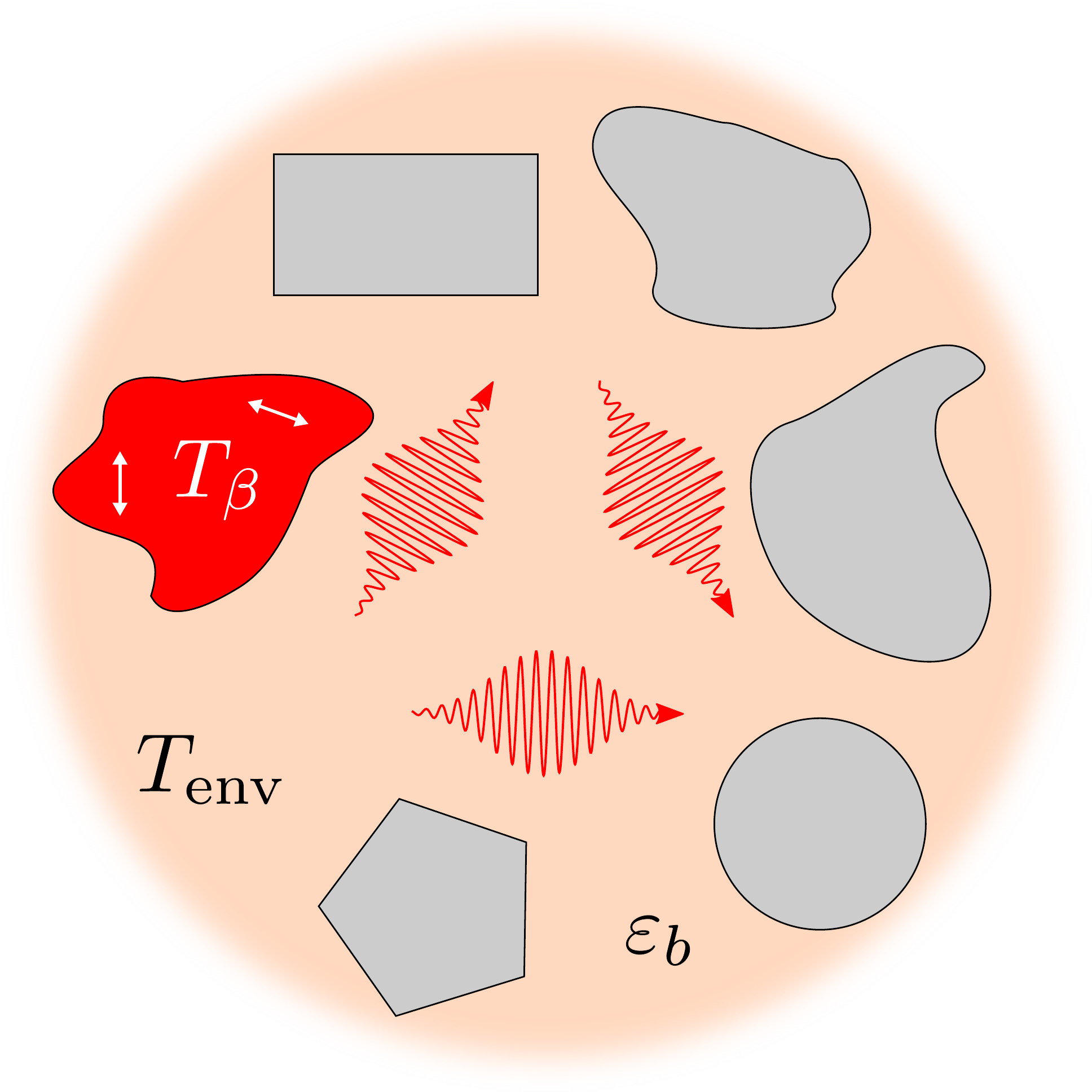}
\caption{\label{fig:Self_emission} The ``self''-emission $H_\beta^{(\beta)}$ in a many body system embedded in a passive non-absorbing background medium. The heat radiation emitted by object $\beta$ (red) at temperature $T_\beta$  is backscattered at the residual objects (gray) and partially reabsorbed.}
\end{figure}

In order to give the total heat absorption of object $\beta$ in Eq.\ \eqref{eq:TotalHeat_absorbed}, we also need to specify its heat emission $H_\beta^{(\beta)}$ in the proximity of the $N-1$ other objects.
The situation under consideration is schematically illustrated in Fig.~\ref{fig:Self_emission}.
The ``self''-emission $H_\beta^{(\beta)}$ is given by Eq.\ \eqref{eq:Heat_Transfer_NObjects}, where we need to integrate the operator $\mathbb{O}_\beta\mathbb{R}_\beta\mathbb{O}_\beta^\dagger\G_0^{-1*}(\vct{r},\vct{r})$ over the volume $V_\beta$ of object $\beta$. This time, we split the multiple scattering operator $\mathbb{O}_\beta$ appearing on the right hand side differently, and after repeating the steps before Eq.\ \eqref{eq:HeatTransferRate}, we find
\be
H_\beta^{(\beta)}(T_\beta)=-\frac{2\hbar}{\pi}\int_0^\infty\mathrm{d}\omega\frac{\omega}{e^{\frac{\hbar\omega}{k_BT_\beta}}-1}\Im\Tr[\mathbb{O}_\beta\mathbb{R}_\beta\mathbb{O}_{\beta,s}^\dagger]\,.
\ee
The trace is now taken over the self-part ``s'' of the operator $\mathbb{O}_\beta\mathbb{R}_\beta\mathbb{O}^\dagger_\beta\G_0^{-1*}$ defined correspondingly to the vacuum case \cite{kruger2012trace}. It is the part of $\mathbb{O}_\beta\mathbb{R}_\beta\mathbb{O}_\beta^\dagger\G_0^{-1*}(\vct{r},\vct{r})$ which is finite inside $V_\beta$ (again, the term with $\mathbb{V}_b$ on the rightmost position does not contribute as the background medium is non-absorbing),
\be
\mathbb{O}_\beta\mathbb{R}_\beta\mathbb{O}_{\beta,s}^\dagger=\mathbb{O}_\beta\mathbb{R}_\beta\frac{1}{1-\T_{\overline{\beta}}^*\G_b^*\T_\beta^*\G_b^*}\G_b^{-1*}\,.
\ee
The final representation of the self-emission in a many body system reads as
\begin{widetext}
\be\label{eq:Self_Emission_NObjects}
H_\beta^{(\beta)}=-\frac{2\hbar}{\pi}\int_0^\infty\mathrm{d}\omega\frac{\omega}{e^{\frac{\hbar\omega}{k_BT_\beta}}-1}\Tr\left\{(1+\G_b\T_{\overline{\beta}})\frac{1}{1-\G_b\T_\beta\G_b\T_{\overline{\beta}}}\G_b[\Im[\T_\beta]-\T_\beta\Im[\G_b]\T_\beta^*]\frac{1}{1-\G_b^*\T_{\overline{\beta}}^*\G_b^*\T_\beta^*}\right\}\,.
\ee
\end{widetext}
Note that $-H_\beta^{(\beta)}$ is positive if the object emits energy. By introducing the operator
\be
\mathbb{W}_{\beta}\equiv\G_b^{-1}\frac{1}{1-\G_b\T_\beta\G_b\T_{\overline{\beta}}}\,,
\ee
we can rewrite the function $H_\beta^{(\beta)}$ in a more compact way, where now the Green's function $G_{\overline{\beta}}$ appears \footnote{We note that Eq.\ (66) in Ref.~\cite{kruger2012trace} contains a typo. In this equation, $\Im[\G_1]$ has to be substituted with $\Im[\G_2]$.}
\be
H_\beta^{(\beta)}=-\frac{2\hbar}{\pi}\int_0^\infty\mathrm{d}\omega\frac{\omega}{e^{\frac{\hbar\omega}{k_BT_\beta}}-1}\Tr\{\Im[\G_{\overline{\beta}}]\mathbb{W}_\beta\mathbb{R}_\beta\mathbb{W}_\beta^\dagger\}\,.
\ee
Taking into account that the product of the two positive semidefinite operators $\Im[\G_{\overline{\beta}}]$ and $\mathbb{W}_\beta\mathbb{R}_\beta\mathbb{W}_\beta^\dagger$ is also positive semidefinite, the positivity of the heat radiation in a many body system including the background medium is found,
\be
-H_\beta^{(\beta)}\ge 0 \,.
\ee 
As a final remark, we note again that all formulas derived in this section are also applicable to the case where the objects are embedded in vacuum. For the vacuum case,  the definition of $\T$ in the first line of Table \ref{table:1} should be applied, and the Green's function $\G_b$ should be replaced by the free Green's function $\G_0$.

\section{Partial wave representation}\label{sec:Partial_wave_representation}
In the previous section, we derived trace formulas for heat radiation and heat transfer in a many body system in operator notation. These formulas are basis-independent and hold for any geometry. They are also valid if the potential of the passive background medium depends on space, as could  for example be the case if the objects' surfaces are covered by a wetting film. In this section, we apply the techniques presented in Refs.~\cite{rahi2009scattering,kruger2012trace} to expand the formulas in partial wave bases. Therefore, we assume that the background medium is homogeneous, isotropic and local, to allow for an expansion of the Green's function $\mathbb{G}_b$ in partial waves. In this way, the traces of operators will turn into sums over matrix elements with respect to partial wave indices.

\subsection{Partial wave expansion of the Green's function and the $\T$ operator in the background medium}
For an isotropic, homogeneous, and local background medium, the dielectric permittivity and magnetic permeability tensors reduce to constant scalars, e.g., $\bbeps_b=\varepsilon_b\mathbb{I}$. In these circumstances,  a simple conversion relates the Green's function in the background medium to  the free Green's function:
\be\label{eq:G_background_G0}
\G_b\left(\frac{\omega}{c},\vct{r}-\vct{r}'\right)=\mu_b\G_0\left(\frac{\omega}{c} \sqrt{\varepsilon_b\mu_b},\vct{r}-\vct{r}'\right)\,.
\ee
For a nonmagnetic background medium, i.e.\ $\mu_b=1$,  only the speed of light outside of the objects is redefined, and the expansion of the free Green's function in eigenfunctions of the Helmholtz equation given in Ref.~\cite{kruger2012trace} can be used. For the sake of completeness, we provide this expansion (suppressing the $\omega$ dependence)
\be\label{eq:G0_partialwaves}
\G_0(\vct{r}-\vct{r}')=i\sum_\mu\begin{cases}\vct{E}^\mathrm{out}_\mu(\vct{r})\otimes\vct{E}^\mathrm{reg}_{\sigma(\mu)}(\vct{r}') &\mbox{if } \xi_1(\vct{r})\,>\,\xi_1'(\vct{r}')\,, \\
\vct{E}^\mathrm{reg}_{\sigma(\mu)}(\vct{r})\otimes\vct{E}^\mathrm{out}_\mu(\vct{r}') & \mbox{if } \xi_1(\vct{r})\,<\,\xi_1'(\vct{r}')\,. \end{cases}
\ee
The free Green's function is expanded in terms of ``regular'' and ``outgoing'' waves propagating through vacuum, which are solutions of the wave equation, differing by their behavior at the coordinate origin. While ``regular'' waves are nonsingular, ``outgoing'' waves are typically singular at the origin and obey outgoing boundary conditions for $\xi_1\to\infty$. Recall that $\xi_1$ is the ``radial'' coordinate which gives rise to a distinction of cases, depending on which of the two arguments of the Green's function has a greater value of $\xi_1$ \cite{rahi2009scattering}. In this way, the outgoing waves are always evaluated for the larger argument and do not encounter the divergence at $\xi_1=0$.
The wave index $\mu$ runs over polarization (electric and magnetic), and indices of vector functions depending on the specific basis. The function $\sigma(\mu)$ is a permutation among these indices,
fulfilling $\sigma[\sigma(\mu)]=\mu$ \cite{kruger2012trace}.

Furthermore, we define the matrix elements of the $\T$ operator by
\be\label{eq:T_partialwaves}
\mathcal{T}_{\mu\mu'}=i\int\mathrm{d}^3\vct{r}\int\mathrm{d}^3\vct{r}'\vct{E}^\mathrm{reg}_{b,\sigma(\mu)}(\vct{r})\T(\vct{r},\vct{r}')\vct{E}^\mathrm{reg}_{b,\mu'}(\vct{r}')\,,
\ee
where we apply a regular wave propagating through the background medium (denoted by the index ``b'') to both sides of the $\T$ operator. 
The waves in the homogeneous background medium are related to those in vacuum by
\be\label{eq:Wave_Conversion}
\vct{E}_{b,\mu}\left(\frac{\omega}{c},\vct{r}\right)=\sqrt{\mu_b}\,\vct{E}_{\mu}\left(\frac{\omega}{c}\sqrt{\varepsilon_b\mu_b},\vct{r}\right)\,.
\ee
As in the vacuum case, the matrix elements $\mathcal{T}_{\mu\mu'}$ in Eq.\ \eqref{eq:T_partialwaves} obey 
\be
\mathcal{T}_{\mu\mu'}=\mathcal{T}_{\sigma(\mu')\sigma(\mu)}\,
\ee
due to the symmetry of the $\T$ operator.  Besides, both expansions of $\G_b$ and $\T$, respectively, do not contain complex conjugations of waves, which ensures manifest analyticity of these quantities in the upper complex frequency plane.

\subsection{Heat transfer and ``self''-emission}
The general formulas for the heat transfer rate $H_\alpha^{(\beta)}$ and the self-emission $H_\beta^{(\beta)}$ in an arrangement of $N$ objects in operator notation are given in Eqs.~\eqref{eq:Heat_Transfer_NObjects2} and \eqref{eq:Self_Emission_NObjects}. Applying the techniques of partial wave expansion presented in this section to these two equations, we obtain
\begin{widetext}
\begin{align}\label{eq:Heat_Transfer_NObjects_PartialWave}
\begin{split}
H_\alpha^{(\beta)}&=\frac{2\hbar}{\pi}\int_0^\infty\mathrm{d}\omega \frac{\omega}{e^{\frac{\hbar\omega}{k_B T_\alpha}}-1}\Tr\Bigg\{\Bigg[\frac{\tau_\beta^\dagger+\tau_\beta}{2}+\tau_\beta^\dagger\Pi^\mathrm{pr}\tau_\beta\Bigg]\frac{1}{1-\upsilon\tau_{\overline{\beta}}\upsilon\tau_\beta}(1+\upsilon\tau_{\overline{\alpha\beta}})\frac{1}{1-\upsilon\tau_\alpha\upsilon\tau_{\overline{\alpha\beta}}}\\
&\times\upsilon\Bigg[\frac{\tau_\alpha^\dagger+\tau_\alpha}{2}+\tau_\alpha\Pi^\mathrm{pr}\tau_\alpha^\dagger\Bigg]\frac{1}{1-\upsilon^\dagger\tau_{\overline{\alpha\beta}}^\dagger\upsilon^\dagger\tau_\alpha^\dagger}(1+\upsilon^\dagger\tau_{\overline{\alpha\beta}}^\dagger)\upsilon^\dagger\frac{1}{1-\tau_\beta^\dagger\upsilon^\dagger\tau_{\overline{\beta}}^\dagger\upsilon^\dagger}\Bigg\}\,,
\end{split}\\\notag\\
\label{eq:Self_Emission_NObjects_PartialWave}
\begin{split}
H^{(\beta)}_\beta&=\frac{2\hbar}{\pi}\int^\infty_0\mathrm{d}\omega\frac{\omega}{\mathrm{e}^{\frac{\hbar\omega}{k_B T_\beta}}-1}\Re\Tr\left\{[\upsilon\tau_{\overline{\beta}}\upsilon+\Pi^{\rm pr}]\frac{1}{1-\tau_\beta\upsilon\tau_{\overline{\beta}}\upsilon}\left[\frac{\tau_\beta^\dagger+\tau_\beta}{2}+\tau_\beta\Pi^{\rm pr}\tau_\beta^\dagger\right]\frac{1}{1-\upsilon^\dagger\tau_{\overline{\beta}}^\dagger\upsilon^\dagger\tau_\beta^\dagger}\right\}\,,
\end{split}
\end{align}
\end{widetext}
where we introduced the redefined matrices
\be
\tau_{\mu\mu'}\equiv e^{-i\phi_\mu}\mathcal{T}_{\mu\mu'},\quad \upsilon_{\mu\mu'}\equiv\mathcal{U}_{\mu\mu'}e^{i\phi_{\mu'}}\,.
\ee
The phase factors $e^{{i\phi}_\mu}=e^{i\phi_{\sigma(\mu)}}$ for evanescent wave contribution vary for each basis individually. The matrix
\be 
\Pi^\mathrm{pr}_{\mu\mu'}=\delta_{\mu\mu'}\delta_{\mu\;\mathrm{pr}} 
\ee
acts as a projector on propagating waves. In the spherical basis, the phase factors are unity, and, since no evanescent waves exist, the projector $\Pi^\mathrm{pr}_{\mu\mu'}$ is the identity so that the above formulas simplify in that case.

$\mathcal{U}$ is the translation matrix defined by
\be
\vct{E}^\mathrm{out}_{b,\mu}(\vct{r}_\beta)=\sum_{\mu'}\mathcal{U}^{\alpha\beta}_{\mu'\mu}(\vct{X}_{\alpha\beta})\vct{E}^\mathrm{reg}_{b,\mu'}(\vct{r}_\alpha)\,,
\ee
which expands outgoing waves in terms of regular waves described in a different coordinate system. $\vct{X}_{\alpha\beta}=\vct{r}_\alpha-\vct{r}_\beta$ is the vector connecting the two coordinate origins.

\section{Applications}\label{sec:Applications}
In this section, we give several applications for radiation and transfer for objects in a background medium. In particular, we study the heat radiation of a sphere and the heat transfer between two semi-infinite bodies for different materials. Besides, we investigate a three-body system in vacuum consisting of two plates and a polarizable atom in between. After distributing the atom's position, this atomic system is a microscopical gas model which we compare to the macroscopic approach, where a medium with dielectric constant $\varepsilon_b$ is placed between the plates. Finally, we compare the radiative transfer between two plates to the energy transfer found from kinetic gas theory. Throughout this section, the medium is supposed to be nonmagnetic, i.e.\ $\mu_b=1$ in all applications.

\subsection{Heat radiation of a sphere in the presence of a non-absorbing background medium}
\subsubsection{General formula}
The heat radiation of a homogeneous sphere in vacuum was computed in Ref.~\cite{Kattawar70}, and analyzed in detail in Ref.~\cite{kruger2012trace}. We revisit the heat radiation of a sphere and investigate how it changes due to the presence of a homogeneous background medium of dielectric constant $\varepsilon_b$. The energy emitted by an arbitrary isolated object is obtained by setting ${\cal T}_{\overline{\beta}}=0$ in the general formula for the self-emission in Eq.\ \eqref{eq:Self_Emission_NObjects_PartialWave} (where we include a minus sign to get the emitted energy):
\be\label{eq:Heat_radiation_partialwave}
H=-\frac{2\hbar}{\pi}\int \mathrm{d}\omega\frac{\omega}{e^{\frac{\hbar\omega}{k_B T}}-1}\Tr_\mathrm{pr}\{\Re[\mathcal{T}]+\mathcal{T}\mathcal{T}^\dagger\}\,.
\ee
Note that the index ``pr'' upon the trace indicates that all partial wave indices involved run over propagating modes and therefore all phase factors $e^{i\phi_\mu}$ turn unity. It is naturally appropriate to write the heat radiation of a sphere  in the spherical wave basis, were only propagating modes exist, and the index ``pr'' is dropped,
\be\label{eq:Heat_radiation_sphere}
H_s=-\frac{2\hbar}{\pi}\int_0^\infty \mathrm{d}\omega\frac{\omega}{e^{\frac{\hbar\omega}{k_B T}}-1}\sum_{P,l,m}\left[\Re\mathcal{T}_l^P+\vert\mathcal{T}_l^P\vert^2\right]\,.
\ee
In this equation, the summation is over the polarization $P$ and the quantum numbers $l$ and $m$.
The matrix elements of $\mathcal{T}$ for a homogeneous sphere in a non-absorbing background medium are well known and sometimes referred to as Mie coefficients \cite{bohren2008absorption} (see Appendix~\ref{app:SphericalWaveBasis} for details). They do not depend on $m$. 

\subsubsection{Gold sphere}
We first consider a sphere which is made up of gold using the Drude model \cite{jackson1999classical}
\be
\varepsilon_\mathrm{Au}(\omega)=1-\frac{\omega_p^2}{\omega(\omega+i\omega_\tau)}\,,
\ee
with $\omega_p=\unit[9.03]{eV}$ and $\omega_\tau=\unit[2.67\times10^{-2}]{eV}$.
\begin{figure}[b]
\includegraphics[width=\linewidth]{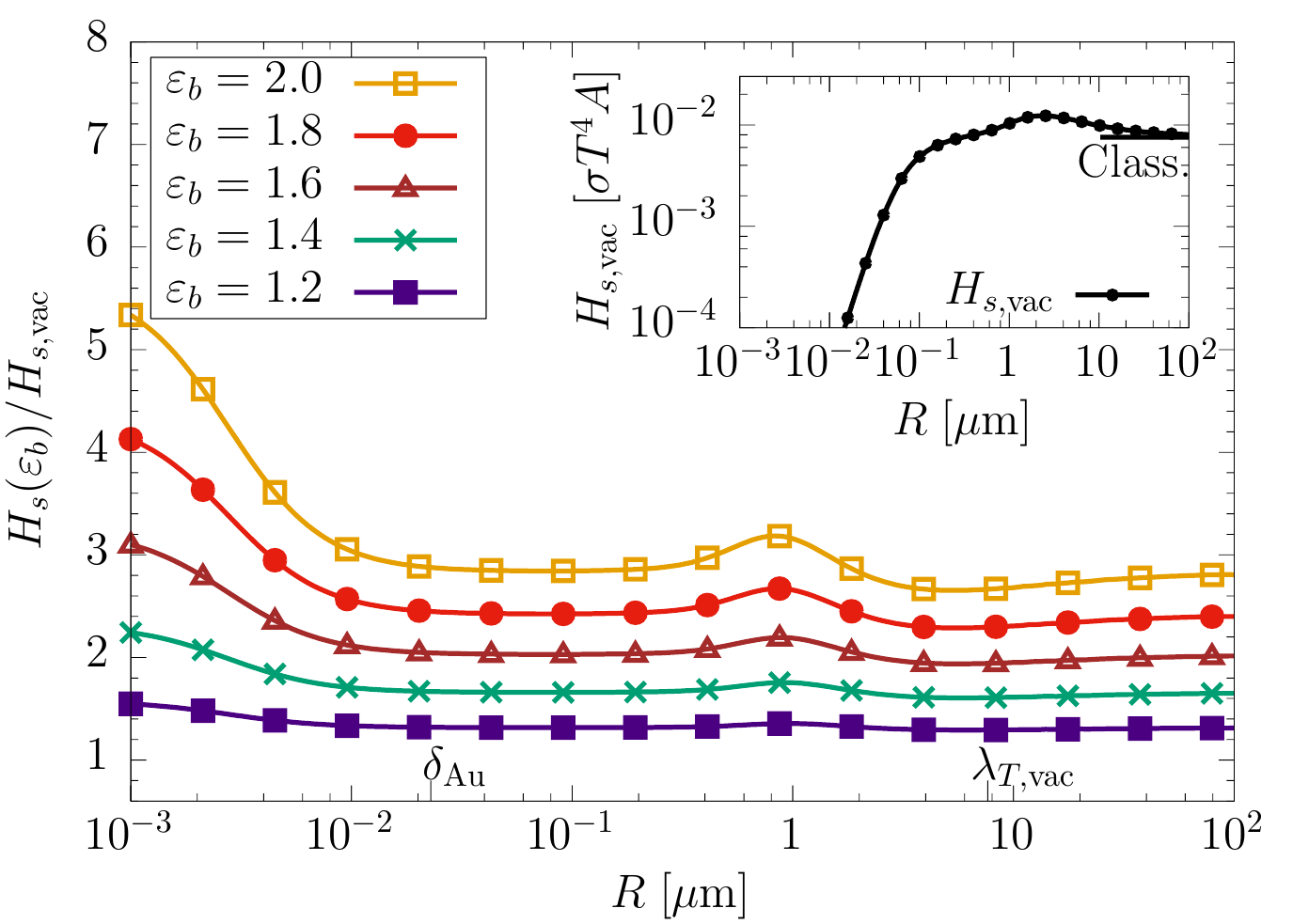}
\caption{\label{fig:Au} Main graph: Heat radiation of a gold sphere at $T=\unit[300]{K}$ in a homogeneous non-absorbing background medium with dielectric constant $\varepsilon_b$ as given, as a function of radius $R$, normalized by the result in vacuum ($\varepsilon_b=1$). The inset graph shows the corresponding result in vacuum, normalized by the Stefan-Boltzmann result (this curve is identical to the one in Fig.~2 of Ref.~\cite{kruger2012trace}).}
\end{figure}
In Fig.~\ref{fig:Au}, we show the change of the heat radiation due to the presence of a background medium by comparing it to the vacuum case. The presence of the background medium yields a strong amplification for the heat radiation of the gold sphere. Physically, this observation can be ascribed to a higher density of propagating waves that can traverse the interface between the two media (compare the angle of total internal reflection for planar surfaces with the modified condition $k_\perp<\frac{\omega}{c}\sqrt{\varepsilon_b}$).  In the limit, where the radius $R$ is the largest scale and much larger than the thermal wavelength $\lambda_T=\hbar c/(\sqrt{\varepsilon_b}k_B T)$ (which is roughly $\unit[8]{\mu m}$ at room temperature in vacuum) and the skin depth $\delta=c/(\Im\sqrt{\varepsilon\mu}\omega)$, the heat radiation of the sphere approaches the classical Stefan-Boltzmann law
\be
\lim_{R\gg\{\lambda_T,\delta\}}H_s=4\pi R^2 \sigma T^4\epsilon(T,\varepsilon_b)=4\pi R^2 \frac{H_p}{A}\,,
\ee
with $\sigma=\pi^2 k_B^4/(60\hbar^3 c^2)$ and emissivity factor $\epsilon(T,\varepsilon_b)$. The linear dependence of $\epsilon$ in the dielectric constant $\varepsilon_b$ in the given range in Fig.~\ref{fig:Au} reflects the increase of propagating modes due to the presence of the background medium for large $R$. In the last step, we introduced the heat radiation $H_p$ per surface area $A$ of a semi-infinite half-space at temperature $T_p$ in the medium given by
\be
\frac{H_p}{A}=\frac{\hbar}{2\pi}\int_0^\infty\mathrm{d}\omega \frac{\omega}{e^{\frac{\hbar\omega}{k_B T_p}}-1}\int_{k_\perp<\frac{\omega}{c}\sqrt{\varepsilon_b}}\frac{d^2k_\perp}{(2\pi)^2}\sum_P[1-|r^P|^2]\,.
\label{eq:plate}
\ee
The Fresnel coefficients $r^P$ for the infinitely thick plate in a medium are given in Appendix~\ref{app:PlaneWaveBasis}. As a  test of the approach presented in this manuscript, we compared Eq.\ \eqref{eq:plate} numerically to the heat transfer between two half-spaces in vacuum given in Eq.\ \eqref{eq:TwoPlates_Medium} below. If, in that formula, the plate separation is set to zero, $d=0$, and the absorbing half-space is assigned the non-absorbing medium's dielectric constant $\varepsilon_b$, the results are numerically identical. 

Interestingly, the radiation, normalized to the corresponding vacuum case, depends pronounced on $R$, and is largest for small $R$. It means that the radiation of nanoparticles is most strongly affected by the presence of the background medium. For radii even smaller than shown in Fig.~\ref{fig:Au}, the curves saturate and reach plateau values. The local maximum around $R=1~\mu$m in Fig.~\ref{fig:Au} shows that, when lowering the contrast $\varepsilon_{\mathrm{Au}}/\varepsilon_b$ between the gold sphere and the background medium, the global maximum of the heat radiation, seen in the graph, is shifted to smaller values of $R$. 

\begin{figure}[t!]
\includegraphics[width=\linewidth]{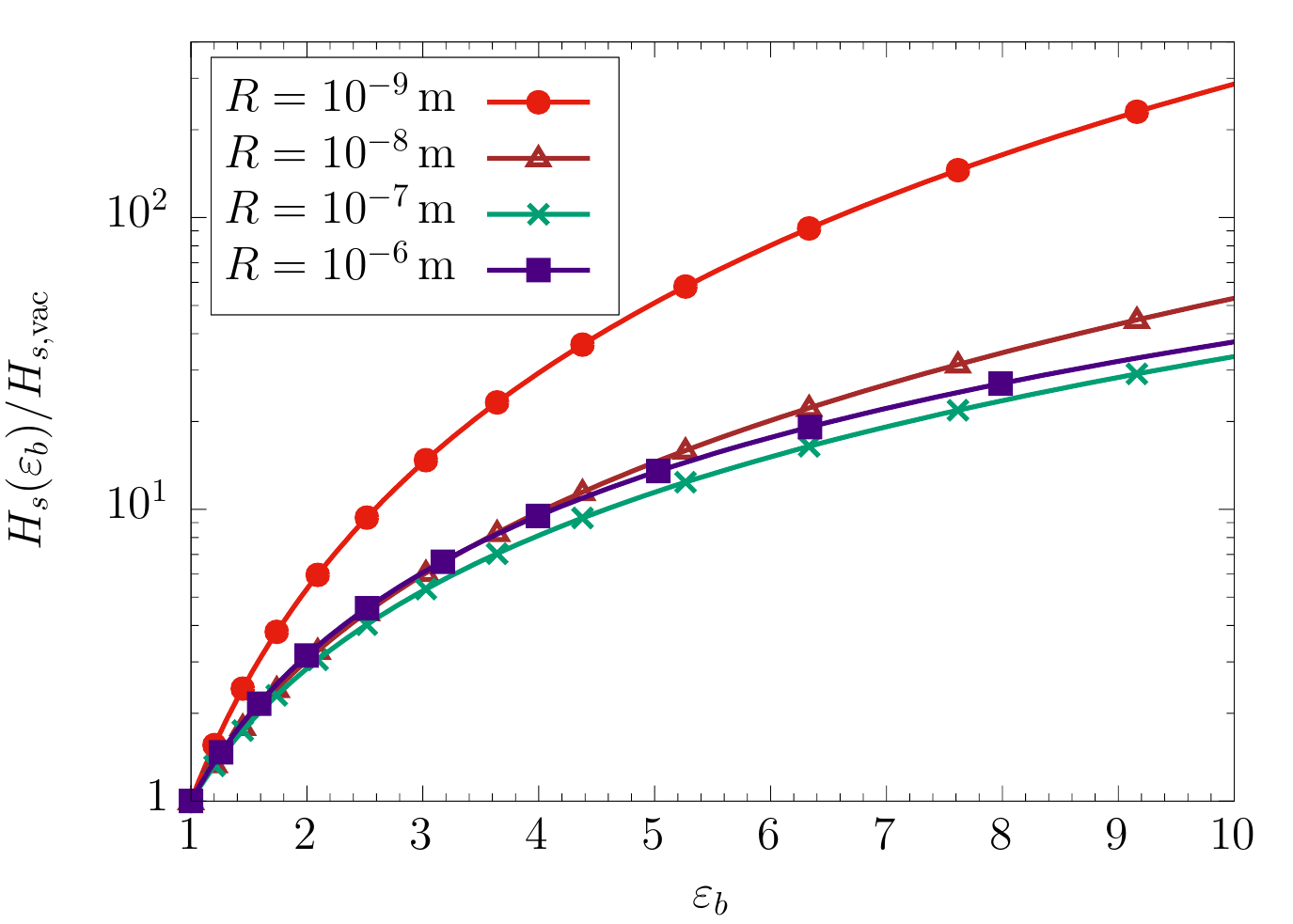}
\caption{\label{fig:Au_Medium2}
 Heat radiation of a gold sphere at $T=\unit[300]{K}$ in a homogeneous non-absorbing background medium as a function of dielectric constant $\varepsilon_b$, normalized by the result in vacuum (see inset graph in Fig.~\ref{fig:Au}).}
\end{figure}

We show the ratio $H_s(\varepsilon_b)/H_{s,\mathrm{vac}}$ as a function of $\varepsilon_b$ in Fig.~\ref{fig:Au_Medium2}. The radiation increases monotonically for increasing $\varepsilon_b$ in the given interval, and for the given values of $R$ (we omit the discussion of practical relevance regarding large and real values of $\varepsilon_b$). Remarkably, for $\varepsilon_b\to\infty$, the curves seem to approach finite values, which may be considered counter-intuitive; For objects in vacuum, the radiation or transfer vanishes for $\varepsilon\to\infty$. It is  insightful to start with a small vacuum gap between object and medium, which is easily done for a plate, where, again, the formula for transfer between two parallel surfaces in vacuum can be used. Then, for a finite vacuum gap, the radiation vanishes for $\varepsilon_b\to\infty$, while it goes to the mentioned finite value if the vacuum gap is set to zero first. Since the limits of vanishing vacuum gap and $\varepsilon_b\to\infty$ do not commute, we expect that (experimental) results for very large $\varepsilon_b$ will depend on the details at the interface between object and medium. 

Numerical accuracy of the truncated sum over multipoles is ensured by requiring a relative error of less than $10^{-3}$. We note that convergence becomes more slow for increasing $\varepsilon_b$, as $\varepsilon_b$ lowers the wavelength in the medium so that the sphere is effectively larger.

\subsubsection{SiO$_2$ sphere}
\begin{figure}[b!]
\includegraphics[width=\linewidth]{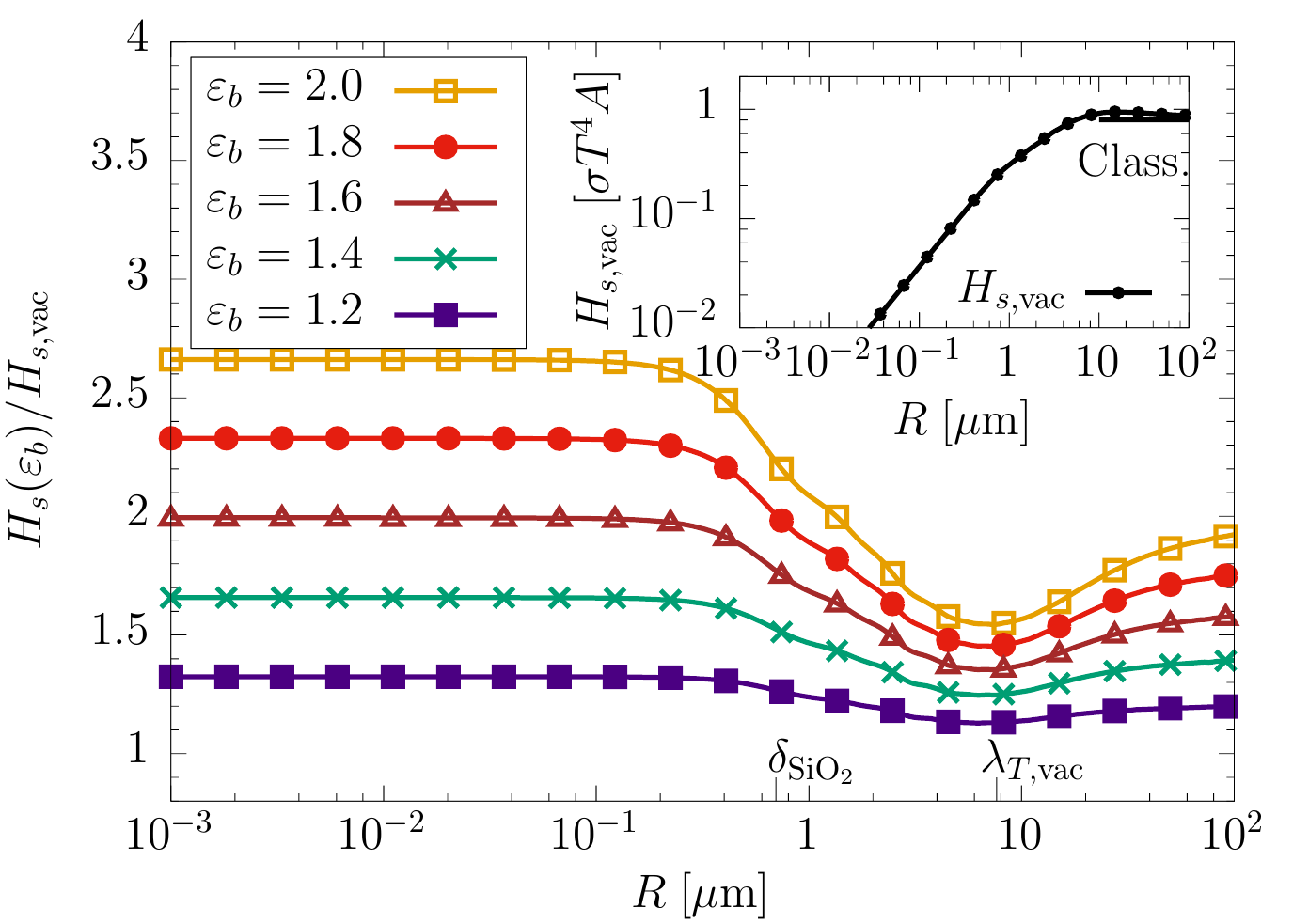}
\caption{\label{fig:SiO2} Main graph: Heat radiation of a SiO$_2$ sphere at $T=\unit[300]{K}$ in a homogeneous non-absorbing background medium with dielectric constant $1\le\varepsilon_b\le 2$ as a function of radius $R$, normalized by the result in vacuum. The inset graph shows the corresponding result in vacuum, normalized by the Stefan-Boltzmann result (this curve is identical to the one in Fig.~2 of Ref.~\cite{kruger2012trace}).}
\end{figure}
As an example for a dielectric, we examine the heat radiation of a SiO$_2$ sphere in a non-absorbing background medium, which has a much larger emissivity in vacuum compared to gold: The  radiation of a SiO$_2$ sphere in vacuum outnumbers the heat radiation of a gold sphere by at least one order of magnitude. In order to analyze the SiO$_2$ sphere in a background medium, see Fig.~\ref{fig:SiO2}, we rely on optical data for $\varepsilon_{\mathrm{SiO}_2}(\omega)$. 

The overall curve is very similar to the one of gold, however it reaches the nanoparticle-plateau already for larger values of $R$. This is because the skin depth $\delta_\mathrm{SiO_2}\approx\unit[0.7]{\mu m}$ is much larger than the one of gold. Again, we note that the amplification due to the background medium is largest for nanoparticles.   

\begin{figure}[t!]
\includegraphics[width=\linewidth]{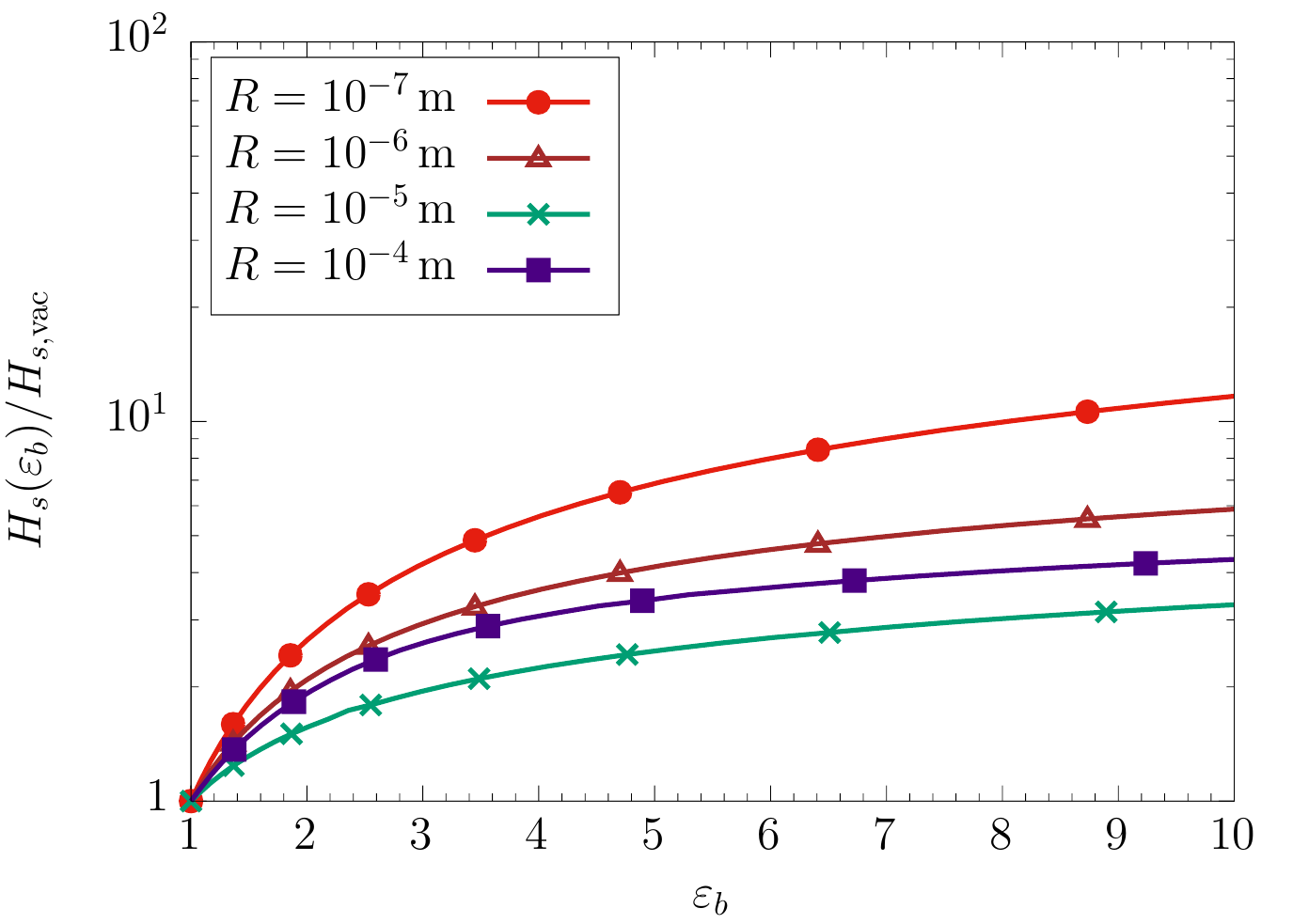}
\caption{\label{fig:SiO2_Medium2} Heat radiation of a SiO$_2$ sphere at $T=\unit[300]{K}$ in a homogeneous non-absorbing background medium as a function of dielectric constant $\varepsilon_b$, normalized by the result in vacuum (see inset graph in Fig.~\ref{fig:SiO2}).}
\end{figure}

In Fig.~\ref{fig:SiO2_Medium2}, we show the amplification factor of the heat radiation as a function of $\varepsilon_b$ for different radii $R$. The curves for the SiO$_2$ sphere are much lower compared to the gold sphere. We note that differences in the contrast $\varepsilon_s/\varepsilon_b$ between the sphere material and the background medium may lead to this observation. It also implies that the radiation of a gold nanoparticle and a glass nanoparticle, which have very different radiation in vacuum, have almost identical values for the radiation if a background medium is present. 
The nanoparticle limit (i.e., the limit where $R$ is small compared to $\delta$ and $\lambda_T$), can be understood by using the form (with $R^*=\frac{\omega R}{c}\sqrt{\varepsilon_b\mu_b}$)
\be\label{eq:Expansion_SmallSpheres}
\lim_{R\ll \{\lambda_T,\delta\}}\mathcal{T}_1^N=i\frac{2(\varepsilon_s-\varepsilon_b)}{3(\varepsilon_s+2\varepsilon_b)}R^{*3}
\ee
and accordingly $\mathcal{T}_1^M$ by substituting the electric responses of sphere and medium, $\varepsilon_s$ and $\varepsilon_b$, with its respective magnetic responses, $\mu_s$ and $\mu_b$. The term in Eq.\ \eqref{eq:Expansion_SmallSpheres} is commonly written in terms of the dipole polarizability of the sphere in a homogeneous background medium \cite{tsang2004scattering},
\be\label{eq:Polarizability}
\alpha_b\equiv\frac{(\varepsilon_s-\varepsilon_b)}{(\varepsilon_s+2\varepsilon_b)}R^3\,.
\ee
For a non-magnetic sphere, i.e.\ $\mu_s=1$, the heat radiation is then given by
\be 
\lim_{R\ll \{\lambda_T,\delta\}} H_s=\frac{4\hbar}{c^3 \pi}\varepsilon_b^{3/2}\int_0^\infty\mathrm{d}\omega \frac{\omega^4}{e^{\hbar\omega/k_B T}-1}\Im \alpha_b\,.
\ee
For small dieletric constants $\varepsilon_b-1\ll 1$, we can expand the dipole polarizability in Eq.\ \eqref{eq:Polarizability} to obtain its vacuum correspondence
\be
\lim_{\varepsilon_b-1\ll 1}\alpha_b=\frac{(\varepsilon_s-1)}{(\varepsilon_s+2)}R^3+\mathcal{O}(\varepsilon_b-1)\,.
\ee
Since $\alpha_b$ approaches a finite value for $\varepsilon_b\to1$, the heat radiation in the medium is then directly proportional to $\varepsilon_b^{3/2}$, which physically accounts for the reduced speed of light in the medium. Indeed, the curves in Fig.~\ref{fig:SiO2} follow a dependence $\propto\varepsilon_b^{3/2}$ for small $R$ well.

\subsection{Heat transfer between two semi-infinite bodies}
The heat transfer between two semi-infinite bodies separated by a vacuum gap has been thoroughly studied by many authors \cite{polder1971theory,levin1980zh,pendry1999radiative,volokitin2001radiative}. Recently, setups consisting of three slabs in vacuum were investigated \cite{messina2012three,messina2014three}. In this section, we use two different approaches to study the transfer between two surfaces in the presence of a non-absorbing medium. We start by ascribing a homogeneous value of $\varepsilon_b$ to the medium between the surfaces. In a second approach, we investigate the situation with a single polarizable atom located inside the gap (which is then homogeneously distributed to mimic a gas). Among other things, we show that the two approaches yield identical results for the dilute limit, where $\varepsilon_b$ is close to unity.    

\subsubsection{Two semi-infinite plates in the presence of a homogeneous background medium}\label{sec:med}
\begin{figure}[b!]
\includegraphics[width=\linewidth]{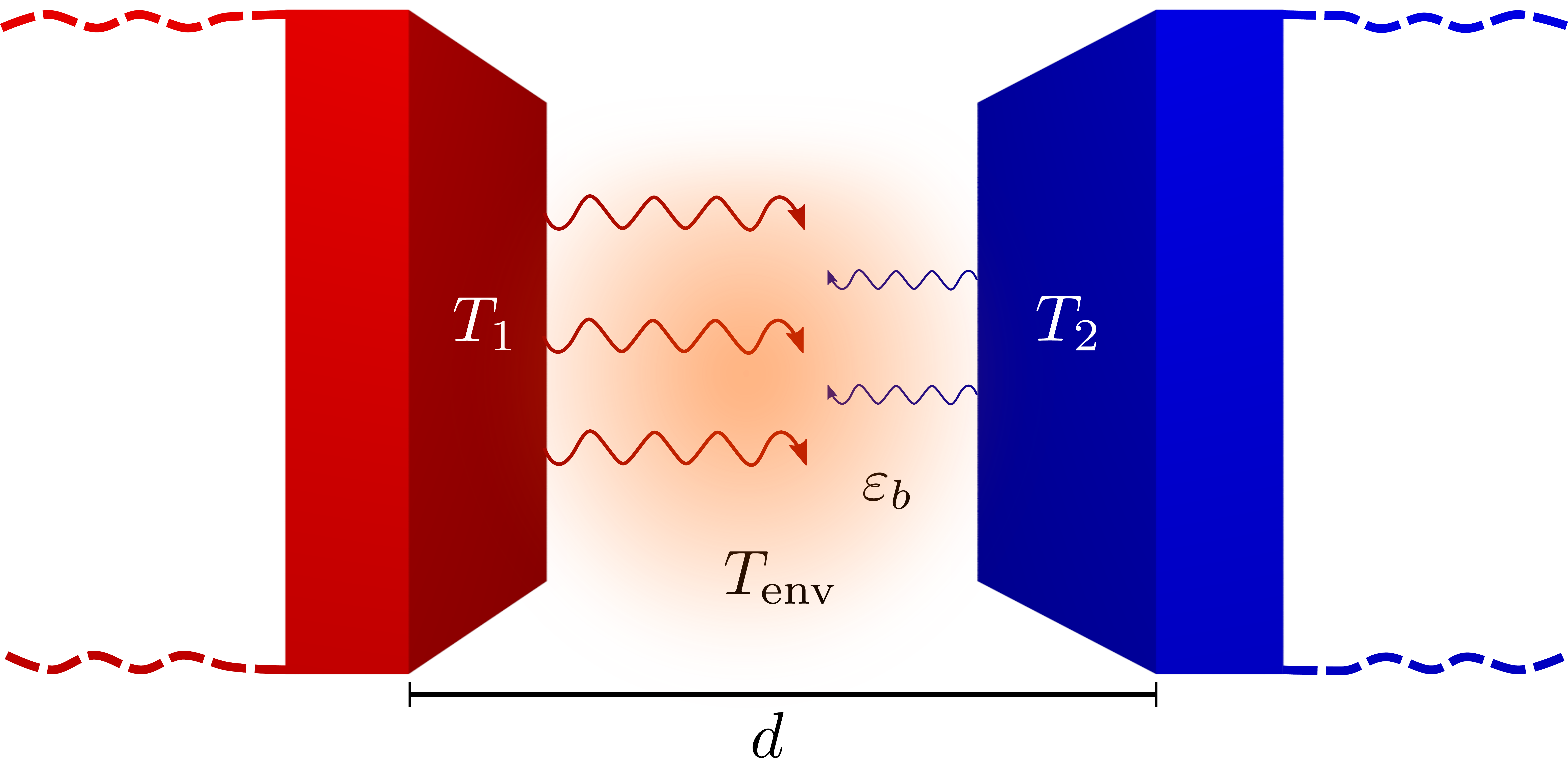}
\caption{\label{fig:TwoPlatesMedium_Setup}Two semi-infinite half-spaces held at different temperatures $T_1> T_2$ and separated by a gap of length $d$. The gap is filled with a non-absorbing homogeneous background medium.}
\end{figure}

\begin{figure}[t!]
\includegraphics[width=\linewidth]{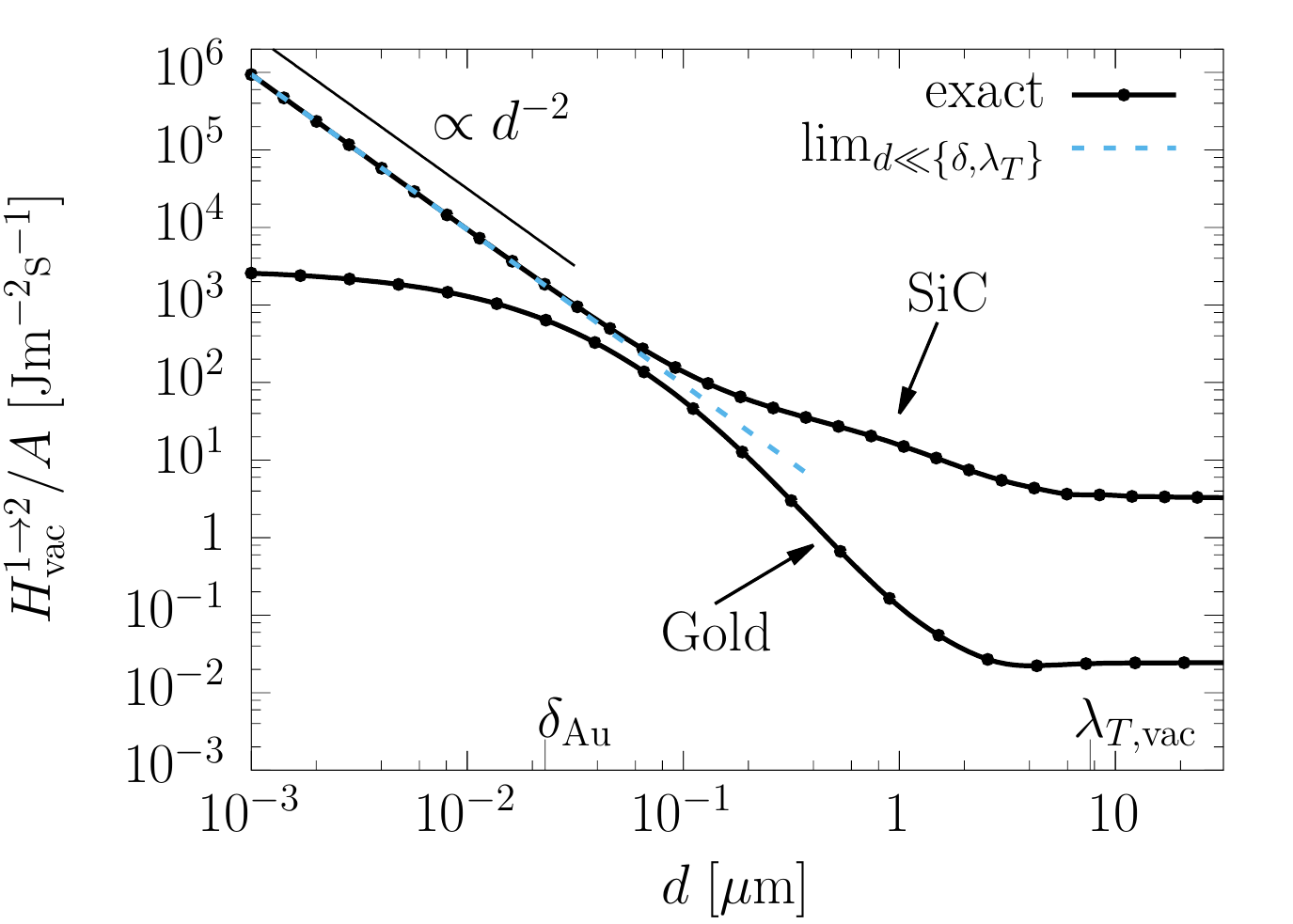}
\caption{\label{fig:HeatTransfer_Vacuum}Heat transfer between two SiC plates (upper curve) and two gold plates (lower curves) embedded in vacuum (i.e., $\varepsilon_b=1$) as a function of the plate distance $d$. The plates are held at the homogeneous temperatures $T_1=\unit[301]{K}$ and $T_2=\unit[300]{K}$. On the submicron scale $d\ll\lambda_{T_1}\approx\unit[8]{\mu m}$, the transfer is strongly enhanced because of photon tunneling processes. The dashed line shows the approximation for two SiC plates in the small distance limit given by Eq.\ \eqref{eq:HeatTransfer_SmallDistances}.}
\end{figure}

The setup of two semi-infinite plates with a medium in between is schematically illustrated in Fig.~\ref{fig:TwoPlatesMedium_Setup}.
We are interested in the heat transfer $H^{1\to 2}$ between the two plates, and start from the general formula for the heat transfer rate $H_\alpha^{(\beta)}$ in Eq.\ \eqref{eq:Heat_Transfer_NObjects_PartialWave}, where for two objects  $\tau_{\overline{\alpha\beta}}=0$. The total heat transferred from object $1$ to object $2$ can be written as the difference of the heat transfer rates between the objects, where we already use the symmetry to write 
\be
H^{1\to 2}=H_1^{(2)}(T_1)-H_1^{(2)}(T_2)\,.
\ee
As found in Sec.~\ref{sec:Three_or_more_objects_heat_absorption}, we saw that the heat transfer in the non-absorbing background medium is given by a redefinition of the involved (scattering) operators compared to the vacuum case.
The heat transfer between two plates in vacuum is well known \cite{polder1971theory,levin1980zh,pendry1999radiative,volokitin2001radiative}, and we can adopt its form to describe the radiative transfer through the non-absorbing medium by redefining the Fresnel coefficients and the ranges of propagating and evanescent waves,
\begin{widetext}
\be\label{eq:TwoPlates_Medium}
\frac{H^{1\to 2}}{A}=\frac{\hbar}{8\pi^3}\int_0^\infty\mathrm{d}\omega\omega \left(n_1(\omega)-n_2(\omega) \right) \sum_P\int\mathrm{d}^2k_\perp\left\{\frac{(1-|r_1^P|^2)(1-|r_2^P|^2)}{|1-e^{2i k_zd}r_1^Pr_2^P|^2}\theta_\mathrm{pr}+\frac{4\Im[r_1^P]\Im[r_2^P]e^{-2|k_z|d}}{|1-e^{-2|k_z|d}r_1^Pr_2^P|^2}\theta_\mathrm{ev}\right\}\,.
\ee
\end{widetext}
We introduced the step functions $\theta_\mathrm{pr}=\theta\left(\frac{\omega}{c}\sqrt{\varepsilon_b}-k_\perp\right)$ and $\theta_\mathrm{ev}=\theta(k_\perp-\frac{\omega}{c}\sqrt{\varepsilon_b})$ to distinguish between propagating and evanescent wave contributions in a medium, as well as $n_\alpha(\omega)=(e^{\frac{\hbar\omega}{k_BT_\alpha}}-1)^{-1}$. Note that the wave vector $\vct{k}_\perp$ perpendicular to the symmetry axes of the plates is measured within the background medium. Its absolute value is related to the vacuum wave vector  by $k_\perp=k_{\perp}^\mathrm{vac}\sqrt{\varepsilon_b}$. 
The Fresnel reflection coefficients $r^P$, given in Appendix~\ref{app:PlaneWaveBasis}, describe waves moving from the medium to the respective plate.
For completeness of our presentation, we show in Fig.~\ref{fig:HeatTransfer_Vacuum} the well-known heat transfer for two SiC plates and two gold plates in vacuum, i.e., for $\varepsilon_b=1$. Optical properties of SiC are taken to be \cite{spitzer1959infrared}
\begin{figure}[t!]
\includegraphics[width=\linewidth]{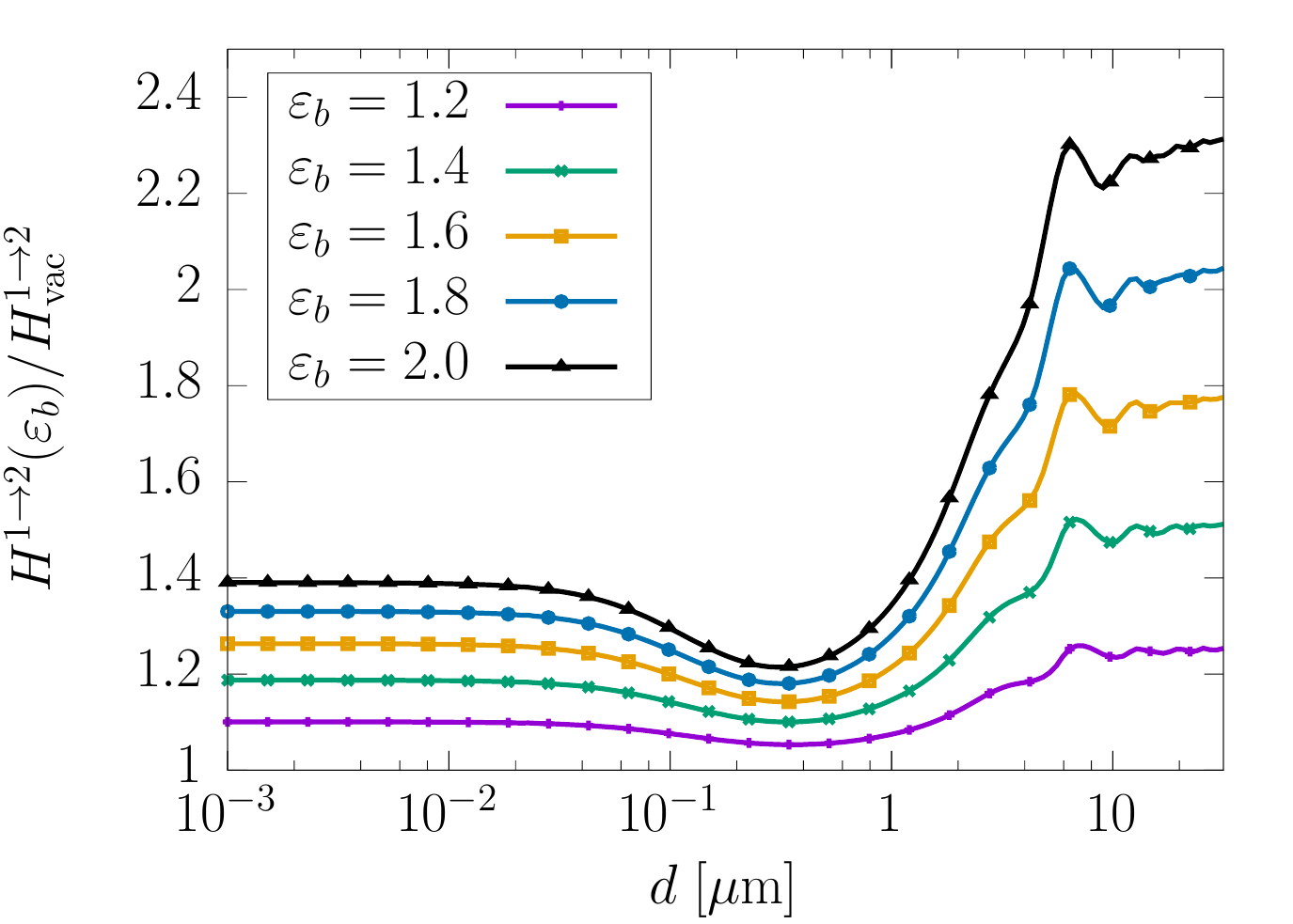}
\caption{\label{fig:HeatTransfer_SiCMedium}Heat transfer between two SiC plates at the homogeneous temperatures $T_1=\unit[301]{K}$ and $T_2=\unit[300]{K}$ embedded in a non-absorbing medium as a function of the plate distance $d$. The transfer is normalized by the result in vacuum (see upper curve in Fig.~\ref{fig:HeatTransfer_Vacuum}).}
\end{figure}
\be
\varepsilon_\mathrm{SiC}=\varepsilon_\infty \frac{\omega^2-\omega_\mathrm{LO}^2+i\omega\gamma}{\omega^2-\omega^2_\mathrm{TO}+i\omega\gamma}\,
\ee
with $\varepsilon_\infty=6.7$, $\omega_\mathrm{LO}=\unit[0.12]{eV}$, $\omega_\mathrm{TO}=\unit[0.098]{eV}$, and $\gamma=\unit[5.88\times 10^{-4}]{eV}$.
For any plate separation $d$, the heat transfer between the SiC plates exceeds the transfer between the gold plates. 
In the near-field $d\ll \lambda_T$, the evanescent wave contribution dominates due to photon tunneling processes \cite{volokitin2007nearfield}, and the transfer scales as $1/d^2$ (not seen yet for gold). The known near-field enhancement is also valid if the vacuum gap is filled with a homogeneous and non-absorbing medium. We find the following approximation that appears to be asymptotically exact in the limit where $d$ is the smallest length scale (cf.~the dashed line for $\varepsilon_b=1$ in Fig.~\ref{fig:HeatTransfer_Vacuum}) 
\begin{align}\label{eq:HeatTransfer_SmallDistances}
&\lim_{d\ll\{\delta,\lambda_T\}}\frac{H^{1\to 2}}{A}=\frac{4\hbar}{d^2\pi^2}\int_0^\infty\mathrm{d}
\omega(n_1(\omega)-n_2(\omega))\omega\int_0^\infty\mathrm{d}\tilde{k}_z\notag\\
&\times \tilde{k}_z\frac{\varepsilon_b^2\Im[\varepsilon_1]\Im[\varepsilon_2]e^{-2 \tilde{k}_z}}{|(\varepsilon_1+\varepsilon_b)(\varepsilon_2+\varepsilon_b)-e^{-2\tilde{k}_z }(\varepsilon_1-\varepsilon_b)(\varepsilon_2-\varepsilon_b)|^2}\,,
\end{align}
where we restricted Eq.\ \eqref{eq:TwoPlates_Medium} to the electric polarization $P=N$ of the evanescent wave contribution, and used the following expansion of the Fresnel reflection coefficient for large wave vectors $k_z$
\begin{align}
r_\alpha^N(k_z,\omega)=\frac{\varepsilon_\alpha(\omega)-\varepsilon_b}{\varepsilon_\alpha(\omega)-\varepsilon_b}+\mathcal{O}\left(\frac{1}{k_z^2}\right)\,.
\end{align}
Fig.~\ref{fig:HeatTransfer_SiCMedium} shows the result for SiC with the medium present.
Generally, we note a strong enhancement of the heat transfer compared to the vacuum case, increasing with increasing value of $\varepsilon_b$. In the small distance regime, where the evanescent wave contribution dominates, we can enhance the heat radiation between $10\%$ and $40\%$ compared to the vacuum case for a little change as  $\varepsilon_b\le 2$. For large distances $d\gg \lambda_T$, we find a larger enhancement up to $230\%$ of the value in vacuum for $\varepsilon_b\le 2$.
\begin{figure}[t!]
\includegraphics[width=\linewidth]{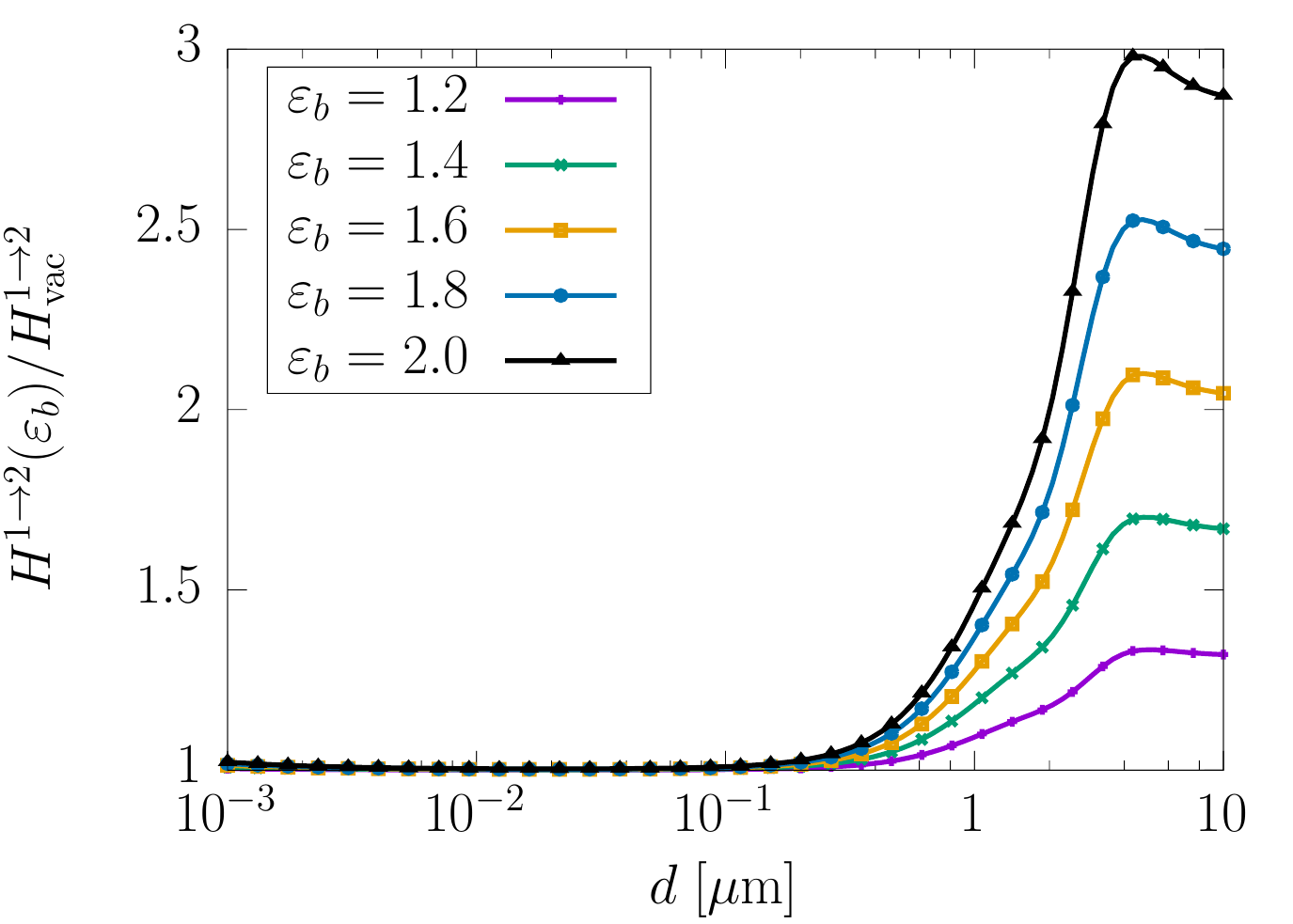}
\caption{\label{fig:HeatTransfer_AuMedium}Heat transfer between two gold plates at the homogeneous temperatures $T_1=\unit[301]{K}$ and $T_2=\unit[300]{K}$ embedded in a non-absorbing medium as a function of the plate distance $d$. The transfer is normalized by the result in vacuum (see lower curve in Fig.~\ref{fig:HeatTransfer_Vacuum}).}
\end{figure}
\begin{figure}[t!]
\includegraphics[width=\linewidth]{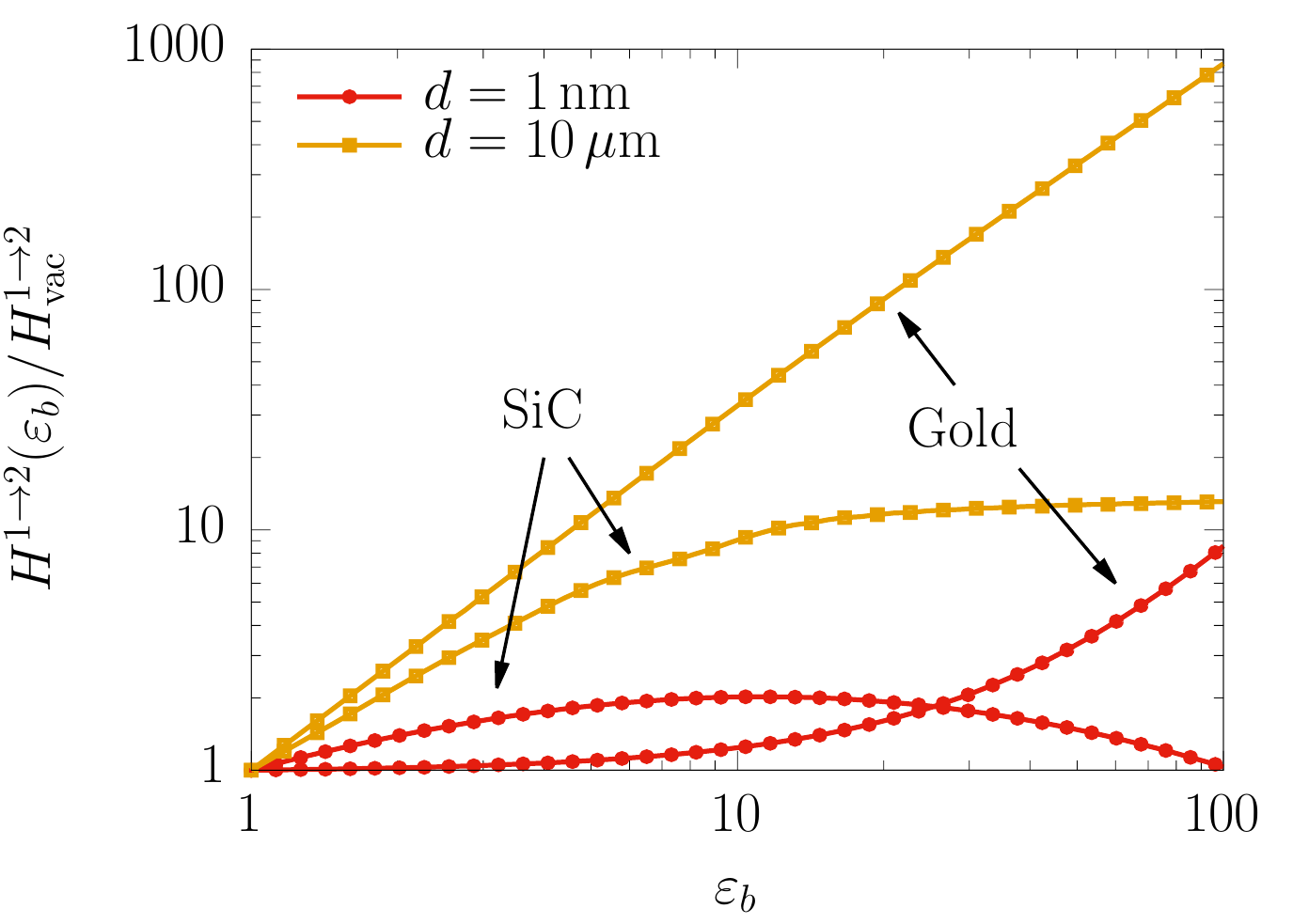}
\caption{\label{fig:HeatTransfer_Limit} Enhancement factor of the heat transfer between two SiC and two gold plates, respectively, in a background medium compared to the vacuum case. Circles symbolize the transfer in the near-field ($d=\unit[1]{nm}$), while squares show plate separations $d=\unit[10]{\mu m}$ (far-field). The temperatures of the plates were chosen identically to Figs.~\ref{fig:HeatTransfer_SiCMedium} and \ref{fig:HeatTransfer_AuMedium}.}
\end{figure}
In Fig.~\ref{fig:HeatTransfer_AuMedium}, we depict the same graph for the case of two gold plates. In contrast to the SiC plates, the amplification factor for small distances is marginally small, i.e., around $2\%$ for $\varepsilon_b=2$. In the regime of the thermal wavelength $d\approx\lambda_T$, we find a strong enhancement of almost $300\%$ in the given interval of $\varepsilon_b$ compared to the vacuum transfer. 

In Fig.~\ref{fig:HeatTransfer_Limit}, we finally show the amplification factor for SiC and gold plates as a function of $\varepsilon_b$, in the near-field $d\ll \lambda_T$, and in the far-field $d>\lambda_T$.

In the former case, we find a maximum for SiC plates at $\varepsilon_b\approx 10$, where the heat transfer is twice as large as in vacuum. For gold in the near-field, the transfer is enhanced significantly only for larger values of $\varepsilon_b$, but then grows strongly. In contrast to SiC, the enhancement factor for gold increases monotonically in the given interval and reaches values of almost ten times the vacuum transfer for $\varepsilon_b<100$. In the far-field, the amplification factor is generally larger than in the near-field for both materials. 

\subsubsection{Two semi-infinite plates and an atom}\label{sec:atom}
\begin{figure}[b!]
\includegraphics[width=\linewidth]{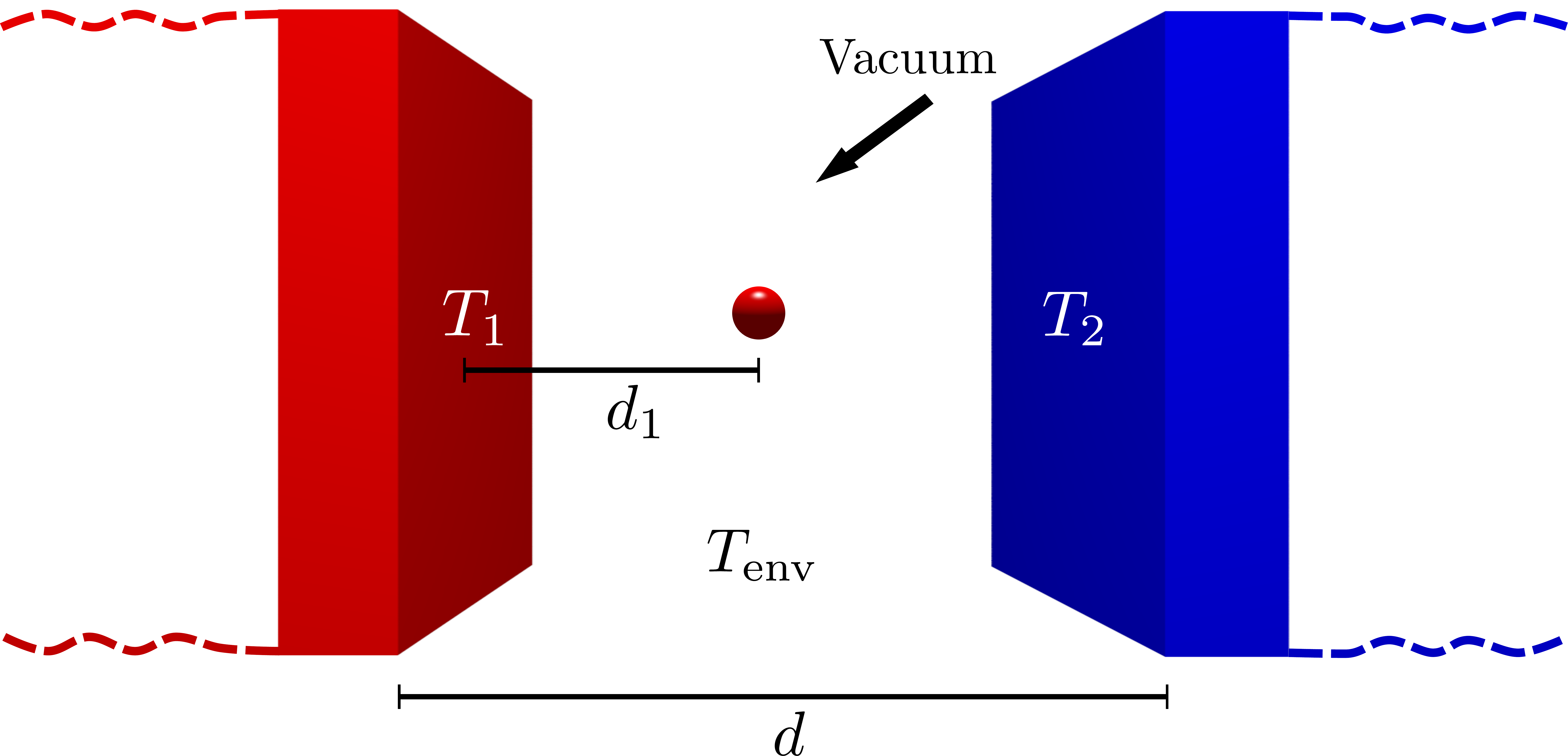}
\caption{\label{fig:TwoPlates_Atom}The arrangement of two semi-infinite half-spaces and an atom in vacuum. The half-spaces are kept at different temperatures $T_1> T_2$ and separated by a gap of length $d$.  The absorptivity of the atom (or nanoparticle) is assumed to be negligible, such that it only scatters the (near- and far-) fields radiated by the plates.}
\end{figure}

The setup under study in this subsection consists of two semi-infinite plates separated by a gap of width $d$ and containing a polarizable particle (an atom) as schematically illustrated in Fig.~\ref{fig:TwoPlates_Atom}. The heat transfer in this three-body configuration can be also calculated using the theory developed in \cite{messina2014three}, where an example of the non-equilibrium Casimir-Polder force acting on the atom has been presented.
Here, we evaluate the heat transfer $H^{1\to 2}$ between the two plates in the presence of the atom (labeled as object 3) from Eq.\ \eqref{eq:Heat_Transfer_NObjects_PartialWave}, where the composite scattering operator ${\cal T}_{13}$ of object $1$ and $3$  appears. This matrix can be expressed through the individual scattering matrices of the objects (see Appendix~\ref{app:Composite_T_operator} for details).
Since the scattering by the atom is generally weak, we linearize Eq.\ \eqref{eq:Heat_Transfer_NObjects_PartialWave} with respect to the scattering matrix $\mathcal{T}$ of the atom. We then find a term additional to the heat transfer between two semi-infinite plates in vacuum 
\be\label{eq:Heat_transfer_TwoPlates_Atom}
H^{1\to 2}=H_\mathrm{vac}^{1\to 2}+\Delta H^{1\to 2}\,.
\ee
Recall that the vacuum heat transfer $H_\mathrm{vac}^{1\to 2}/A$ per surface area between two plates is obtained by setting $\varepsilon_b=1$ in Eq.\ \eqref{eq:TwoPlates_Medium}. The additional term $\Delta H^{1\to 2}$ represents the change of transfer due to the presence of the  atom. Splitting it up into propagating and evanescent parts, we find (recall that $\alpha$ is the (real and non-dispersive) dipole polarizability of the atom)
\begin{widetext}
\begin{align}
\begin{split}\label{eq:DeltaH_Atom_pr}
\Delta H^{1\to 2}_\mathrm{pr}=&\frac{\hbar}{4\pi^3}\int_0^\infty\mathrm{d}\omega (n_1(\omega)-n_2(\omega))\omega\sum_P\int\mathrm{d}^2k_\perp\frac{(1-|r_1^P|^2)(1-|r_2^P|^2)}{|1-e^{2ik_z d}r_1^Pr_2^P|^2}\Re\Bigg[\frac{\alpha}{1-e^{2ik_zd}r_1^Pr_2^P}\Bigg(e^{2ik_zd_1}r_1^Pr_3^P\\
&+e^{2ik_z(d-d_1)}r_3^Pr_2^P+t_3^P+e^{2ik_zd}r_1^Pt_3^Pr_2^P\Bigg)\Bigg]\theta_\mathrm{pr}\,,
\end{split}\\\notag\\
\begin{split}\label{eq:DeltaH_Atom_ev}
\Delta H^{1\to 2}_\mathrm{ev}=&\frac{\hbar}{\pi^3}\int_0^\infty\mathrm{d}\omega (n_1(\omega)-n_2(\omega))\omega\sum_P\int\mathrm{d}^2k_\perp\frac{\Im[r_1^P]\Im[r_2^P]}{|1-e^{-2|k_z| d}r_1^Pr_2^P|^2}\Re\Bigg[\frac{\alpha}{1-e^{-2|k_z|d}r_1^Pr_2^P}\Bigg(e^{-2|k_z|(d+d_1)}r_1^Pr_3^P\\
&+e^{-2|k_z|(2d-d_1)}r_3^Pr_2^P+e^{-4|k_z|d}r_1^Pt_3^Pr_2^P+e^{-2|k_z|d}t_3^P\Bigg)\Bigg]\theta_\mathrm{ev}\,.
\end{split}
\end{align}
\end{widetext}
Again, these terms are linearized with respect to $\alpha$. We provide the expressions for the reflection and transmission coefficients $r_3$ and $t_3$ of the atom in Appendix~\ref{app:PlaneWaveBasis}. In Fig.~\ref{fig:SiC_Atom_NearField}, we illustrate the correction term $\Delta H^{1\to 2}$ for two SiC plates at fixed separation $d$ in the near-field $d=\unit[10]{nm}\ll \lambda_T$ plotted over the atom position $d_1$. The two plates are held at the constant temperatures $T_1=\unit[301]{K}$ and $T_2=\unit[300]{K}$.
We choose the two plate temperatures very closely (still with a huge temperature gradient for small gap widths $d$), to allow comparison of $\Delta H^{1\to 2}$ to kinetic gas theory below. In the graph, the correction term is normalized by the polarizability $\alpha$ of the atom. The symmetry of the heat transfer (see Sec.~\ref{sec:Three_or_more_objects_heat_absorption}) is reflected by the fact that  $\Delta H^{1\to 2}$ is an even function of the atom position $d_1$, where the symmetry axis is located at the center of the gap (only given for identical plates). The influence of the atom on the heat transfer between the two plates is maximal when the atom is in close proximity to one of the plates.
This can be explained by  examining the evanescent contribution in Eq.\ \eqref{eq:DeltaH_Atom_ev}. In the near-field, the correction term is dominated by the electric polarization $P=N$ of the reflection coefficient $r_3^P$ of the atom [see Eq.\ \eqref{eq:Atom_ReflectionCoeff}]. Physically, the impact of the atom on the heat transfer is maximized by minimizing the light path of the electromagnetic waves emitted by one of the plates and reflected by the atom. In Eq.\ \eqref{eq:DeltaH_Atom_ev}, this is achieved by minimizing $d+d_1$ or $2d-d_1$ (see the distance dependence in the exponential functions of the two terms carrying the reflection coefficient $r_3^P$) so that the maxima are at $d_1=0$ and $d_1=d$. 
\begin{figure}[t!]
\includegraphics[width=\linewidth]{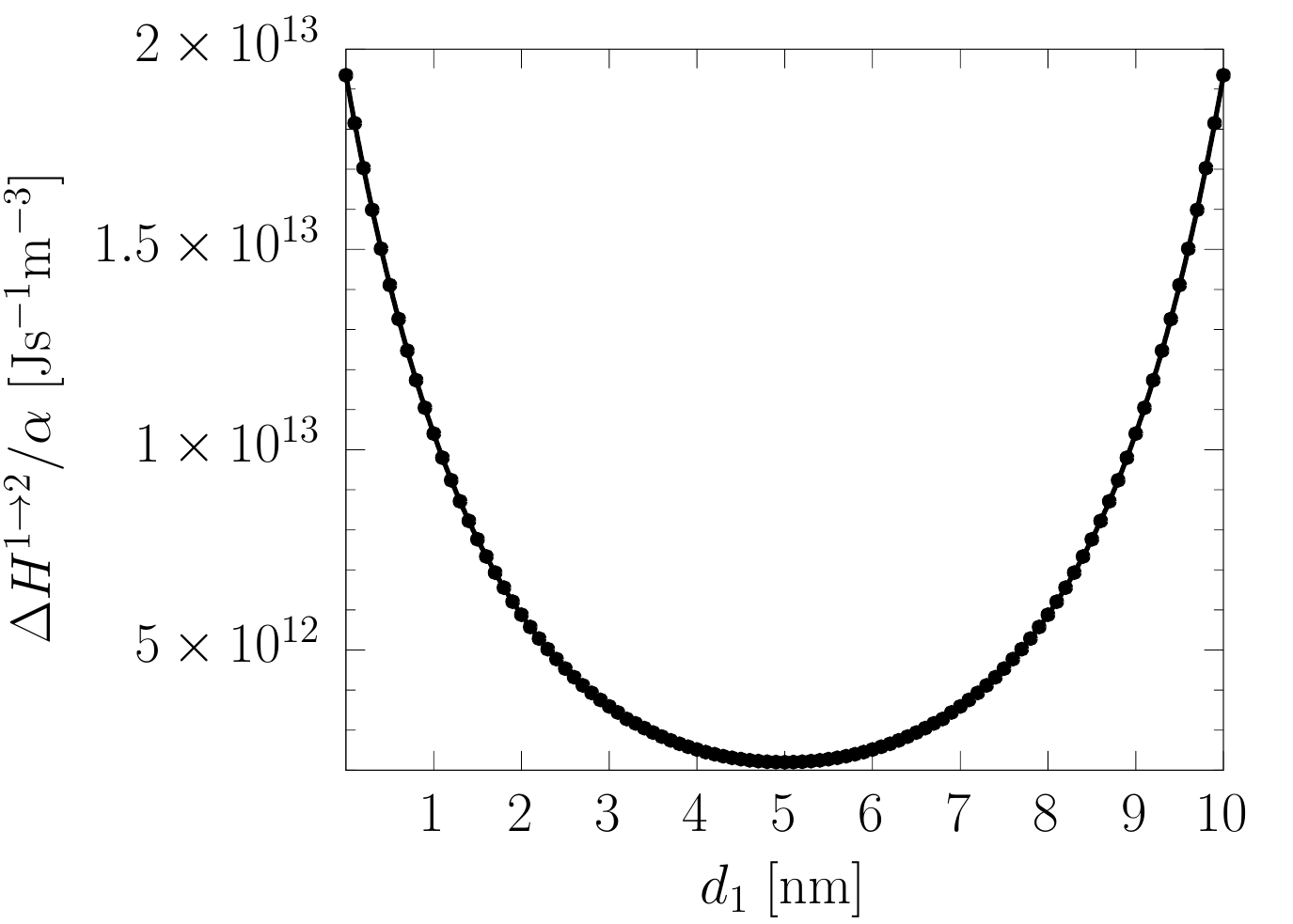}
\caption{\label{fig:SiC_Atom_NearField}Correction term $\Delta H^{1\to 2}$ to the heat transfer between two SiC plates in vacuum at distance $d=\unit[10]{nm}$ due to the presence of a polarizable atom over the atom position $d_1$. The plates are held at the homogeneous temperatures $T_1=\unit[301]{K}$ and $T_2=\unit[300]{K}$. The correction is normalized by the polarizability $\alpha$ of the atom, which is assumed to be a non-dispersive real quantity.}
\end{figure}

\begin{figure}[t]
\includegraphics[width=\linewidth]{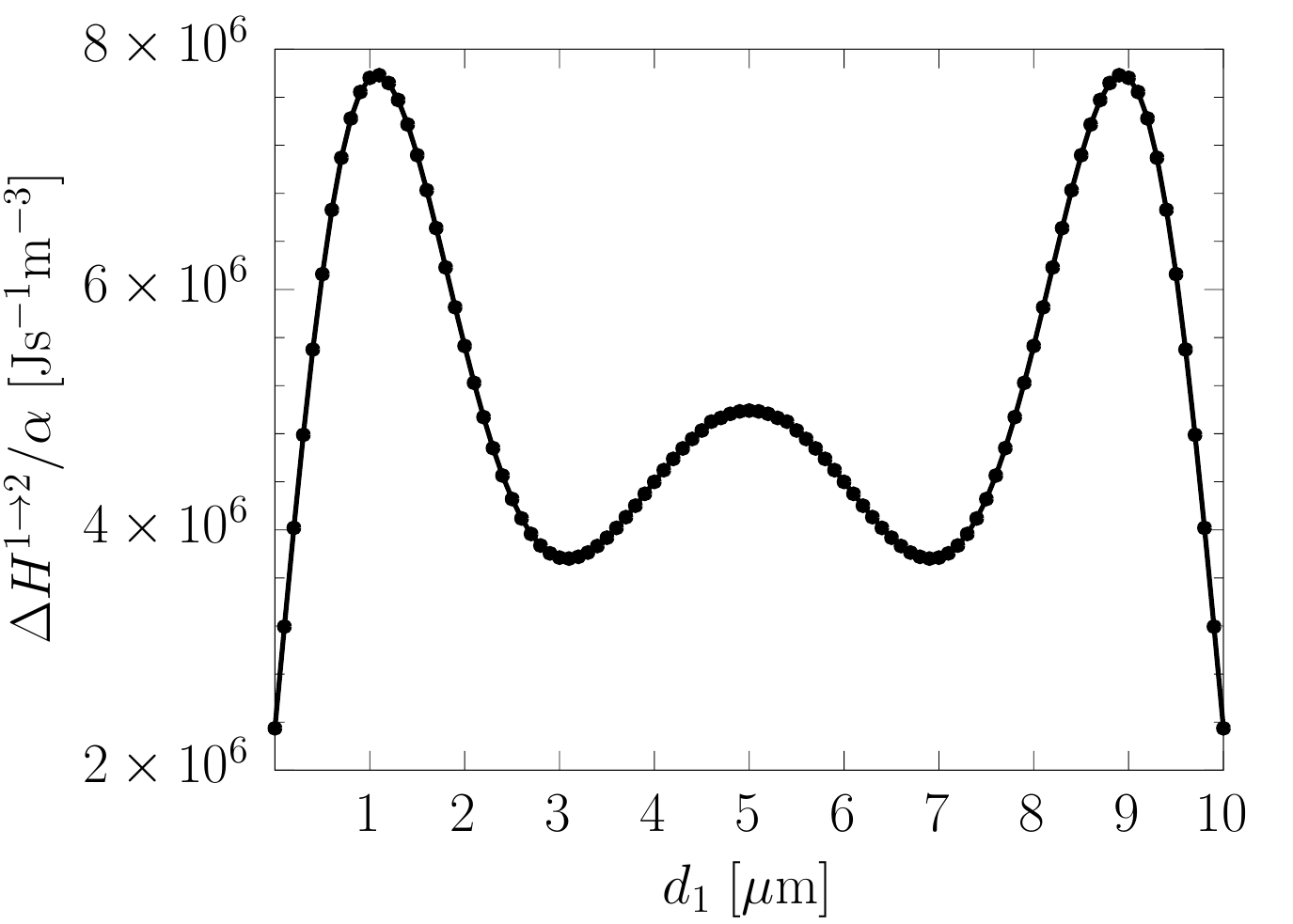}
\caption{\label{fig:SiC_Atom_FarField}Correction term $\Delta H^{1\to 2}$ to the heat transfer between two SiC plates in vacuum at distance $d=\unit[10]{\mu m}$ due to the presence of a polarizable atom over the atom position $d_1$. The other parameters are chosen identically to Fig.~\ref{fig:SiC_Atom_NearField}.}
\end{figure}
In Fig.~\ref{fig:SiC_Atom_FarField}, we show the corresponding graph for a fixed plate separation $d=\unit[10]{\mu m}>\lambda_T$. In this regime, the correction term is dominated by the propagating wave modes in Eq.\ \eqref{eq:DeltaH_Atom_pr} and is more than six orders of magnitude smaller than in the near-field. The correction term is now minimized when the atom is located at one of the plates. Moving the atom away from the plate leads to a global maximum at roughly $d_1\approx \unit[1]{\mu m}$. In the center of the gap we observe another local maximum. The oscillatory behavior of the curve is due to interference effects of the reflected and transmitted waves.

\subsubsection{Comparison of the approaches of subsections \ref{sec:med} and \ref{sec:atom} in the dilute limit}
In subsection \ref{sec:med}, we regarded the case of a homogeneous background medium filling the gap between the plates, while in subsection \ref{sec:atom}, we analyzed the situation of a single atom between the two plates.
These cases are expected to share the limit of dilution, as microscopically, a diluted medium consists of  atoms. We integrate the results for the atom in Eqs.~\eqref{eq:DeltaH_Atom_pr} and \eqref{eq:DeltaH_Atom_ev} over the position $d_1$ of the atom,  assuming a homogeneous distribution. We then multiply the result by a finite particle number density, as dictated by the theory of Clausius-Mossotti (i.e., assuming the effect of the atoms to be additive), 
\be
\varepsilon_b-1=4\pi n_b\alpha_b\,.
\ee

\begin{figure}[t]
\includegraphics[width=\linewidth]{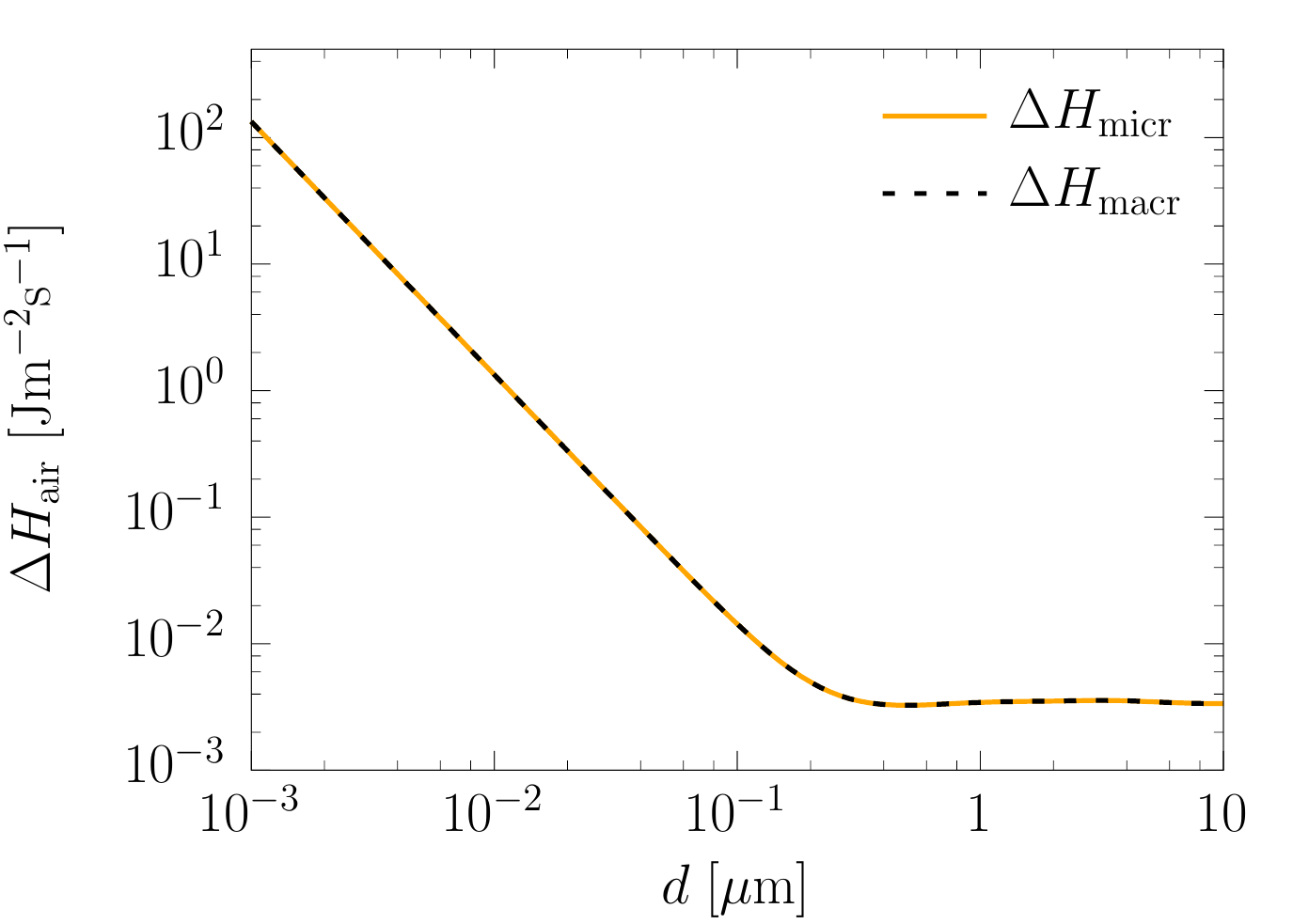}
\caption{\label{fig:DeltaHAir} Comparison between the microscopic and the macroscopic model for the change in the heat transfer due to the presence of air between two SiC half-spaces at temperatures $T_1=\unit[301]{K}$ and $T_2=\unit[300]{K}$. The completely different approaches yield the same result in the dilute limit.}
\end{figure}

With this relation, the two approaches can now be compared, as shown in Fig.~\ref{fig:DeltaHAir} for the case of air (where we use $\varepsilon_\mathrm{air}=1.00059$ and $n_\mathrm{air}=\unit[2.55\times 10^{19}]{cm^{-3}}$, which are typical numbers at sea level \cite{hippel1954dielectric}). The two curves are (numerically) identical for any plate distance $d$. Coming from two completely different starting points, we have thus demonstrated that the two approaches are consistent and yield the same result in the dilute limit as expected.

\subsubsection{Comparison to heat transfer via kinetic gas theory}
With a medium present, the energy transferred via electromagnetic radiation is in competition to other mechanisms of thermal conduction, i.e., by phonons or collisions  of gas atoms. For the case of a gas, such transfer can be expressed by simple formulas: For the regime where the distance between the plates is small compared to the mean free path $l$ of the gas molecules, $d\ll l$, we can assume the particles of the medium to traverse the gap between the two plates unhindered and therefore use a simple model from kinetic gas theory to describe the energy flux evoked by collisions of the atoms with the two surfaces,
\be\label{eq:Hkin_ballistic}
\lim_{d\ll l}\frac{H_\mathrm{kin}}{A}\approx\frac{3}{2}k_B \Delta T n_b \bar{v}_x\,. 
\ee
In this equation, $\Delta T=T_1-T_2$ is the temperature difference between the two plates, $n_b$ is the particle number density of the dilute gas, and $\bar{v}_x=\sqrt{\frac{k_B T}{m}}$ is the mean velocity of the particles pointing in the direction normal to the plates, with molecule mass $m$. One may use $T=T_1$ or $T=T_2$ for an estimate, as $T_1$ and $T_2$ are almost equal.  On the other hand, for distances $d$ much larger than the mean free path, i.e.\ $d\gg l$, the kinetic part is described by Fourier's law for thermal conduction, being proportional to $d^{-1}$ for the case of two plates,
\be\label{eq:Hkin_fourier}
\lim_{d\gg l}\frac{H_\mathrm{kin}}{A}=\kappa\frac{\Delta T}{d}\,.
\ee
\begin{figure}[t]
\includegraphics[width=\linewidth]{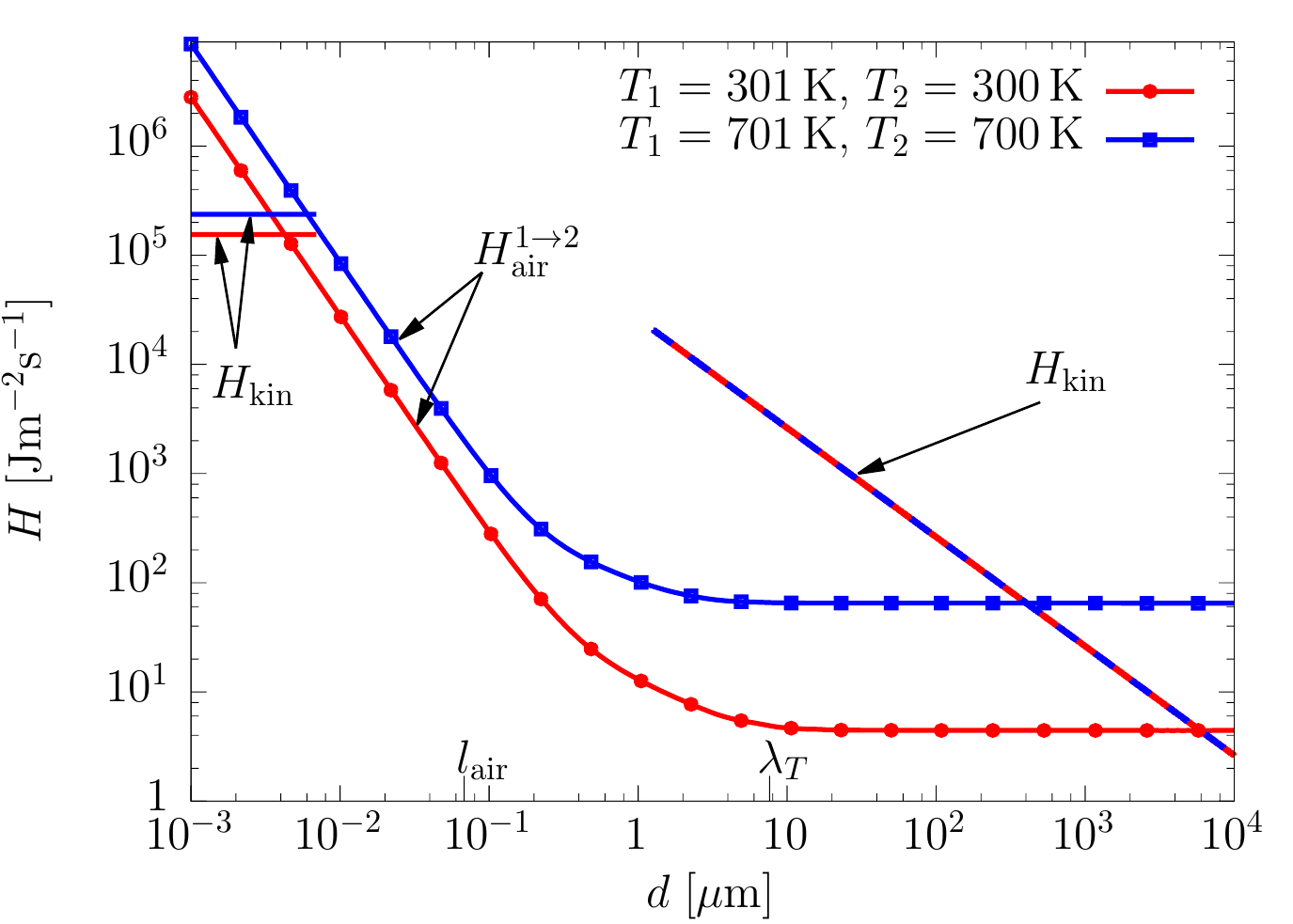}
\caption{\label{fig:SiO2_HeatTransferAir} Energy fluxes in the system of two plates embedded in air. The radiative component $H^{1\to 2}_\mathrm{air}$ surpasses the kinetic part  $H_\mathrm{kin}$ both in the near and in the far-field.  Increasing the temperature of the plates shifts the point of intersection of the two contributions to larger plate separations $d$ in the near-field and to smaller plate separations in the far-field, respectively.}
\end{figure}
Here, $\kappa$ is the thermal conductivity. Since for air, $l\approx \unit[68]{nm}$ \cite{jennings1988the}, the optical near-field should be compared to Eq.\ \eqref{eq:Hkin_ballistic}, while the optical far-field should be compared to Eq.\ \eqref{eq:Hkin_fourier}.  Recently, it was shown that the radiative heat exchange between two closely spaced bodies in vacuum can in principle surpass conductive heat transfer through air at room temperature \cite{miller2015shape}. Using the derived formula for the heat transfer in a background medium, we can explicitly analyze the case of  two SiO$_2$ plates surrounded by air (again $\varepsilon_\mathrm{air}=1.00059$ \cite{hippel1954dielectric}). In Fig.~\ref{fig:SiO2_HeatTransferAir}, we show both the kinetic part $H_\mathrm{kin}$ and the radiative part of energy transfer $H^{1\to 2}_\mathrm{air}$. In general, the radiative part exceeds the kinetic part in both limits for small and large $d$.  In the near-field, the radiative energy flux grows as $d^{-2}$, and exceeds the kinetic part, which in the regime of Eq.\ \eqref{eq:Hkin_ballistic} is independent of $d$. For the parameters chosen, the cross over is at roughly 10~nm. For larger plate temperatures, it generally shifts to larger distances, as the radiative part grows stronger with temperature than the kinetic part. 

In the far-field, radiation also eventually dominates, as the kinetic part vanishes with $1/d$. Here, depending on the temperatures, we find the cross over at $d\approx\unit[400]{\mu m}$ or $d\approx\unit[6]{mm}$. (We neglect the temperature dependence of $\kappa_\mathrm{air}=\unit[0.0262]{W/(m\;K)}$.) 

While the numbers we have found are already promising concerning experimental detection, we believe that, e.g.\ using metamaterials, interesting scenarios regarding competition of radiation and other transport mechanisms can be proposed in the future.  

\section{Summary}
Closed formulas for heat radiation and radiative heat transfer in a many body system embedded in a non-absorbing background medium (described by real magnetic and electric permeabilities $\mu_b$ and $\varepsilon_b$) were derived. In contrast to the case of objects in vacuum ($\mu_b=\varepsilon_b=1$), these formulas contain the classical scattering operators which describe scattering of objects in the corresponding background medium. The formulas show a number of general properties (such as positivity and symmetry). The presence of a non-absorbing background medium can have very strong effects on the resulting radiation: Both the radiation of a sphere, as well as the heat transfer between two parallel plates can be enhanced by orders of magnitude, this enhancement effect being especially strong for metals such as gold: For example, the radiation of a gold nanosphere in a medium with $\varepsilon_b\approx 6.5$ is a hundred times larger compared to vacuum, an effect which can be relevant for composite materials. The enhancement is generally smaller for near-field quantities, but can still be pronounced: The near-field transfer between two SiC plates is doubled at $\varepsilon_b\approx 8$. For very large values of $\varepsilon_b$, the found results are expected to depend on the details of the interface between object and medium, an expectation which should be explored further in the future. Concerning near- and
far-field transfer through a gas like air, both a microscopic model (based on gas particles) and a macroscopic model (using a dielectric constant) can be used with equal accuracy. The radiative transfer through air exceeds the energy transfer from kinetic gas theory for both very small and very large distances.

The generality of the derived formalism provides incentive for future studies of heat radiation and radiative heat transfer in complex environments.

The theory derived in this manuscript can also be useful for possible implementation and improvement of the performances of the recently proposed non-equilibrium-based many-body quantum thermal machines \cite{doyeux2016quantum} and configurations to create and/or protect entanglement \cite{bellomo2015nonequilibrium}.

\begin{acknowledgments}
BM, RI and MK were supported by MIT-Germany Seed Fund Grant No. 2746830 and Deutsche Forschungsgemeinschaft (DFG) Grant No. KR 3844/2-1.
\end{acknowledgments}

\begin{appendix}
\section{Non-equilibrium field fluctuations}\label{app:Electric_field_correlator}
\subsection{The spectral density}
Random processes of spontaneous fluctuations can in general be described by means of correlation functions. For any two field operators $\hat{A}$ and $\hat{B}$ in the Heisenberg picture, we employ the symmetrized expectation value \cite{ekstein1955quantum}
\be
\langle \hat{A}(t,\vct{r})\hat{B}(t',\vct{r}')\rangle_s\equiv\frac{1}{2}\langle\hat{A}(t,\vct{r})\hat{B}(t',\vct{r}')+\hat{B}(t',\vct{r}')\hat{A}(t,\vct{r})\rangle\,.
\ee
In stationary conditions, the correlation function is invariant under time translation and depends only on the time difference $t-t'$. Accordingly, we define the spectral density of fluctuations $\langle A(\vct{r})B^*(\vct{r}')\rangle_\omega$ by \cite{landau2013course}
\be
\langle \hat{A}(t,\vct{r})\hat{B}(t',\vct{r}')\rangle_s=\int_{-\infty}^\infty\frac{\mathrm{d}\omega}{2\pi}e^{-i\omega(t-t')}\langle A(\vct{r})B^*(\vct{r}')\rangle_\omega\,.
\ee
Note that the spectral density satisfies the reality condition
\be\label{eq:Reality_condition}
\langle A(\vct{r})B^*(\vct{r}')\rangle_\omega^*=\langle A(\vct{r})B^*(\vct{r}')\rangle_{-\omega}\,.
\ee

\subsection{Electric field correlator}
The spectral density of the electric field $\vct{E}$ at points $\vct{r}$ and $\vct{r}'$ is given by
\be
C_{ij} \equiv \langle E_i(\vct{r})E_j^*(\vct{r}')\rangle_\omega\,.
\ee
In global thermal equilibrium, the spectral density of the electric field is related to the imaginary part of the dyadic retarded Green's function of the system via the fluctuation dissipation theorem \cite{eckhardt1984macroscopic,rytov1989elements}:
\be\label{eq:FDT}
C^\mathrm{eq}_{ij}(T;\vct{r},\vct{r}')=[a(T)+a_0]\Im[G_{ij}(\vct{r},\vct{r}')]\,.
\ee
The two contributions are identified with the zero-point and thermal fluctuations with amplitude
\begin{align}\label{eq:Amplitude_T}
a(T)&\equiv\mathrm{sgn}(\omega)\frac{8\pi\hbar\omega^2}{c^2}[\exp(\hbar\vert\omega\vert/k_BT)-1]^{-1}\,,\\\label{eq:Amplitude_Zero}
a_0&\equiv\mathrm{sgn}(\omega)\frac{4\pi\hbar\omega^2}{c^2}\,.
\end{align}
Note that $c$ is the vacuum speed of light, $\hbar$ is the reduced Planck constant, $k_B$ is the Boltzmann constant, and $T$ is the temperature of the equilibrium system.

\section{Redefined operators}\label{app:Redefinied_operators}
\subsection{Radiation operator}
The radiation operator $\mathbb{R}_\alpha$ specifies the radiation of an arbitrary object in isolation. Corresponding to the vacuum case in Ref.~\cite{kruger2012trace}, we redefine this operator by
\begin{align}
\mathbb{C}_\alpha(T_\alpha)&=a(T_\alpha)\mathbb{R}_\alpha\,,\\
\mathbb{R}_\alpha&\equiv\G_b\left[\Im[\T_\alpha]-\T_\alpha\Im[\G_b]\T_\alpha^*\right]\G_b^*\,.
\end{align}
Note that $\mathbb{C}_\alpha$ is the field correlator of the isolated object~$\alpha$ as defined in Eq.\ \eqref{eq:Corr_Sc} (where the Green's function $\G_\alpha$ appears) and $\T_\alpha$ is the individual scattering operator of this object.

\subsection{Multiple scattering operator}
The multiple scattering operator $\mathbb{O}_\alpha$ describes the scattering of the field $\vct{E}^\mathrm{iso}_{b,\alpha}$ radiated by the isolated object $\alpha$ at all other objects in the system. For objects in a background medium, this operator is found to be
\begin{align}
\vct{E}^\mathrm{sc}_{b,\alpha}&=\mathbb{O}_\alpha\vct{E}^\mathrm{iso}_{b,\alpha}\,,\\\label{eq:Multiple_scattering}
\mathbb{O}_\alpha &=(1+\G_b\T_{\overline{\alpha}})\frac{1}{1-\G_b\T_\alpha\G_b\T_{\overline{\alpha}}}\,.
\end{align}
For a detailed derivation of this important operator we refer the reader to Ref.~\cite{kruger2012trace}. Note that $\T_\alpha$ is the scattering operator of the single object $\alpha$, while $\T_{\overline{\alpha}}$ is the composite $T$ operator of the residual objects.

\section{Non-absorbing background medium}\label{app:Nonabsorbing_background_medium}
We want to prove that the third term $\mathbb{O}_\alpha\mathbb{R}_\alpha\mathbb{O}_\alpha^\dagger\mathbb{V}_b^*$ in Eq.\ \eqref{eq:SplitUp} does not contribute to the heat transfer rate $H_\alpha^{(\beta)}$ given in Eq.\ \eqref{eq:Heat_Transfer_NObjects} for the case of a non-absorbing background potential. The derivation of the ``self''-emission $H_\beta^{(\beta)}$ leads to a similar term that can be treated in the same way as presented in this appendix. Since the volume integral in Eq.\ \eqref{eq:Heat_Transfer_NObjects} is restricted to the volume $V_\beta$ of object $\beta$, and we do not allow the potential $\mathbb{V}_b$ to connect points in the background medium and the objects, we can introduce a redefined background potential
\be
\tilde{\mathbb{V}}_b(\vct{r},\vct{r}')=\theta(\vct{r}_\beta-\vct{r})\theta(\vct{r}_\beta-\vct{r}')\mathbb{V}_b(\vct{r},\vct{r}')\,.
\ee
Note that $\theta(\vct{r}_\beta-\vct{r})$ is the Heaviside step function, where $\vct{r}_\beta$ is the coordinate attached to the surface of object~$\beta$. Because $\tilde{\mathbb{V}}_b(\vct{r},\vct{r}')$ is only nonzero if both arguments are located within the volume $V_\beta$, the range of integration can be extended to all space to obtain the following operator trace
\be\label{eq:Contribution_Medium}
-\frac{2\hbar}{\pi}\int_0^\infty\mathrm{d}\omega\frac{\omega}{e^{\frac{\hbar\omega}{k_BT_\alpha}}-1}\Im\Tr[\mathbb{O}_\alpha\mathbb{R}_\alpha\mathbb{O}_\alpha^\dagger\tilde{\mathbb{V}}_b^*]\,.
\ee
As discussed in the main text, the potential of the non-absorbing background medium is Hermitian, $\tilde{\mathbb{V}}_b=\tilde{\mathbb{V}}_b^\dagger$. Since $\mathbb{R}_\alpha$ is a Hermitian operator as well, which can be directly seen from its definition in Eq.\ \eqref{eq:Radiation_operator}, the operator to be traced in the preceding equation consists of a product of two Hermitian operators. By using the properties of the trace, we can prove that the trace of a product of two Hermitian operators is always real, although the product is not necessarily Hermitian in general. For two arbitrary Hermitian operators $\mathbb{A}$ and $\mathbb{B}$ we have \cite{zhang2011matrix}
\be
\Tr\{\mathbb{A}\mathbb{B}\}=\Tr\{(\mathbb{B}\mathbb{A})^\dagger\}=\Tr\{\mathbb{B}\mathbb{A}\}^*=\Tr\{\mathbb{A}\mathbb{B}\}^*\,,
\ee
where we used the invariance of the trace under transposition in the second step and the cyclicity of the trace in the last step. The last equality shows that $\Tr\{\mathbb{A}\mathbb{B}\}$ is a real number and we conclude $\Im\Tr[\mathbb{A}\mathbb{B}]=0$. 
As a consequence, the imaginary part of the trace in Eq.\ \eqref{eq:Contribution_Medium} is identically zero for any non-absorbing background medium with Hermitian potential $\mathbb{V}_b$.

\section{Positivity and symmetry of transfer}\label{app:Positivity_and_symmetry_of_transfer}

Eq.\ \eqref{eq:Heat_Transfer_NObjects2} gives the general formula for the heat transfer rate $H_\alpha^{(\beta)}$ in the presence of $N$ objects embedded in a passive non-absorbing background medium. It is straightforward to prove the positivity and symmetry of this expression. For this purpose, we define the operator
\be
\mathbb{W}_{\alpha\beta}=\G_b^{-1}\frac{1}{1-\G_b\T_{\overline{\beta}}\G_b\T_\beta}(1+\G_b\T_{\overline{\alpha\beta}})\frac{1}{1-\G_b\T_\alpha\G_b\T_{\overline{\alpha\beta}}}\,.
\ee
Eq.\ \eqref{eq:Heat_Transfer_NObjects2} can then be rewritten in compact form as
\be\label{eq:Heat_Transfer_NObjects_compact}
H_\alpha^{(\beta)}=\frac{2\hbar}{\pi}\int_0^\infty\mathrm{d}\omega\frac{\omega}{e^{\frac{\hbar\omega}{k_BT_\alpha}}-1}\Tr\{\mathbb{R}_\beta^*\mathbb{W}_{\alpha\beta}\mathbb{R}_\alpha\mathbb{W}_{\alpha\beta}^\dagger\}\,.
\ee
For the positivity of transfer, we need to show that the radiation operator $\mathbb{R}_\alpha$ as defined in Eq.\ \eqref{eq:Radiation_operator} is positive semidefinite. By virtue of the symmetry relation for the potential difference
\be
\Delta\mathbb{V}_\alpha(-\omega)=\Delta\mathbb{V}_\alpha^*(\omega)\,,
\ee
we restrict ourselves to positive frequencies $\omega$. For any body made of passive material, the imaginary part of the potential difference is positive semidefinite, i.e.
\be
\Im[\Delta\mathbb{V}_\alpha]\ge 0\,.
\ee
Recalling that for any positive semidefinite operator $A$, the product $BAB^\dagger$ is also positive semidefinite, we can write
\be
\G_b^{-1}\G_\alpha\Im[\Delta\mathbb{V}_\alpha]\G_\alpha^*\G_b^{*-1}\ge 0\,.
\ee
Application of Eq.\ \eqref{eq:TOperator_medium} and \eqref{eq:GreenBackground_T_Medium} then yields the desired property for the radiation operator $\mathbb{R}_\alpha$
\be
\mathbb{R}_\alpha=\G_b[\Im[\T_\alpha]-\T_\alpha\Im[\G_b]\T_\alpha^*]\G_b^*\ge 0\,.
\ee
Finally, as the product of the two positive semidefinite operators, $\mathbb{R}_\beta^*$ and $\mathbb{W}_{\alpha\beta}\mathbb{R}_\alpha\mathbb{W}_{\alpha\beta}^\dagger$, is also positive semidefinite, we conclude that the heat transfer rate $H_\alpha^{(\beta)}$ in a many body system including the background medium in Eq.\ \eqref{eq:Heat_Transfer_NObjects_compact} is non-negative,
\be
H_\alpha^{(\beta)}\ge 0\,.
\ee
In order to prove the symmetry of the heat transfer, i.e.\ $H_\alpha^{(\beta)}(T)=H_\beta^{(\alpha)}(T)$, we need to show the following equality of operators
\be
\mathbb{W}_{\alpha\beta}^\dagger=\mathbb{W}_{\beta\alpha}^*\,.
\ee
This relation can be straightforwardly proven by making use of the symmetry of $\G_b$ and $\T$ as well as the decomposition of the composite scattering operator $\T_{\overline{\beta}}$ (see Appendix~\ref{app:Composite_T_operator}). As a result, we can write
\begin{align}
\begin{split}
H_\alpha^{(\beta)}(T_\alpha)&=\frac{2\hbar}{\pi}\int_0^\infty\mathrm{d}\omega\frac{\omega}{e^{\frac{\hbar\omega}{k_BT_\alpha}}-1}\Tr\{\mathbb{R}_\beta^*\mathbb{W}_{\alpha\beta}\mathbb{R}_\alpha\mathbb{W}_{\beta\alpha}^*\}\,\\
&=H_\beta^{(\alpha)}(T_\alpha)\,.
\end{split}
\end{align}
In the second line we have used the cyclic property of the trace, and furthermore, we have taken the complex conjugate of the expression since it is real. The symmetry and positivity of the heat transfer in the presence of an arbitrary number of other objects can be used to show that the net heat flux between a warm object $\alpha$ and a cold object $\beta$ is always positive, i.e.\ that heat is always transferred from the warmer object to the colder one:
\be
H_\alpha^{(\beta)}(T_\alpha)-H_\alpha^{(\beta)}(T_\beta)\ge 0 \quad \mathrm{if\;} T_\alpha\ge T_\beta\,.
\ee
This relation holds because of the monotonic increase of $(\exp[\hbar\omega/k_B T]-1)^{-1}$ with $T$ for any $\omega$. And it also relies on the assumption that the objects' potentials $\{\mathbb{V}_\alpha\}$ do not depend on temperature. 

\section{Spherical wave basis}\label{app:SphericalWaveBasis}
\subsection{Partial waves and Green's function}
In the main text, we stated that the Green's function $\G_b$, describing an homogeneous, isotropic and local background medium, can be traced back to the partial wave expansion of the free Green's function $\G_0$ given by the relation in Eq.\ \eqref{eq:G_background_G0}. For explicit knowledge of the expansion of the free Green's function in spherical waves, we refer the interested reader to Ref.~\cite{kruger2012trace}.

\subsection{$\mathcal{T}$ matrix of a sphere}
The scattering problem of a homogeneous sphere of radius $R$ is an exactly solvable problem. The matrix elements of $\mathcal{T}$ are well known \cite{bohren2008absorption} and sometimes referred to as Mie coefficients. Considering spheres with isotropic and local $\varepsilon$ and $\mu$, renders the matrix diagonal and independent of $m$, $\mathcal{T}^{P'P}_{l'lm'm}=\mathcal{T}^P_l\delta_{PP'}\delta_{ll'}\delta_{mm'}$. The matrix elements can be conveniently written in terms of $R^*=\sqrt{\varepsilon_b\mu_b}R\omega/c$ and $\tilde{R}^*=\sqrt{\varepsilon \mu}R\omega/c$ as
\be
\mathcal{T}_l^N=-\frac{\frac{\varepsilon}{\varepsilon_b}j_l(\tilde{R}^*)\frac{\mathrm{d}}{\mathrm{d}R^*}[R^* j_l(R^*)]-j_l(R^*)\frac{\mathrm{d}}{\mathrm{d}\tilde{R}^*}[\tilde{R}^* j_l(\tilde{R}^*)]}{\frac{\varepsilon}{\varepsilon_b}j_l(\tilde{R}^*)\frac{\mathrm{d}}{\mathrm{d}R^*}[R^* h_l(R^*)]-h_l(R^*)\frac{\mathrm{d}}{\mathrm{d}\tilde{R}^*}[\tilde{R}^* j_l(\tilde{R}^*)]}\,.
\ee
$j_l$ is the spherical Bessel function of order $l$, and $h_l$ is the spherical Hankel function of the first kind of order $l$. $\mathcal{T}^M_l$ follows from $\mathcal{T}^N_l$ by interchanging $\varepsilon$ and $\mu$ as well as $\varepsilon_b$ and $\mu_b$. Note that the matrix elements for a sphere in vacuum are restored by taking the limit $\varepsilon_b=\mu_b=1$.

\section{Plane wave basis}\label{app:PlaneWaveBasis}
\subsection{Green's function and $\mathcal{T}$ matrix of a plate}
The expansion of the Green's function $\G_b$ in plane waves is found with the help of Eq.\ \eqref{eq:G_background_G0} and the representation of the vacuum waves given in Ref.~\cite{kruger2012trace}. For the $\mathcal{T}$ matrix of a plate in a homogeneous, isotropic and local background medium, we may resort to its representation in vacuum \cite{kruger2012trace}, where we only need to adjust the Fresnel coefficients as given below for the case of an infinitely thick plate.

\subsection{Fresnel coefficients of the plate}
The Fresnel reflection coefficients for an infinitely thick plate embedded in a background medium are given in Ref.~\cite{jackson1999classical}:
\be
r^N(k_\perp,\omega)=\frac{\frac{\varepsilon}{\varepsilon_b}\sqrt{\frac{\omega^2}{c^2}\varepsilon_b\mu_b-k_\perp^2}-\sqrt{\varepsilon\mu\frac{\omega^2}{c^2}-k_\perp^2}}{\frac{\varepsilon}{\varepsilon_b}\sqrt{\frac{\omega^2}{c^2}\varepsilon_b\mu_b-k_\perp^2}+\sqrt{\varepsilon\mu\frac{\omega^2}{c^2}-k_\perp^2}}
\ee
$r^M$ is obtained from $r^N$ by interchanging $\varepsilon$ and $\mu$ as well as $\varepsilon_b$ and $\mu_b$. In order to regain the Fresnel coefficients in the vacuum case, one has to set $\varepsilon_b=\mu_b=1$.

\subsection{Fresnel coefficients of the atom}
The Fresnel reflection and transmission coefficients of a polarizable atom in vacuum with dipole polarizability $\alpha$ are straightforwardly derived from the definition of the $\mathcal{T}$ matrix in Eq.\ \eqref{eq:T_partialwaves}. For the reflection coefficients we obtain
\begin{align}\label{eq:Atom_ReflectionCoeff}
r^N&=2\pi i\frac{\omega^2}{c^2}\frac{1}{k_z}\bigg(\frac{2c^2k_\perp^2}{\omega^2}-1\bigg)\,,\\
r^M&=2\pi i\frac{\omega^2}{c^2}\frac{1}{k_z}\,,
\end{align}
and for the transmission coefficients we find
\begin{align}
t^N&=2\pi i\frac{\omega^2}{c^2}\frac{1}{k_z}\,,\\
t^M&=t^N\,.
\end{align}

\section{Composite scattering operators}\label{app:Composite_T_operator}
\subsection{$\T$ operator for two or more objects}
The $\T_{12}$ operator, describing the composite scattering operator for two objects embedded in a background medium, can be straightforwardly expanded in terms of the single scattering operators $\T_1$ and $\T_2$.
The desired representation is found by starting from the single object $1$ in the medium with $\G_1=(1+\G_b\T_1)\G_b$ [see Eq.\ \eqref{eq:GreenBackground_T_Medium}], and inserting object $2$ by use of the operator $\mathbb{O}_1$ in Eq.\ \eqref{eq:Multiple_scattering}, as
\be
\G=(1+\G_b\T_2)\frac{1}{1-\G_b\T_1\G_b\T_2}(1+\G_b\T_1)\G_b\,.
\ee
On the other hand, we can introduce the $\T_{12}$ operator by writing
\be
\G=(1+\G_b\T_{12})\G_b\,.
\ee
Solving these two equations for $\T_{12}$, we obtain
\begin{align}\label{eq:Composite_TOperator_TwoObjects}
\T_{12}&=(\T_1+\T_2\G_b\T_1)\frac{1}{1-\G_b\T_2\G_b\T_1}\notag\\
&+(\T_2+\T_1\G_b\T_2)\frac{1}{1-\G_b\T_1\G_0\T_2}\,.
\end{align}
Note that this operator is symmetric and we can write $\T_{12}=\T_{21}$. The composite scattering operator for more than two objects is then obtained by iteratively substituting the composite scattering operator in Eq.\ \eqref{eq:Composite_TOperator_TwoObjects} for either of the two single scattering operators $\T_1$ and $\T_2$, respectively. 

\subsection{Expansion for $N$ objects}
An expansion for the $\T_{\alpha\beta\gamma\dots}$ operator for $N$ objects in terms of the single scattering operators \cite{boyce1968probabilistic,fest2008modeling} can be given by starting from the definition in Eq.\ \eqref{eq:TOperator_medium}. The expansion of the inverse operator in that equation into a power series, yields
\be
\T_{\alpha\beta\gamma\dots}=\Delta\mathbb{V}\sum_{i=0}^\infty[\G_b\Delta\mathbb{V}]^i\,.
\ee
$\Delta\mathbb{V}$ is the potential difference between the collection of objects and the background medium, i.e.
\be
\Delta\mathbb{V}=\sum_{i=1}^N \Delta\mathbb{V}_i\,.
\ee
Insertion of this relation into the power series leads to
\begin{align}
\T_{\alpha\beta\gamma\dots}=&\sum_{i=1}^N\Delta\mathbb{V}_i+\sum_{i=1,j=1}^{N,N}\Delta\mathbb{V}_i\mathbb{G}_b\Delta\mathbb{V}_j\notag\\
&+\sum_{i=1,j=1,k=1}^{N,N,N}\Delta\mathbb{V}_i\G_b\Delta\mathbb{V}_j\G_b\Delta\mathbb{V}_k+\dots\,,
\end{align}
where the dots stand for the remaining infinite terms. Notice that in the higher order terms the same index of the potential introduced by object $\alpha$ may be repeated several times. By resummation of the infinite terms, we finally arrive at the expansion of $\T$ in terms of the single scattering operators, reading as
\begin{align}
\T_{\alpha\beta\gamma\dots}=&\sum_{i=1}^N\T_i+\sum_{i=1,j=1,i\neq j}^{N,N}\T_i\mathbb{G}_b\T_j\notag\\
&+\sum_{i=1,j=1,k=1,i\neq j,j\neq k}^{N,N,N}\T_i\G_b\T_j\G_b\T_k+\dots
\end{align}
Again the dots represent the remaining higher order terms. From this representation, it becomes apparent that the scattering operator for an arbitrary number of objects is always symmetric with respect to a permutation of indices, e.g.\ $\T_{\alpha\beta\gamma\dots}=\T_{\gamma\beta\alpha\dots}$. Note that for the composite $\T$ operator in vacuum, one has to exchange $\G_b$ with the free Green's function $\G_0$ and use the definition of the $\T$ operator in vacuum given in the first line of Table~\ref{table:1}.
\end{appendix}


\begin{thebibliography}{73}%
\makeatletter
\providecommand \@ifxundefined [1]{%
 \@ifx{#1\undefined}
}%
\providecommand \@ifnum [1]{%
 \ifnum #1\expandafter \@firstoftwo
 \else \expandafter \@secondoftwo
 \fi
}%
\providecommand \@ifx [1]{%
 \ifx #1\expandafter \@firstoftwo
 \else \expandafter \@secondoftwo
 \fi
}%
\providecommand \natexlab [1]{#1}%
\providecommand \enquote  [1]{``#1''}%
\providecommand \bibnamefont  [1]{#1}%
\providecommand \bibfnamefont [1]{#1}%
\providecommand \citenamefont [1]{#1}%
\providecommand \href@noop [0]{\@secondoftwo}%
\providecommand \href [0]{\begingroup \@sanitize@url \@href}%
\providecommand \@href[1]{\@@startlink{#1}\@@href}%
\providecommand \@@href[1]{\endgroup#1\@@endlink}%
\providecommand \@sanitize@url [0]{\catcode `\\12\catcode `\$12\catcode
  `\&12\catcode `\#12\catcode `\^12\catcode `\_12\catcode `\%12\relax}%
\providecommand \@@startlink[1]{}%
\providecommand \@@endlink[0]{}%
\providecommand \url  [0]{\begingroup\@sanitize@url \@url }%
\providecommand \@url [1]{\endgroup\@href {#1}{\urlprefix }}%
\providecommand \urlprefix  [0]{URL }%
\providecommand \Eprint [0]{\href }%
\providecommand \doibase [0]{http://dx.doi.org/}%
\providecommand \selectlanguage [0]{\@gobble}%
\providecommand \bibinfo  [0]{\@secondoftwo}%
\providecommand \bibfield  [0]{\@secondoftwo}%
\providecommand \translation [1]{[#1]}%
\providecommand \BibitemOpen [0]{}%
\providecommand \bibitemStop [0]{}%
\providecommand \bibitemNoStop [0]{.\EOS\space}%
\providecommand \EOS [0]{\spacefactor3000\relax}%
\providecommand \BibitemShut  [1]{\csname bibitem#1\endcsname}%
\let\auto@bib@innerbib\@empty
%</preamble>
\bibitem [{\citenamefont {Planck}(1901)}]{planck1901law}%
  \BibitemOpen
  \bibfield  {author} {\bibinfo {author} {\bibfnamefont {M.}~\bibnamefont
  {Planck}},\ }\href@noop {} {\bibfield  {journal} {\bibinfo  {journal} {Ann.
  Phys}\ }\textbf {\bibinfo {volume} {4}},\ \bibinfo {pages} {553} (\bibinfo
  {year} {1901})}\BibitemShut {NoStop}%
\bibitem [{\citenamefont {Rytov}(1958)}]{rytov1958correlation}%
  \BibitemOpen
  \bibfield  {author} {\bibinfo {author} {\bibfnamefont {S.}~\bibnamefont
  {Rytov}},\ }\href@noop {} {\bibfield  {journal} {\bibinfo  {journal} {Soviet
  Journal of Experimental and Theoretical Physics}\ }\textbf {\bibinfo {volume}
  {6}},\ \bibinfo {pages} {130} (\bibinfo {year} {1958})}\BibitemShut {NoStop}%
\bibitem [{\citenamefont {Rytov}\ \emph {et~al.}(1989)\citenamefont {Rytov},
  \citenamefont {Kravtsov},\ and\ \citenamefont
  {Tatarskii}}]{rytov1989elements}%
  \BibitemOpen
  \bibfield  {author} {\bibinfo {author} {\bibfnamefont {S.}~\bibnamefont
  {Rytov}}, \bibinfo {author} {\bibfnamefont {Y.~A.}\ \bibnamefont {Kravtsov}},
  \ and\ \bibinfo {author} {\bibfnamefont {V.}~\bibnamefont {Tatarskii}},\
  }\href@noop {} {\enquote {\bibinfo {title} {Elements of random fields
  (principles of statistical radiophysics vol 3)},}\ } (\bibinfo {year}
  {1989})\BibitemShut {NoStop}%
\bibitem [{\citenamefont {Casimir}(1948)}]{casimir1948attraction}%
  \BibitemOpen
  \bibfield  {author} {\bibinfo {author} {\bibfnamefont {H.~B.~G.}\
  \bibnamefont {Casimir}},\ }\href@noop {} {\bibfield  {journal} {\bibinfo
  {journal} {Proc. K. Ned. Akad. Wet. B}\ }\textbf {\bibinfo {volume} {51}},\
  \bibinfo {pages} {793} (\bibinfo {year} {1948})}\BibitemShut {NoStop}%
\bibitem [{\citenamefont {Casimir}\ and\ \citenamefont
  {Polder}(1948)}]{casimir1948influence}%
  \BibitemOpen
  \bibfield  {author} {\bibinfo {author} {\bibfnamefont {H.~B.~G.}\
  \bibnamefont {Casimir}}\ and\ \bibinfo {author} {\bibfnamefont
  {D.}~\bibnamefont {Polder}},\ }\href@noop {} {\bibfield  {journal} {\bibinfo
  {journal} {Phys. Rev.}\ }\textbf {\bibinfo {volume} {73}},\ \bibinfo {pages}
  {360} (\bibinfo {year} {1948})}\BibitemShut {NoStop}%
\bibitem [{\citenamefont {Dzyaloshinskii}\ \emph {et~al.}(1961)\citenamefont
  {Dzyaloshinskii}, \citenamefont {Lifshitz},\ and\ \citenamefont
  {Pitaevskii}}]{dzyaloshinskii1961the}%
  \BibitemOpen
  \bibfield  {author} {\bibinfo {author} {\bibfnamefont {I.}~\bibnamefont
  {Dzyaloshinskii}}, \bibinfo {author} {\bibfnamefont {E.}~\bibnamefont
  {Lifshitz}}, \ and\ \bibinfo {author} {\bibfnamefont {L.}~\bibnamefont
  {Pitaevskii}},\ }\href@noop {} {\bibfield  {journal} {\bibinfo  {journal}
  {Advances in Physics}\ }\textbf {\bibinfo {volume} {10}},\ \bibinfo {pages}
  {165} (\bibinfo {year} {1961})}\BibitemShut {NoStop}%
\bibitem [{\citenamefont {Henkel}\ \emph {et~al.}(2002)\citenamefont {Henkel},
  \citenamefont {Joulain}, \citenamefont {Mulet},\ and\ \citenamefont
  {Greffet}}]{henkel2002radiation}%
  \BibitemOpen
  \bibfield  {author} {\bibinfo {author} {\bibfnamefont {C.}~\bibnamefont
  {Henkel}}, \bibinfo {author} {\bibfnamefont {K.}~\bibnamefont {Joulain}},
  \bibinfo {author} {\bibfnamefont {J.-P.}\ \bibnamefont {Mulet}}, \ and\
  \bibinfo {author} {\bibfnamefont {J.-J.}\ \bibnamefont {Greffet}},\
  }\href@noop {} {\bibfield  {journal} {\bibinfo  {journal} {Journal of Optics
  A: Pure and Applied Optics}\ }\textbf {\bibinfo {volume} {4}},\ \bibinfo
  {pages} {S109} (\bibinfo {year} {2002})}\BibitemShut {NoStop}%
\bibitem [{\citenamefont {Antezza}\ \emph {et~al.}(2005)\citenamefont
  {Antezza}, \citenamefont {Pitaevskii},\ and\ \citenamefont
  {Stringari}}]{antezza2005new}%
  \BibitemOpen
  \bibfield  {author} {\bibinfo {author} {\bibfnamefont {M.}~\bibnamefont
  {Antezza}}, \bibinfo {author} {\bibfnamefont {L.~P.}\ \bibnamefont
  {Pitaevskii}}, \ and\ \bibinfo {author} {\bibfnamefont {S.}~\bibnamefont
  {Stringari}},\ }\href@noop {} {\bibfield  {journal} {\bibinfo  {journal}
  {Phys. Rev. Lett.}\ }\textbf {\bibinfo {volume} {95}},\ \bibinfo {pages}
  {113202} (\bibinfo {year} {2005})}\BibitemShut {NoStop}%
\bibitem [{\citenamefont {Antezza}\ \emph {et~al.}(2008)\citenamefont
  {Antezza}, \citenamefont {Pitaevskii}, \citenamefont {Stringari},\ and\
  \citenamefont {Svetovoy}}]{antezza2008casimir}%
  \BibitemOpen
  \bibfield  {author} {\bibinfo {author} {\bibfnamefont {M.}~\bibnamefont
  {Antezza}}, \bibinfo {author} {\bibfnamefont {L.~P.}\ \bibnamefont
  {Pitaevskii}}, \bibinfo {author} {\bibfnamefont {S.}~\bibnamefont
  {Stringari}}, \ and\ \bibinfo {author} {\bibfnamefont {V.~B.}\ \bibnamefont
  {Svetovoy}},\ }\href@noop {} {\bibfield  {journal} {\bibinfo  {journal}
  {Phys. Rev. A}\ }\textbf {\bibinfo {volume} {77}},\ \bibinfo {pages} {022901}
  (\bibinfo {year} {2008})}\BibitemShut {NoStop}%
\bibitem [{\citenamefont {Bimonte}(2009)}]{bimonte2009scattering}%
  \BibitemOpen
  \bibfield  {author} {\bibinfo {author} {\bibfnamefont {G.}~\bibnamefont
  {Bimonte}},\ }\href@noop {} {\bibfield  {journal} {\bibinfo  {journal} {Phys.
  Rev. A}\ }\textbf {\bibinfo {volume} {80}},\ \bibinfo {pages} {042102}
  (\bibinfo {year} {2009})}\BibitemShut {NoStop}%
\bibitem [{\citenamefont {Messina}\ and\ \citenamefont
  {Antezza}(2011{\natexlab{a}})}]{messina2011casimir}%
  \BibitemOpen
  \bibfield  {author} {\bibinfo {author} {\bibfnamefont {R.}~\bibnamefont
  {Messina}}\ and\ \bibinfo {author} {\bibfnamefont {M.}~\bibnamefont
  {Antezza}},\ }\href@noop {} {\bibfield  {journal} {\bibinfo  {journal} {EPL
  (Europhysics Letters)}\ }\textbf {\bibinfo {volume} {95}},\ \bibinfo {pages}
  {61002} (\bibinfo {year} {2011}{\natexlab{a}})}\BibitemShut {NoStop}%
\bibitem [{\citenamefont {Kr\"uger}\ \emph
  {et~al.}(2011{\natexlab{a}})\citenamefont {Kr\"uger}, \citenamefont {Emig},
  \citenamefont {Bimonte},\ and\ \citenamefont {Kardar}}]{krueger2011non}%
  \BibitemOpen
  \bibfield  {author} {\bibinfo {author} {\bibfnamefont {M.}~\bibnamefont
  {Kr\"uger}}, \bibinfo {author} {\bibfnamefont {T.}~\bibnamefont {Emig}},
  \bibinfo {author} {\bibfnamefont {G.}~\bibnamefont {Bimonte}}, \ and\
  \bibinfo {author} {\bibfnamefont {M.}~\bibnamefont {Kardar}},\ }\href@noop {}
  {\bibfield  {journal} {\bibinfo  {journal} {EPL (Europhysics Letters)}\
  }\textbf {\bibinfo {volume} {95}},\ \bibinfo {pages} {21002} (\bibinfo {year}
  {2011}{\natexlab{a}})}\BibitemShut {NoStop}%
\bibitem [{\citenamefont {Messina}\ and\ \citenamefont
  {Antezza}(2011{\natexlab{b}})}]{messina2011scattering}%
  \BibitemOpen
  \bibfield  {author} {\bibinfo {author} {\bibfnamefont {R.}~\bibnamefont
  {Messina}}\ and\ \bibinfo {author} {\bibfnamefont {M.}~\bibnamefont
  {Antezza}},\ }\href@noop {} {\bibfield  {journal} {\bibinfo  {journal} {Phys.
  Rev. A}\ }\textbf {\bibinfo {volume} {84}},\ \bibinfo {pages} {042102}
  (\bibinfo {year} {2011}{\natexlab{b}})}\BibitemShut {NoStop}%
\bibitem [{\citenamefont {Kr\"uger}\ \emph {et~al.}(2012)\citenamefont
  {Kr\"uger}, \citenamefont {Bimonte}, \citenamefont {Emig},\ and\
  \citenamefont {Kardar}}]{kruger2012trace}%
  \BibitemOpen
  \bibfield  {author} {\bibinfo {author} {\bibfnamefont {M.}~\bibnamefont
  {Kr\"uger}}, \bibinfo {author} {\bibfnamefont {G.}~\bibnamefont {Bimonte}},
  \bibinfo {author} {\bibfnamefont {T.}~\bibnamefont {Emig}}, \ and\ \bibinfo
  {author} {\bibfnamefont {M.}~\bibnamefont {Kardar}},\ }\href@noop {}
  {\bibfield  {journal} {\bibinfo  {journal} {Phys. Rev. B}\ }\textbf {\bibinfo
  {volume} {86}},\ \bibinfo {pages} {115423} (\bibinfo {year}
  {2012})}\BibitemShut {NoStop}%
\bibitem [{\citenamefont {Golyk}\ \emph {et~al.}(2012)\citenamefont {Golyk},
  \citenamefont {Kr\"uger}, \citenamefont {Reid},\ and\ \citenamefont
  {Kardar}}]{golyk2012casimir}%
  \BibitemOpen
  \bibfield  {author} {\bibinfo {author} {\bibfnamefont {V.~A.}\ \bibnamefont
  {Golyk}}, \bibinfo {author} {\bibfnamefont {M.}~\bibnamefont {Kr\"uger}},
  \bibinfo {author} {\bibfnamefont {M.~T.~H.}\ \bibnamefont {Reid}}, \ and\
  \bibinfo {author} {\bibfnamefont {M.}~\bibnamefont {Kardar}},\ }\href@noop {}
  {\bibfield  {journal} {\bibinfo  {journal} {Phys. Rev. D}\ }\textbf {\bibinfo
  {volume} {85}},\ \bibinfo {pages} {065011} (\bibinfo {year}
  {2012})}\BibitemShut {NoStop}%
\bibitem [{\citenamefont {Narayanaswamy}\ and\ \citenamefont
  {Zheng}(2014)}]{narayanaswamy2014a}%
  \BibitemOpen
  \bibfield  {author} {\bibinfo {author} {\bibfnamefont {A.}~\bibnamefont
  {Narayanaswamy}}\ and\ \bibinfo {author} {\bibfnamefont {Y.}~\bibnamefont
  {Zheng}},\ }\href@noop {} {\bibfield  {journal} {\bibinfo  {journal} {Journal
  of Quantitative Spectroscopy and Radiative Transfer}\ }\textbf {\bibinfo
  {volume} {132}},\ \bibinfo {pages} {12 } (\bibinfo {year} {2014})},\ \bibinfo
  {note} {special Issue on Micro- and Nano-Scale Radiative
  Transfer}\BibitemShut {NoStop}%
\bibitem [{\citenamefont {Messina}\ and\ \citenamefont
  {Antezza}(2014)}]{messina2014three}%
  \BibitemOpen
  \bibfield  {author} {\bibinfo {author} {\bibfnamefont {R.}~\bibnamefont
  {Messina}}\ and\ \bibinfo {author} {\bibfnamefont {M.}~\bibnamefont
  {Antezza}},\ }\href@noop {} {\bibfield  {journal} {\bibinfo  {journal} {Phys.
  Rev. A}\ }\textbf {\bibinfo {volume} {89}},\ \bibinfo {pages} {052104}
  (\bibinfo {year} {2014})}\BibitemShut {NoStop}%
\bibitem [{\citenamefont {M\"uller}\ and\ \citenamefont
  {Kr\"uger}(2016)}]{mueller2016anisotropic}%
  \BibitemOpen
  \bibfield  {author} {\bibinfo {author} {\bibfnamefont {B.}~\bibnamefont
  {M\"uller}}\ and\ \bibinfo {author} {\bibfnamefont {M.}~\bibnamefont
  {Kr\"uger}},\ }\href@noop {} {\bibfield  {journal} {\bibinfo  {journal}
  {Phys. Rev. A}\ }\textbf {\bibinfo {volume} {93}},\ \bibinfo {pages} {032511}
  (\bibinfo {year} {2016})}\BibitemShut {NoStop}%
\bibitem [{\citenamefont {Bimonte}\ \emph {et~al.}(2016)\citenamefont
  {Bimonte}, \citenamefont {Emig}, \citenamefont {Kardar},\ and\ \citenamefont
  {Kr{\"u}ger}}]{bimonte2016non}%
  \BibitemOpen
  \bibfield  {author} {\bibinfo {author} {\bibfnamefont {G.}~\bibnamefont
  {Bimonte}}, \bibinfo {author} {\bibfnamefont {T.}~\bibnamefont {Emig}},
  \bibinfo {author} {\bibfnamefont {M.}~\bibnamefont {Kardar}}, \ and\ \bibinfo
  {author} {\bibfnamefont {M.}~\bibnamefont {Kr{\"u}ger}},\ }\href@noop {} {\
  (\bibinfo {year} {2016})},\ \Eprint {http://arxiv.org/abs/1606.03740}
  {arXiv:1606.03740} \BibitemShut {NoStop}%
%%CITATION = ARXIV:1606.03740;%%
\bibitem [{\citenamefont {Antezza}\ \emph {et~al.}(2004)\citenamefont
  {Antezza}, \citenamefont {Pitaevskii},\ and\ \citenamefont
  {Stringari}}]{antezza2004effect}%
  \BibitemOpen
  \bibfield  {author} {\bibinfo {author} {\bibfnamefont {M.}~\bibnamefont
  {Antezza}}, \bibinfo {author} {\bibfnamefont {L.~P.}\ \bibnamefont
  {Pitaevskii}}, \ and\ \bibinfo {author} {\bibfnamefont {S.}~\bibnamefont
  {Stringari}},\ }\href@noop {} {\bibfield  {journal} {\bibinfo  {journal}
  {Phys. Rev. A}\ }\textbf {\bibinfo {volume} {70}},\ \bibinfo {pages} {053619}
  (\bibinfo {year} {2004})}\BibitemShut {NoStop}%
\bibitem [{\citenamefont {Obrecht}\ \emph {et~al.}(2007)\citenamefont
  {Obrecht}, \citenamefont {Wild}, \citenamefont {Antezza}, \citenamefont
  {Pitaevskii}, \citenamefont {Stringari},\ and\ \citenamefont
  {Cornell}}]{obrecht2007measurement}%
  \BibitemOpen
  \bibfield  {author} {\bibinfo {author} {\bibfnamefont {J.~M.}\ \bibnamefont
  {Obrecht}}, \bibinfo {author} {\bibfnamefont {R.~J.}\ \bibnamefont {Wild}},
  \bibinfo {author} {\bibfnamefont {M.}~\bibnamefont {Antezza}}, \bibinfo
  {author} {\bibfnamefont {L.~P.}\ \bibnamefont {Pitaevskii}}, \bibinfo
  {author} {\bibfnamefont {S.}~\bibnamefont {Stringari}}, \ and\ \bibinfo
  {author} {\bibfnamefont {E.~A.}\ \bibnamefont {Cornell}},\ }\href@noop {}
  {\bibfield  {journal} {\bibinfo  {journal} {Phys. Rev. Lett.}\ }\textbf
  {\bibinfo {volume} {98}},\ \bibinfo {pages} {063201} (\bibinfo {year}
  {2007})}\BibitemShut {NoStop}%
\bibitem [{\citenamefont {Bimonte}(2015)}]{bimonte2015observing}%
  \BibitemOpen
  \bibfield  {author} {\bibinfo {author} {\bibfnamefont {G.}~\bibnamefont
  {Bimonte}},\ }\href@noop {} {\bibfield  {journal} {\bibinfo  {journal} {Phys.
  Rev. A}\ }\textbf {\bibinfo {volume} {92}},\ \bibinfo {pages} {032116}
  (\bibinfo {year} {2015})}\BibitemShut {NoStop}%
\bibitem [{\citenamefont {Kr\"uger}\ \emph
  {et~al.}(2011{\natexlab{b}})\citenamefont {Kr\"uger}, \citenamefont {Emig},\
  and\ \citenamefont {Kardar}}]{kruger2011nonequilibrium}%
  \BibitemOpen
  \bibfield  {author} {\bibinfo {author} {\bibfnamefont {M.}~\bibnamefont
  {Kr\"uger}}, \bibinfo {author} {\bibfnamefont {T.}~\bibnamefont {Emig}}, \
  and\ \bibinfo {author} {\bibfnamefont {M.}~\bibnamefont {Kardar}},\
  }\href@noop {} {\bibfield  {journal} {\bibinfo  {journal} {Phys. Rev. Lett.}\
  }\textbf {\bibinfo {volume} {106}},\ \bibinfo {pages} {210404} (\bibinfo
  {year} {2011}{\natexlab{b}})}\BibitemShut {NoStop}%
\bibitem [{\citenamefont {Polder}\ and\ \citenamefont
  {Van~Hove}(1971)}]{polder1971theory}%
  \BibitemOpen
  \bibfield  {author} {\bibinfo {author} {\bibfnamefont {D.}~\bibnamefont
  {Polder}}\ and\ \bibinfo {author} {\bibfnamefont {M.}~\bibnamefont
  {Van~Hove}},\ }\href@noop {} {\bibfield  {journal} {\bibinfo  {journal}
  {Phys. Rev. B}\ }\textbf {\bibinfo {volume} {4}},\ \bibinfo {pages} {3303}
  (\bibinfo {year} {1971})}\BibitemShut {NoStop}%
\bibitem [{\citenamefont {Hargreaves}(1969)}]{hargreaves1969anomalous}%
  \BibitemOpen
  \bibfield  {author} {\bibinfo {author} {\bibfnamefont {C.}~\bibnamefont
  {Hargreaves}},\ }\href@noop {} {\bibfield  {journal} {\bibinfo  {journal}
  {Physics Letters A}\ }\textbf {\bibinfo {volume} {30}},\ \bibinfo {pages}
  {491} (\bibinfo {year} {1969})}\BibitemShut {NoStop}%
\bibitem [{\citenamefont {Volokitin}\ and\ \citenamefont
  {Persson}(2001)}]{volokitin2001radiative}%
  \BibitemOpen
  \bibfield  {author} {\bibinfo {author} {\bibfnamefont {A.~I.}\ \bibnamefont
  {Volokitin}}\ and\ \bibinfo {author} {\bibfnamefont {B.~N.~J.}\ \bibnamefont
  {Persson}},\ }\href@noop {} {\bibfield  {journal} {\bibinfo  {journal} {Phys.
  Rev. B}\ }\textbf {\bibinfo {volume} {63}},\ \bibinfo {pages} {205404}
  (\bibinfo {year} {2001})}\BibitemShut {NoStop}%
\bibitem [{\citenamefont {Volokitin}\ and\ \citenamefont
  {Persson}(2007)}]{volokitin2007nearfield}%
  \BibitemOpen
  \bibfield  {author} {\bibinfo {author} {\bibfnamefont {A.~I.}\ \bibnamefont
  {Volokitin}}\ and\ \bibinfo {author} {\bibfnamefont {B.~N.~J.}\ \bibnamefont
  {Persson}},\ }\href@noop {} {\bibfield  {journal} {\bibinfo  {journal} {Rev.
  Mod. Phys.}\ }\textbf {\bibinfo {volume} {79}},\ \bibinfo {pages} {1291}
  (\bibinfo {year} {2007})}\BibitemShut {NoStop}%
\bibitem [{\citenamefont {Rodriguez}\ \emph {et~al.}(2011)\citenamefont
  {Rodriguez}, \citenamefont {Ilic}, \citenamefont {Bermel}, \citenamefont
  {Celanovic}, \citenamefont {Joannopoulos}, \citenamefont {Solja\ifmmode
  \check{c}\else \v{c}\fi{}i\ifmmode~\acute{c}\else \'{c}\fi{}},\ and\
  \citenamefont {Johnson}}]{rodriguez2011frequency}%
  \BibitemOpen
  \bibfield  {author} {\bibinfo {author} {\bibfnamefont {A.~W.}\ \bibnamefont
  {Rodriguez}}, \bibinfo {author} {\bibfnamefont {O.}~\bibnamefont {Ilic}},
  \bibinfo {author} {\bibfnamefont {P.}~\bibnamefont {Bermel}}, \bibinfo
  {author} {\bibfnamefont {I.}~\bibnamefont {Celanovic}}, \bibinfo {author}
  {\bibfnamefont {J.~D.}\ \bibnamefont {Joannopoulos}}, \bibinfo {author}
  {\bibfnamefont {M.}~\bibnamefont {Solja\ifmmode \check{c}\else
  \v{c}\fi{}i\ifmmode~\acute{c}\else \'{c}\fi{}}}, \ and\ \bibinfo {author}
  {\bibfnamefont {S.~G.}\ \bibnamefont {Johnson}},\ }\href@noop {} {\bibfield
  {journal} {\bibinfo  {journal} {Phys. Rev. Lett.}\ }\textbf {\bibinfo
  {volume} {107}},\ \bibinfo {pages} {114302} (\bibinfo {year}
  {2011})}\BibitemShut {NoStop}%
\bibitem [{\citenamefont {McCauley}\ \emph {et~al.}(2012)\citenamefont
  {McCauley}, \citenamefont {Reid}, \citenamefont {Kr\"uger},\ and\
  \citenamefont {Johnson}}]{mccauley2012modeling}%
  \BibitemOpen
  \bibfield  {author} {\bibinfo {author} {\bibfnamefont {A.~P.}\ \bibnamefont
  {McCauley}}, \bibinfo {author} {\bibfnamefont {M.~T.~H.}\ \bibnamefont
  {Reid}}, \bibinfo {author} {\bibfnamefont {M.}~\bibnamefont {Kr\"uger}}, \
  and\ \bibinfo {author} {\bibfnamefont {S.~G.}\ \bibnamefont {Johnson}},\
  }\href@noop {} {\bibfield  {journal} {\bibinfo  {journal} {Phys. Rev. B}\
  }\textbf {\bibinfo {volume} {85}},\ \bibinfo {pages} {165104} (\bibinfo
  {year} {2012})}\BibitemShut {NoStop}%
\bibitem [{\citenamefont {Rodriguez}\ \emph {et~al.}(2013)\citenamefont
  {Rodriguez}, \citenamefont {Reid},\ and\ \citenamefont
  {Johnson}}]{rodriguez2013fluctuating}%
  \BibitemOpen
  \bibfield  {author} {\bibinfo {author} {\bibfnamefont {A.~W.}\ \bibnamefont
  {Rodriguez}}, \bibinfo {author} {\bibfnamefont {M.~T.~H.}\ \bibnamefont
  {Reid}}, \ and\ \bibinfo {author} {\bibfnamefont {S.~G.}\ \bibnamefont
  {Johnson}},\ }\href@noop {} {\bibfield  {journal} {\bibinfo  {journal} {Phys.
  Rev. B}\ }\textbf {\bibinfo {volume} {88}},\ \bibinfo {pages} {054305}
  (\bibinfo {year} {2013})}\BibitemShut {NoStop}%
\bibitem [{\citenamefont {Polimeridis}\ \emph {et~al.}(2015)\citenamefont
  {Polimeridis}, \citenamefont {Reid}, \citenamefont {Jin}, \citenamefont
  {Johnson}, \citenamefont {White},\ and\ \citenamefont
  {Rodriguez}}]{polimeridis2015fluctuating}%
  \BibitemOpen
  \bibfield  {author} {\bibinfo {author} {\bibfnamefont {A.~G.}\ \bibnamefont
  {Polimeridis}}, \bibinfo {author} {\bibfnamefont {M.~T.~H.}\ \bibnamefont
  {Reid}}, \bibinfo {author} {\bibfnamefont {W.}~\bibnamefont {Jin}}, \bibinfo
  {author} {\bibfnamefont {S.~G.}\ \bibnamefont {Johnson}}, \bibinfo {author}
  {\bibfnamefont {J.~K.}\ \bibnamefont {White}}, \ and\ \bibinfo {author}
  {\bibfnamefont {A.~W.}\ \bibnamefont {Rodriguez}},\ }\href@noop {} {\bibfield
   {journal} {\bibinfo  {journal} {Phys. Rev. B}\ }\textbf {\bibinfo {volume}
  {92}},\ \bibinfo {pages} {134202} (\bibinfo {year} {2015})}\BibitemShut
  {NoStop}%
\bibitem [{\citenamefont {Eckhardt}(1984)}]{eckhardt1984macroscopic}%
  \BibitemOpen
  \bibfield  {author} {\bibinfo {author} {\bibfnamefont {W.}~\bibnamefont
  {Eckhardt}},\ }\href@noop {} {\bibfield  {journal} {\bibinfo  {journal}
  {Phys. Rev. A}\ }\textbf {\bibinfo {volume} {29}},\ \bibinfo {pages} {1991}
  (\bibinfo {year} {1984})}\BibitemShut {NoStop}%
\bibitem [{\citenamefont {Messina}\ \emph {et~al.}(2012)\citenamefont
  {Messina}, \citenamefont {Antezza},\ and\ \citenamefont
  {Ben-Abdallah}}]{messina2012three}%
  \BibitemOpen
  \bibfield  {author} {\bibinfo {author} {\bibfnamefont {R.}~\bibnamefont
  {Messina}}, \bibinfo {author} {\bibfnamefont {M.}~\bibnamefont {Antezza}}, \
  and\ \bibinfo {author} {\bibfnamefont {P.}~\bibnamefont {Ben-Abdallah}},\
  }\href@noop {} {\bibfield  {journal} {\bibinfo  {journal} {Phys. Rev. Lett.}\
  }\textbf {\bibinfo {volume} {109}},\ \bibinfo {pages} {244302} (\bibinfo
  {year} {2012})}\BibitemShut {NoStop}%
\bibitem [{\citenamefont {Messina}\ \emph {et~al.}(2016)\citenamefont
  {Messina}, \citenamefont {Ben-Abdallah}, \citenamefont {Guizal},
  \citenamefont {Antezza},\ and\ \citenamefont
  {Biehs}}]{messina2016hyperbolic}%
  \BibitemOpen
  \bibfield  {author} {\bibinfo {author} {\bibfnamefont {R.}~\bibnamefont
  {Messina}}, \bibinfo {author} {\bibfnamefont {P.}~\bibnamefont
  {Ben-Abdallah}}, \bibinfo {author} {\bibfnamefont {B.}~\bibnamefont
  {Guizal}}, \bibinfo {author} {\bibfnamefont {M.}~\bibnamefont {Antezza}}, \
  and\ \bibinfo {author} {\bibfnamefont {S.-A.}\ \bibnamefont {Biehs}},\
  }\href@noop {} {\bibfield  {journal} {\bibinfo  {journal} {Phys. Rev. B}\
  }\textbf {\bibinfo {volume} {94}},\ \bibinfo {pages} {104301} (\bibinfo
  {year} {2016})}\BibitemShut {NoStop}%
\bibitem [{\citenamefont {Chapuis}\ \emph {et~al.}(2008)\citenamefont
  {Chapuis}, \citenamefont {Volz}, \citenamefont {Henkel}, \citenamefont
  {Joulain},\ and\ \citenamefont {Greffet}}]{chapuis2008effects}%
  \BibitemOpen
  \bibfield  {author} {\bibinfo {author} {\bibfnamefont {P.-O.}\ \bibnamefont
  {Chapuis}}, \bibinfo {author} {\bibfnamefont {S.}~\bibnamefont {Volz}},
  \bibinfo {author} {\bibfnamefont {C.}~\bibnamefont {Henkel}}, \bibinfo
  {author} {\bibfnamefont {K.}~\bibnamefont {Joulain}}, \ and\ \bibinfo
  {author} {\bibfnamefont {J.-J.}\ \bibnamefont {Greffet}},\ }\href@noop {}
  {\bibfield  {journal} {\bibinfo  {journal} {Phys. Rev. B}\ }\textbf {\bibinfo
  {volume} {77}},\ \bibinfo {pages} {035431} (\bibinfo {year}
  {2008})}\BibitemShut {NoStop}%
\bibitem [{\citenamefont {Biehs}\ \emph {et~al.}(2011)\citenamefont {Biehs},
  \citenamefont {Rosa},\ and\ \citenamefont
  {Ben-Abdallah}}]{biehs2011modulation}%
  \BibitemOpen
  \bibfield  {author} {\bibinfo {author} {\bibfnamefont {S.-A.}\ \bibnamefont
  {Biehs}}, \bibinfo {author} {\bibfnamefont {F.~S.~S.}\ \bibnamefont {Rosa}},
  \ and\ \bibinfo {author} {\bibfnamefont {P.}~\bibnamefont {Ben-Abdallah}},\
  }\href@noop {} {\bibfield  {journal} {\bibinfo  {journal} {Applied Physics
  Letters}\ }\textbf {\bibinfo {volume} {98}},\ \bibinfo {eid} {243102}
  (\bibinfo {year} {2011})}\BibitemShut {NoStop}%
\bibitem [{\citenamefont {Lussange}\ \emph {et~al.}(2012)\citenamefont
  {Lussange}, \citenamefont {Gu\'erout}, \citenamefont {Rosa}, \citenamefont
  {Greffet}, \citenamefont {Lambrecht},\ and\ \citenamefont
  {Reynaud}}]{lussange2012radiative}%
  \BibitemOpen
  \bibfield  {author} {\bibinfo {author} {\bibfnamefont {J.}~\bibnamefont
  {Lussange}}, \bibinfo {author} {\bibfnamefont {R.}~\bibnamefont {Gu\'erout}},
  \bibinfo {author} {\bibfnamefont {F.~S.~S.}\ \bibnamefont {Rosa}}, \bibinfo
  {author} {\bibfnamefont {J.-J.}\ \bibnamefont {Greffet}}, \bibinfo {author}
  {\bibfnamefont {A.}~\bibnamefont {Lambrecht}}, \ and\ \bibinfo {author}
  {\bibfnamefont {S.}~\bibnamefont {Reynaud}},\ }\href@noop {} {\bibfield
  {journal} {\bibinfo  {journal} {Phys. Rev. B}\ }\textbf {\bibinfo {volume}
  {86}},\ \bibinfo {pages} {085432} (\bibinfo {year} {2012})}\BibitemShut
  {NoStop}%
\bibitem [{\citenamefont {Gu\'erout}\ \emph {et~al.}(2012)\citenamefont
  {Gu\'erout}, \citenamefont {Lussange}, \citenamefont {Rosa}, \citenamefont
  {Hugonin}, \citenamefont {Dalvit}, \citenamefont {Greffet}, \citenamefont
  {Lambrecht},\ and\ \citenamefont {Reynaud}}]{guerout2012enhanced}%
  \BibitemOpen
  \bibfield  {author} {\bibinfo {author} {\bibfnamefont {R.}~\bibnamefont
  {Gu\'erout}}, \bibinfo {author} {\bibfnamefont {J.}~\bibnamefont {Lussange}},
  \bibinfo {author} {\bibfnamefont {F.~S.~S.}\ \bibnamefont {Rosa}}, \bibinfo
  {author} {\bibfnamefont {J.-P.}\ \bibnamefont {Hugonin}}, \bibinfo {author}
  {\bibfnamefont {D.~A.~R.}\ \bibnamefont {Dalvit}}, \bibinfo {author}
  {\bibfnamefont {J.-J.}\ \bibnamefont {Greffet}}, \bibinfo {author}
  {\bibfnamefont {A.}~\bibnamefont {Lambrecht}}, \ and\ \bibinfo {author}
  {\bibfnamefont {S.}~\bibnamefont {Reynaud}},\ }\href@noop {} {\bibfield
  {journal} {\bibinfo  {journal} {Phys. Rev. B}\ }\textbf {\bibinfo {volume}
  {85}},\ \bibinfo {pages} {180301} (\bibinfo {year} {2012})}\BibitemShut
  {NoStop}%
\bibitem [{\citenamefont {Incardone}\ \emph {et~al.}(2014)\citenamefont
  {Incardone}, \citenamefont {Emig},\ and\ \citenamefont
  {Kr{\"u}ger}}]{incardone2014heat}%
  \BibitemOpen
  \bibfield  {author} {\bibinfo {author} {\bibfnamefont {R.}~\bibnamefont
  {Incardone}}, \bibinfo {author} {\bibfnamefont {T.}~\bibnamefont {Emig}}, \
  and\ \bibinfo {author} {\bibfnamefont {M.}~\bibnamefont {Kr{\"u}ger}},\
  }\href@noop {} {\bibfield  {journal} {\bibinfo  {journal} {Europhysics
  Letters}\ }\textbf {\bibinfo {volume} {106}},\ \bibinfo {pages} {41001}
  (\bibinfo {year} {2014})}\BibitemShut {NoStop}%
\bibitem [{\citenamefont {Zhu}\ and\ \citenamefont {Fan}(2014)}]{zhu2014near}%
  \BibitemOpen
  \bibfield  {author} {\bibinfo {author} {\bibfnamefont {L.}~\bibnamefont
  {Zhu}}\ and\ \bibinfo {author} {\bibfnamefont {S.}~\bibnamefont {Fan}},\
  }\href@noop {} {\bibfield  {journal} {\bibinfo  {journal} {Phys. Rev. B}\
  }\textbf {\bibinfo {volume} {90}},\ \bibinfo {pages} {220301} (\bibinfo
  {year} {2014})}\BibitemShut {NoStop}%
\bibitem [{\citenamefont {Guo}\ and\ \citenamefont
  {Jacob}(2014)}]{guo2014fluctuational}%
  \BibitemOpen
  \bibfield  {author} {\bibinfo {author} {\bibfnamefont {Y.}~\bibnamefont
  {Guo}}\ and\ \bibinfo {author} {\bibfnamefont {Z.}~\bibnamefont {Jacob}},\
  }\href@noop {} {\bibfield  {journal} {\bibinfo  {journal} {Journal of Applied
  Physics}\ }\textbf {\bibinfo {volume} {115}},\ \bibinfo {eid} {234306}
  (\bibinfo {year} {2014})}\BibitemShut {NoStop}%
\bibitem [{\citenamefont {Chen}\ \emph {et~al.}(2015)\citenamefont {Chen},
  \citenamefont {Santhanam}, \citenamefont {Sandhu}, \citenamefont {Zhu},\ and\
  \citenamefont {Fan}}]{chen2015heat}%
  \BibitemOpen
  \bibfield  {author} {\bibinfo {author} {\bibfnamefont {K.}~\bibnamefont
  {Chen}}, \bibinfo {author} {\bibfnamefont {P.}~\bibnamefont {Santhanam}},
  \bibinfo {author} {\bibfnamefont {S.}~\bibnamefont {Sandhu}}, \bibinfo
  {author} {\bibfnamefont {L.}~\bibnamefont {Zhu}}, \ and\ \bibinfo {author}
  {\bibfnamefont {S.}~\bibnamefont {Fan}},\ }\href@noop {} {\bibfield
  {journal} {\bibinfo  {journal} {Phys. Rev. B}\ }\textbf {\bibinfo {volume}
  {91}},\ \bibinfo {pages} {134301} (\bibinfo {year} {2015})}\BibitemShut
  {NoStop}%
\bibitem [{\citenamefont {Chen}\ \emph {et~al.}(2016)\citenamefont {Chen},
  \citenamefont {Santhanam},\ and\ \citenamefont {Fan}}]{chen2016near}%
  \BibitemOpen
  \bibfield  {author} {\bibinfo {author} {\bibfnamefont {K.}~\bibnamefont
  {Chen}}, \bibinfo {author} {\bibfnamefont {P.}~\bibnamefont {Santhanam}}, \
  and\ \bibinfo {author} {\bibfnamefont {S.}~\bibnamefont {Fan}},\ }\href@noop
  {} {\bibfield  {journal} {\bibinfo  {journal} {Phys. Rev. Applied}\ }\textbf
  {\bibinfo {volume} {6}},\ \bibinfo {pages} {024014} (\bibinfo {year}
  {2016})}\BibitemShut {NoStop}%
\bibitem [{\citenamefont {Kittel}\ \emph {et~al.}(2005)\citenamefont {Kittel},
  \citenamefont {M\"uller-Hirsch}, \citenamefont {Parisi}, \citenamefont
  {Biehs}, \citenamefont {Reddig},\ and\ \citenamefont
  {Holthaus}}]{kittel2005near}%
  \BibitemOpen
  \bibfield  {author} {\bibinfo {author} {\bibfnamefont {A.}~\bibnamefont
  {Kittel}}, \bibinfo {author} {\bibfnamefont {W.}~\bibnamefont
  {M\"uller-Hirsch}}, \bibinfo {author} {\bibfnamefont {J.}~\bibnamefont
  {Parisi}}, \bibinfo {author} {\bibfnamefont {S.-A.}\ \bibnamefont {Biehs}},
  \bibinfo {author} {\bibfnamefont {D.}~\bibnamefont {Reddig}}, \ and\ \bibinfo
  {author} {\bibfnamefont {M.}~\bibnamefont {Holthaus}},\ }\href@noop {}
  {\bibfield  {journal} {\bibinfo  {journal} {Phys. Rev. Lett.}\ }\textbf
  {\bibinfo {volume} {95}},\ \bibinfo {pages} {224301} (\bibinfo {year}
  {2005})}\BibitemShut {NoStop}%
\bibitem [{\citenamefont {Shen}\ \emph {et~al.}(2009)\citenamefont {Shen},
  \citenamefont {Narayanaswamy},\ and\ \citenamefont {Chen}}]{shen2009surface}%
  \BibitemOpen
  \bibfield  {author} {\bibinfo {author} {\bibfnamefont {S.}~\bibnamefont
  {Shen}}, \bibinfo {author} {\bibfnamefont {A.}~\bibnamefont {Narayanaswamy}},
  \ and\ \bibinfo {author} {\bibfnamefont {G.}~\bibnamefont {Chen}},\
  }\href@noop {} {\bibfield  {journal} {\bibinfo  {journal} {Nano letters}\
  }\textbf {\bibinfo {volume} {9}},\ \bibinfo {pages} {2909} (\bibinfo {year}
  {2009})}\BibitemShut {NoStop}%
\bibitem [{\citenamefont {Rousseau}\ \emph {et~al.}(2009)\citenamefont
  {Rousseau}, \citenamefont {Siria}, \citenamefont {Jourdan}, \citenamefont
  {Volz}, \citenamefont {Comin}, \citenamefont {Chevrier},\ and\ \citenamefont
  {Greffet}}]{rousseau2009radiative}%
  \BibitemOpen
  \bibfield  {author} {\bibinfo {author} {\bibfnamefont {E.}~\bibnamefont
  {Rousseau}}, \bibinfo {author} {\bibfnamefont {A.}~\bibnamefont {Siria}},
  \bibinfo {author} {\bibfnamefont {G.}~\bibnamefont {Jourdan}}, \bibinfo
  {author} {\bibfnamefont {S.}~\bibnamefont {Volz}}, \bibinfo {author}
  {\bibfnamefont {F.}~\bibnamefont {Comin}}, \bibinfo {author} {\bibfnamefont
  {J.}~\bibnamefont {Chevrier}}, \ and\ \bibinfo {author} {\bibfnamefont
  {J.-J.}\ \bibnamefont {Greffet}},\ }\href@noop {} {\bibfield  {journal}
  {\bibinfo  {journal} {Nature Photonics}\ }\textbf {\bibinfo {volume} {3}},\
  \bibinfo {pages} {514} (\bibinfo {year} {2009})}\BibitemShut {NoStop}%
\bibitem [{\citenamefont {Ottens}\ \emph {et~al.}(2011)\citenamefont {Ottens},
  \citenamefont {Quetschke}, \citenamefont {Wise}, \citenamefont {Alemi},
  \citenamefont {Lundock}, \citenamefont {Mueller}, \citenamefont {Reitze},
  \citenamefont {Tanner},\ and\ \citenamefont {Whiting}}]{ottens2011nearfield}%
  \BibitemOpen
  \bibfield  {author} {\bibinfo {author} {\bibfnamefont {R.~S.}\ \bibnamefont
  {Ottens}}, \bibinfo {author} {\bibfnamefont {V.}~\bibnamefont {Quetschke}},
  \bibinfo {author} {\bibfnamefont {S.}~\bibnamefont {Wise}}, \bibinfo {author}
  {\bibfnamefont {A.~A.}\ \bibnamefont {Alemi}}, \bibinfo {author}
  {\bibfnamefont {R.}~\bibnamefont {Lundock}}, \bibinfo {author} {\bibfnamefont
  {G.}~\bibnamefont {Mueller}}, \bibinfo {author} {\bibfnamefont {D.~H.}\
  \bibnamefont {Reitze}}, \bibinfo {author} {\bibfnamefont {D.~B.}\
  \bibnamefont {Tanner}}, \ and\ \bibinfo {author} {\bibfnamefont {B.~F.}\
  \bibnamefont {Whiting}},\ }\href@noop {} {\bibfield  {journal} {\bibinfo
  {journal} {Phys. Rev. Lett.}\ }\textbf {\bibinfo {volume} {107}},\ \bibinfo
  {pages} {014301} (\bibinfo {year} {2011})}\BibitemShut {NoStop}%
\bibitem [{\citenamefont {Kajihara}\ \emph {et~al.}(2011)\citenamefont
  {Kajihara}, \citenamefont {Kosaka},\ and\ \citenamefont
  {Komiyama}}]{yusuke2011infrared}%
  \BibitemOpen
  \bibfield  {author} {\bibinfo {author} {\bibfnamefont {Y.}~\bibnamefont
  {Kajihara}}, \bibinfo {author} {\bibfnamefont {K.}~\bibnamefont {Kosaka}}, \
  and\ \bibinfo {author} {\bibfnamefont {S.}~\bibnamefont {Komiyama}},\
  }\href@noop {} {\bibfield  {journal} {\bibinfo  {journal} {Opt. Express}\
  }\textbf {\bibinfo {volume} {19}},\ \bibinfo {pages} {7695} (\bibinfo {year}
  {2011})}\BibitemShut {NoStop}%
\bibitem [{\citenamefont {Kim}\ \emph {et~al.}(2015)\citenamefont {Kim},
  \citenamefont {Song}, \citenamefont {Fern{\'a}ndez-Hurtado}, \citenamefont
  {Lee}, \citenamefont {Jeong}, \citenamefont {Cui}, \citenamefont {Thompson},
  \citenamefont {Feist}, \citenamefont {Reid}, \citenamefont
  {Garc{\'\i}a-Vidal} \emph {et~al.}}]{kim2015radiative}%
  \BibitemOpen
  \bibfield  {author} {\bibinfo {author} {\bibfnamefont {K.}~\bibnamefont
  {Kim}}, \bibinfo {author} {\bibfnamefont {B.}~\bibnamefont {Song}}, \bibinfo
  {author} {\bibfnamefont {V.}~\bibnamefont {Fern{\'a}ndez-Hurtado}}, \bibinfo
  {author} {\bibfnamefont {W.}~\bibnamefont {Lee}}, \bibinfo {author}
  {\bibfnamefont {W.}~\bibnamefont {Jeong}}, \bibinfo {author} {\bibfnamefont
  {L.}~\bibnamefont {Cui}}, \bibinfo {author} {\bibfnamefont {D.}~\bibnamefont
  {Thompson}}, \bibinfo {author} {\bibfnamefont {J.}~\bibnamefont {Feist}},
  \bibinfo {author} {\bibfnamefont {M.~H.}\ \bibnamefont {Reid}}, \bibinfo
  {author} {\bibfnamefont {F.~J.}\ \bibnamefont {Garc{\'\i}a-Vidal}},  \emph
  {et~al.},\ }\href@noop {} {\bibfield  {journal} {\bibinfo  {journal}
  {Nature}\ } (\bibinfo {year} {2015})}\BibitemShut {NoStop}%
\bibitem [{\citenamefont {Kirchhoff}(1860)}]{kirchhoff1860ueber}%
  \BibitemOpen
  \bibfield  {author} {\bibinfo {author} {\bibfnamefont {G.}~\bibnamefont
  {Kirchhoff}},\ }\href@noop {} {\bibfield  {journal} {\bibinfo  {journal}
  {Annalen der Physik}\ }\textbf {\bibinfo {volume} {185}},\ \bibinfo {pages}
  {275} (\bibinfo {year} {1860})}\BibitemShut {NoStop}%
\bibitem [{\citenamefont {Antezza}(2006)}]{antezza2006surface}%
  \BibitemOpen
  \bibfield  {author} {\bibinfo {author} {\bibfnamefont {M.}~\bibnamefont
  {Antezza}},\ }\href@noop {} {\bibfield  {journal} {\bibinfo  {journal}
  {Journal of Physics A: Mathematical and General}\ }\textbf {\bibinfo {volume}
  {39}},\ \bibinfo {pages} {6117} (\bibinfo {year} {2006})}\BibitemShut
  {NoStop}%
\bibitem [{\citenamefont {Narayana}\ \emph {et~al.}(2013)\citenamefont
  {Narayana}, \citenamefont {Savo},\ and\ \citenamefont
  {Sato}}]{narayana2013transient}%
  \BibitemOpen
  \bibfield  {author} {\bibinfo {author} {\bibfnamefont {S.}~\bibnamefont
  {Narayana}}, \bibinfo {author} {\bibfnamefont {S.}~\bibnamefont {Savo}}, \
  and\ \bibinfo {author} {\bibfnamefont {Y.}~\bibnamefont {Sato}},\ }\href@noop
  {} {\bibfield  {journal} {\bibinfo  {journal} {Applied Physics Letters}\
  }\textbf {\bibinfo {volume} {102}},\ \bibinfo {pages} {201904} (\bibinfo
  {year} {2013})}\BibitemShut {NoStop}%
\bibitem [{\citenamefont {Davis}\ and\ \citenamefont
  {Hussein}(2014)}]{davisn2014nanophonic}%
  \BibitemOpen
  \bibfield  {author} {\bibinfo {author} {\bibfnamefont {B.~L.}\ \bibnamefont
  {Davis}}\ and\ \bibinfo {author} {\bibfnamefont {M.~I.}\ \bibnamefont
  {Hussein}},\ }\href@noop {} {\bibfield  {journal} {\bibinfo  {journal} {Phys.
  Rev. Lett.}\ }\textbf {\bibinfo {volume} {112}},\ \bibinfo {pages} {055505}
  (\bibinfo {year} {2014})}\BibitemShut {NoStop}%
\bibitem [{\citenamefont {Jackson}(1999)}]{jackson1999classical}%
  \BibitemOpen
  \bibfield  {author} {\bibinfo {author} {\bibfnamefont {J.~D.}\ \bibnamefont
  {Jackson}},\ }\href@noop {} {\emph {\bibinfo {title} {Classical
  electrodynamics}}}\ (\bibinfo  {publisher} {Wiley},\ \bibinfo {year}
  {1999})\BibitemShut {NoStop}%
\bibitem [{\citenamefont {Tsang}\ \emph {et~al.}(2004)\citenamefont {Tsang},
  \citenamefont {Kong},\ and\ \citenamefont {Ding}}]{tsang2004scattering}%
  \BibitemOpen
  \bibfield  {author} {\bibinfo {author} {\bibfnamefont {L.}~\bibnamefont
  {Tsang}}, \bibinfo {author} {\bibfnamefont {J.~A.}\ \bibnamefont {Kong}}, \
  and\ \bibinfo {author} {\bibfnamefont {K.-H.}\ \bibnamefont {Ding}},\
  }\href@noop {} {\emph {\bibinfo {title} {Scattering of Electromagnetic Waves,
  Theories and Applications}}},\ Vol.~\bibinfo {volume} {27}\ (\bibinfo
  {publisher} {John Wiley \& Sons},\ \bibinfo {year} {2004})\BibitemShut
  {NoStop}%
\bibitem [{\citenamefont {Rahi}\ \emph {et~al.}(2009)\citenamefont {Rahi},
  \citenamefont {Emig}, \citenamefont {Graham}, \citenamefont {Jaffe},\ and\
  \citenamefont {Kardar}}]{rahi2009scattering}%
  \BibitemOpen
  \bibfield  {author} {\bibinfo {author} {\bibfnamefont {S.~J.}\ \bibnamefont
  {Rahi}}, \bibinfo {author} {\bibfnamefont {T.}~\bibnamefont {Emig}}, \bibinfo
  {author} {\bibfnamefont {N.}~\bibnamefont {Graham}}, \bibinfo {author}
  {\bibfnamefont {R.~L.}\ \bibnamefont {Jaffe}}, \ and\ \bibinfo {author}
  {\bibfnamefont {M.}~\bibnamefont {Kardar}},\ }\href@noop {} {\bibfield
  {journal} {\bibinfo  {journal} {Phys. Rev. D}\ }\textbf {\bibinfo {volume}
  {80}},\ \bibinfo {pages} {085021} (\bibinfo {year} {2009})}\BibitemShut
  {NoStop}%
\bibitem [{\citenamefont {Stancil}(2012)}]{stancil2012theory}%
  \BibitemOpen
  \bibfield  {author} {\bibinfo {author} {\bibfnamefont {D.~D.}\ \bibnamefont
  {Stancil}},\ }\href@noop {} {\emph {\bibinfo {title} {Theory of magnetostatic
  waves}}}\ (\bibinfo  {publisher} {Springer Science \& Business Media},\
  \bibinfo {year} {2012})\BibitemShut {NoStop}%
\bibitem [{\citenamefont {Lippmann}\ and\ \citenamefont
  {Schwinger}(1950)}]{lippmann1950variational}%
  \BibitemOpen
  \bibfield  {author} {\bibinfo {author} {\bibfnamefont {B.~A.}\ \bibnamefont
  {Lippmann}}\ and\ \bibinfo {author} {\bibfnamefont {J.}~\bibnamefont
  {Schwinger}},\ }\href@noop {} {\bibfield  {journal} {\bibinfo  {journal}
  {Phys. Rev.}\ }\textbf {\bibinfo {volume} {79}},\ \bibinfo {pages} {469}
  (\bibinfo {year} {1950})}\BibitemShut {NoStop}%
\bibitem [{\citenamefont {Kattawar}\ and\ \citenamefont
  {Eisner}(1970)}]{Kattawar70}%
  \BibitemOpen
  \bibfield  {author} {\bibinfo {author} {\bibfnamefont {G.~W.}\ \bibnamefont
  {Kattawar}}\ and\ \bibinfo {author} {\bibfnamefont {M.}~\bibnamefont
  {Eisner}},\ }\href@noop {} {\bibfield  {journal} {\bibinfo  {journal} {Appl.
  Opt.}\ }\textbf {\bibinfo {volume} {9}},\ \bibinfo {pages} {2685} (\bibinfo
  {year} {1970})}\BibitemShut {NoStop}%
\bibitem [{\citenamefont {Bohren}\ and\ \citenamefont
  {Huffman}(2008)}]{bohren2008absorption}%
  \BibitemOpen
  \bibfield  {author} {\bibinfo {author} {\bibfnamefont {C.~F.}\ \bibnamefont
  {Bohren}}\ and\ \bibinfo {author} {\bibfnamefont {D.~R.}\ \bibnamefont
  {Huffman}},\ }\href@noop {} {\emph {\bibinfo {title} {Absorption and
  scattering of light by small particles}}}\ (\bibinfo  {publisher} {John Wiley
  \& Sons},\ \bibinfo {year} {2008})\BibitemShut {NoStop}%
\bibitem [{\citenamefont {Levin}\ \emph {et~al.}(1980)\citenamefont {Levin},
  \citenamefont {Polevoy},\ and\ \citenamefont {Rytov}}]{levin1980zh}%
  \BibitemOpen
  \bibfield  {author} {\bibinfo {author} {\bibfnamefont {M.}~\bibnamefont
  {Levin}}, \bibinfo {author} {\bibfnamefont {V.}~\bibnamefont {Polevoy}}, \
  and\ \bibinfo {author} {\bibfnamefont {S.}~\bibnamefont {Rytov}},\
  }\href@noop {} {\bibfield  {journal} {\bibinfo  {journal} {JETP}\ }\textbf
  {\bibinfo {volume} {52}},\ \bibinfo {pages} {1054} (\bibinfo {year}
  {1980})}\BibitemShut {NoStop}%
\bibitem [{\citenamefont {Pendry}(1999)}]{pendry1999radiative}%
  \BibitemOpen
  \bibfield  {author} {\bibinfo {author} {\bibfnamefont {J.}~\bibnamefont
  {Pendry}},\ }\href@noop {} {\bibfield  {journal} {\bibinfo  {journal}
  {Journal of Physics: Condensed Matter}\ }\textbf {\bibinfo {volume} {11}},\
  \bibinfo {pages} {6621} (\bibinfo {year} {1999})}\BibitemShut {NoStop}%
\bibitem [{\citenamefont {Spitzer}\ \emph {et~al.}(1959)\citenamefont
  {Spitzer}, \citenamefont {Kleinman},\ and\ \citenamefont
  {Walsh}}]{spitzer1959infrared}%
  \BibitemOpen
  \bibfield  {author} {\bibinfo {author} {\bibfnamefont {W.~G.}\ \bibnamefont
  {Spitzer}}, \bibinfo {author} {\bibfnamefont {D.}~\bibnamefont {Kleinman}}, \
  and\ \bibinfo {author} {\bibfnamefont {D.}~\bibnamefont {Walsh}},\
  }\href@noop {} {\bibfield  {journal} {\bibinfo  {journal} {Phys. Rev.}\
  }\textbf {\bibinfo {volume} {113}},\ \bibinfo {pages} {127} (\bibinfo {year}
  {1959})}\BibitemShut {NoStop}%
\bibitem [{\citenamefont {Hippel}\ \emph {et~al.}(1954)\citenamefont {Hippel}
  \emph {et~al.}}]{hippel1954dielectric}%
  \BibitemOpen
  \bibfield  {author} {\bibinfo {author} {\bibfnamefont {A.~v.}\ \bibnamefont
  {Hippel}} \emph {et~al.},\ }\href@noop {} {\bibfield  {journal} {\bibinfo
  {journal} {London: Artech House}\ } (\bibinfo {year} {1954})}\BibitemShut
  {NoStop}%
\bibitem [{\citenamefont {Jennings}(1988)}]{jennings1988the}%
  \BibitemOpen
  \bibfield  {author} {\bibinfo {author} {\bibfnamefont {S.}~\bibnamefont
  {Jennings}},\ }\href@noop {} {\bibfield  {journal} {\bibinfo  {journal}
  {Journal of Aerosol Science}\ }\textbf {\bibinfo {volume} {19}},\ \bibinfo
  {pages} {159 } (\bibinfo {year} {1988})}\BibitemShut {NoStop}%
\bibitem [{\citenamefont {Miller}\ \emph {et~al.}(2015)\citenamefont {Miller},
  \citenamefont {Johnson},\ and\ \citenamefont {Rodriguez}}]{miller2015shape}%
  \BibitemOpen
  \bibfield  {author} {\bibinfo {author} {\bibfnamefont {O.~D.}\ \bibnamefont
  {Miller}}, \bibinfo {author} {\bibfnamefont {S.~G.}\ \bibnamefont {Johnson}},
  \ and\ \bibinfo {author} {\bibfnamefont {A.~W.}\ \bibnamefont {Rodriguez}},\
  }\href@noop {} {\bibfield  {journal} {\bibinfo  {journal} {Phys. Rev. Lett.}\
  }\textbf {\bibinfo {volume} {115}},\ \bibinfo {pages} {204302} (\bibinfo
  {year} {2015})}\BibitemShut {NoStop}%
\bibitem [{\citenamefont {Doyeux}\ \emph {et~al.}(2016)\citenamefont {Doyeux},
  \citenamefont {Leggio}, \citenamefont {Messina},\ and\ \citenamefont
  {Antezza}}]{doyeux2016quantum}%
  \BibitemOpen
  \bibfield  {author} {\bibinfo {author} {\bibfnamefont {P.}~\bibnamefont
  {Doyeux}}, \bibinfo {author} {\bibfnamefont {B.}~\bibnamefont {Leggio}},
  \bibinfo {author} {\bibfnamefont {R.}~\bibnamefont {Messina}}, \ and\
  \bibinfo {author} {\bibfnamefont {M.}~\bibnamefont {Antezza}},\ }\href@noop
  {} {\bibfield  {journal} {\bibinfo  {journal} {Phys. Rev. E}\ }\textbf
  {\bibinfo {volume} {93}},\ \bibinfo {pages} {022134} (\bibinfo {year}
  {2016})}\BibitemShut {NoStop}%
\bibitem [{\citenamefont {Bellomo}\ and\ \citenamefont
  {Antezza}(2015)}]{bellomo2015nonequilibrium}%
  \BibitemOpen
  \bibfield  {author} {\bibinfo {author} {\bibfnamefont {B.}~\bibnamefont
  {Bellomo}}\ and\ \bibinfo {author} {\bibfnamefont {M.}~\bibnamefont
  {Antezza}},\ }\href@noop {} {\bibfield  {journal} {\bibinfo  {journal} {Phys.
  Rev. A}\ }\textbf {\bibinfo {volume} {91}},\ \bibinfo {pages} {042124}
  (\bibinfo {year} {2015})}\BibitemShut {NoStop}%
\bibitem [{\citenamefont {Ekstein}\ and\ \citenamefont
  {Rostoker}(1955)}]{ekstein1955quantum}%
  \BibitemOpen
  \bibfield  {author} {\bibinfo {author} {\bibfnamefont {H.}~\bibnamefont
  {Ekstein}}\ and\ \bibinfo {author} {\bibfnamefont {N.}~\bibnamefont
  {Rostoker}},\ }\href@noop {} {\bibfield  {journal} {\bibinfo  {journal}
  {Phys. Rev.}\ }\textbf {\bibinfo {volume} {100}},\ \bibinfo {pages} {1023}
  (\bibinfo {year} {1955})}\BibitemShut {NoStop}%
\bibitem [{\citenamefont {Landau}\ and\ \citenamefont
  {Lifshitz}(2013)}]{landau2013course}%
  \BibitemOpen
  \bibfield  {author} {\bibinfo {author} {\bibfnamefont {L.~D.}\ \bibnamefont
  {Landau}}\ and\ \bibinfo {author} {\bibfnamefont {E.~M.}\ \bibnamefont
  {Lifshitz}},\ }\href@noop {} {\emph {\bibinfo {title} {Course of theoretical
  physics}}}\ (\bibinfo  {publisher} {Elsevier},\ \bibinfo {year}
  {2013})\BibitemShut {NoStop}%
\bibitem [{\citenamefont {Zhang}(2011)}]{zhang2011matrix}%
  \BibitemOpen
  \bibfield  {author} {\bibinfo {author} {\bibfnamefont {F.}~\bibnamefont
  {Zhang}},\ }\href@noop {} {\emph {\bibinfo {title} {Matrix theory: basic
  results and techniques}}}\ (\bibinfo  {publisher} {Springer Science \&
  Business Media},\ \bibinfo {year} {2011})\BibitemShut {NoStop}%
\bibitem [{\citenamefont {Frisch}(1968)}]{boyce1968probabilistic}%
  \BibitemOpen
  \bibinfo {editor} {\bibfnamefont {U.}~\bibnamefont {Frisch}},\ ed.,\
  \href@noop {} {\emph {\bibinfo {title} {``Wave propagation in random media''
  in Probabilistic methods in applied mathematics.}}},\ Vol.~\bibinfo {volume}
  {1}\ (\bibinfo  {publisher} {Acad. Pr.},\ \bibinfo {year} {1968})\BibitemShut
  {NoStop}%
\bibitem [{\citenamefont {Fest}(2008)}]{fest2008modeling}%
  \BibitemOpen
  \bibfield  {author} {\bibinfo {author} {\bibfnamefont {E.~C.}\ \bibnamefont
  {Fest}},\ }in\ \href@noop {} {\emph {\bibinfo {booktitle} {Optical
  Engineering+ Applications}}}\ (\bibinfo {organization} {International Society
  for Optics and Photonics},\ \bibinfo {year} {2008})\ pp.\ \bibinfo {pages}
  {70650B--70650B}\BibitemShut {NoStop}%
\end{thebibliography}
\end{document}